\documentclass[camera,letterpaper,nomarginnotes,nonarrowgutter]{jpaper}
\newif\ifdraft
\draftfalse

\usepackage{tikz}
\usepackage{mathptmx} 
\usepackage{amsmath,amssymb,amsfonts}
\usepackage{comment}
\usepackage{graphics}
\usepackage{fancyhdr}
\usepackage{booktabs}
\usepackage{pifont}
\usepackage[normalem]{ulem}
\usepackage{cite}
\usepackage{hhline}
\usepackage{xcolor}
\usepackage{setspace}
\usepackage{textcomp}
\usepackage{enumitem}
\usepackage{afterpage}
\usepackage{graphicx}
\usepackage[htt]{hyphenat}
\usepackage{makecell}
\usepackage{xspace}
\usepackage{listings}
\usepackage{multirow}
\usepackage{balance}
\usepackage{glossaries} 
\usepackage{siunitx}
\usepackage{duckuments} 
\usepackage{dblfloatfix} 
\usepackage{subcaption}
\usepackage{rotating}
\usepackage[us,12hr]{datetime}
\usepackage[en-GB, useregional=numeric]{datetime2}

\usepackage[compact]{titlesec}
\usepackage[colorinlistoftodos,prependcaption,textsize=small]{todonotes} 
\usepackage{setspace}
\usepackage{algorithm2e}
\usepackage{soul}
\usepackage{bm}

\usepackage[bookmarks=true,breaklinks=true,hidelinks]{hyperref}

\newif\ifarxiv 
\arxivfalse

\newif\ifcameraready
\camerareadyfalse

\newif\ifrev
\revfalse

\newif\ifpagenumbers
\pagenumberstrue

\newcounter{version}
\ifcameraready
\else
\fi

\newcommand{\hpcayear}{2025}

\newcommand{\affilETH}[0]{\textsuperscript{\S}}
\newcommand{\affilETU}[0]{\textsuperscript{$\dagger$}}
\newcommand{\affilUOS}[0]{\textsuperscript{$\ddagger$}}
\author{
{O\u{g}uzhan Canpolat\affilETH\affilETU}\qquad%
{A. Giray Ya\u{g}l{\i}kç{\i}\affilETH}\qquad%
{Geraldo F. Oliveira\affilETH}\qquad%
{Ataberk Olgun\affilETH}\\
{Nisa Bostancı\affilETH}\qquad%
{Ismail Emir Yuksel\affilETH}\qquad%
{Haocong Luo\affilETH}\qquad%
{O\u{g}uz Ergin\affilUOS\affilETU}\qquad%
{Onur Mutlu\affilETH}\\
{\affilETH \emph{ETH Z{\"u}rich}}\qquad{}\affilETU \emph{TOBB University of Economics and Technology}\qquad{}\affilUOS \emph{University of Sharjah}
}

\newcommand{\rfmbo}[0]{RFM+BO}
\newacronym{rfmbo}{\rfmbo{}}{refresh management with back-off support}

\newcommand{\rfmth}[0]{RFM_{th}}
\newcommand{\rfmab}[0]{RFM_{ab}}
\newcommand{\rfmsb}[0]{RFM_{sb}}
\newcommand{\nrh}[0]{N_{RH}}
\newcommand{\aboth}[0]{N_{BO}}
\newcommand{\taboact}[0]{t_{ABO_{ACT}}}
\newcommand{\tbodelay}[0]{t_{BackOffDelay}}
\newcommand{\bonrefs}[0]{N_{Ref}}
\newcommand{\bonacts}[0]{N_{Delay}}

\newcommand{\maxact}[0]{\text{MAX}_{\text{ACT}}}
\newcommand{\maxrfm}[0]{\text{MAX}_{\text{RFM}}}
\newcommand{\dallref}[0]{D_{\text{refresh}}}
\newcommand{\trfmperiod}[0]{T_{\text{RFM}}}

\newcommand{\dbc}[0]{\textit{DBC}}
\newcommand{\anbot}[0]{\textit{BOT}}
\newcommand{\advanal}[0]{P_{\textit{ADV}}}
\newcommand{\advhyp}[0]{P_{\textit{HYP}}}

\newcommand{\trcd}[0]{t_{RCD}}
\newcommand{\tras}[0]{t_{RAS}}
\newcommand{\trp}[0]{t_{RP}}
\newcommand{\trc}[0]{t_{RC}}
\newcommand{\trefi}[0]{t_{REFI}}
\newcommand{\trefw}[0]{t_{REFW}}
\newcommand{\trfc}[0]{t_{RFC}}
\newcommand{\trrd}[0]{t_{RRD}}
\newcommand{\trfm}[0]{t_{RFM}}
\newcommand{\trtp}[0]{t_{RTP}}
\newcommand{\twr}[0]{t_{WR}}

\newacronym{maxact}{$\maxact{}$}{maximum activation commands possible within a refresh window}
\newacronym{maxrfm}{$\maxrfm{}$}{maximum refresh management commands possible within a refresh window}
\newacronym{nrh}{$\nrh{}$}{\emph{minimum hammer count to induce the first bitflip}}
\newacronym{dallref}{$\dallref{}$}{time taken to complete refresh commands within a refresh window}
\newacronym{trfmperiod}{$\trfmperiod{}$}{the minimum time needed between two consecutive \gls{RFM} commands targeting the same bank}

\newacronym{trefw}{$\trefw{}$}{refresh window}
\newacronym{trefi}{$\trefi{}$}{refresh interval}
\newacronym{trfm}{$\trfm{}$}{refresh management latency}
\newacronym{trfc}{$\trfc{}$}{refresh latency}

\newacronym{rfm}{RFM}{refresh management}
\newacronym{prfm}{PRFM}{Periodic RFM}
\newacronym{rfmth}{$\rfmth{}$}{\emph{bank activation threshold to issue an RFM command}}
\newacronym{rfmab}{$\rfmab{}$}{all bank refresh management}
\newacronym{rfmsb}{$\rfmsb{}$}{same bank refresh management}
\newacronym{prac}{PRAC}{Per Row Activation Counting}
\newacronym{abo}{$ABO$}{Alert Back-Off}
\newacronym{aboth}{$\aboth{}$}{the back-off threshold}
\newacronym{taboact}{$\taboact{}$}{\emph{the window of normal traffic}}
\newacronym{tbodelay}{$\tbodelay{}$}{the delay until a new back-off can be initiated}
\newacronym{trc}{$\trc{}$}{row cycle time}

\newacronym{trcd}{$\trcd{}$}{row activation latency}
\newacronym{tras}{$\tras{}$}{charge restoration latency}
\newacronym{trp}{$\trp{}$}{precharge latency}
\newacronym{trrd}{$\trrd{}$}{the minimum time window between two row activation commands targeting different banks}

\newcommand{\tfilter}[0]{T_{Filter}}
\newacronym{tfilter}{$\tfilter{}$}{the filtering threshold}

\newcommand{\act}[0]{ACT}
\newcommand{\pre}[0]{PRE}
\newcommand{\refresh}[0]{REF}
\newcommand{\wri}[0]{\texttt{WR}}
\newcommand{\rd}[0]{\texttt{RD}}

\newcommand{\apa}[0]{\texttt{APA}}

\newcommand{\pum}[0]{\texttt{PuM}}

\usepackage[shortcuts]{extdash}

\newacronym{iqr}{$IQR$}{inter-quartile range}
\newacronym{act}{\act{}}{activate}
\newacronym{pre}{\pre{}}{precharge}
\newacronym{ref}{\refresh{}}{refresh}
\newacronym{wr}{\wri{}}{write}
\newacronym{rd}{\rd{}}{read}
\newacronym{pum}{\pum{}}{Processing-using-Memory}
\newacronym{apa}{\apa{}}{\act{} $\rightarrow$ \pre{} $\rightarrow$ \act{}}
\newacronym{jedec}{JEDEC}{Joint Electron Device Engineering Council}
\newacronym{tcl}{$t_{CL}$}{column access latency}
\newacronym{tcwl}{$t_{CWL}$}{column write latency}

\newacronym{puf}{PUF}{physical unclonable function}
\newacronym{trn}{TRN}{true random number}

\newacronym{taggon}{$t_{AggOn}$}{the time that an aggressor row stays active, {i.e., aggressor row's on-time}}

\definecolor{gfored}{rgb}{0.580, 0.050, 0.211}
\definecolor{ao}{rgb}{0.007, 0.520, 0.867}
\definecolor{moegi}{rgb}{0.357, 0.537, 0.188}
\definecolor{jl}{rgb}{1.0, 0.2, 0.8}
\definecolor{brown(web)}{rgb}{0.65, 0.16, 0.16}
\definecolor{bisque}{rgb}{1.0, 0.89, 0.77}
\definecolor{nbs}{rgb}{0.88, 0.07, 0.37}
\definecolor{yt}{rgb}{0.58, 0.44, 0.86}
\definecolor{iy}{rgb}{0.0, 0.36, 0.05}
\definecolor{burntorange}{rgb}{0.8, 0.33, 0.0}
\definecolor{ouscolor}{rgb}{0.25, 0.41, 0.88}

\newcommand*\circled[1]{\tikz[baseline=(char.base)]{%
  \node[shape=circle,fill,inner sep=0.5pt] (char) {\textcolor{white}{#1}};}%
}%

\newcommand{\ignore}[1]{}

\ifdraft
    
    \newcommand{\param}[1]{\textcolor{red}{#1}} 

\else

\fi

\definecolor{frenchblue}{rgb}{0.19, 0.55, 0.91}

\usepackage[most]{tcolorbox} 
\tcbset{before skip=1.5pt, after skip=4pt}

\newtcolorbox[auto counter]{obsx}[3][]{%
    colframe = #2!45,
    colback  = #2!10,
    coltitle = #2!20!black, 
    colbacktitle=#2!20,
    coltitle=black,
    fonttitle=\bfseries, 
    title=#3~\thetcbcounter.\ ,
    enhanced,
    attach boxed title to top left={yshift=-2.8mm, xshift=0.15cm},
    bottom=-2.2pt,
    #1%
}

\usepackage[most]{tcolorbox} 
\newtcolorbox[auto counter]{tkx}[2][]{%
    enhanced, breakable, center title,
    colframe = #2!45,
    colback  = #2!10,
    colbacktitle=#2!20,
    left=-0.5pt,
    right=-0.5pt,
    bottom=-2pt,
    top=-0.25pt,
    #1%
}
\newcounter{obs}
\definecolor{amber}{rgb}{1.0, 0.49, 0.0}
\definecolor{awesome}{rgb}{1.0, 0.13, 0.32}
\definecolor{dollarbill}{rgb}{0.52,0.73,0.4}
\definecolor{moegi}{rgb}{0.357, 0.537, 0.188}
\definecolor{burgundy}{rgb}{0.5, 0.0, 0.13}
\definecolor{ballblue}{rgb}{0.13, 0.67, 0.8}
\definecolor{ups-truck}{rgb}{0.53, 0.28, 0.21}
\definecolor{airforceblue}{rgb}{0.36, 0.54, 0.66}
\definecolor{cadmiumgreen}{rgb}{0.0, 0.42, 0.24}
\definecolor{darkcyan}{rgb}{0.0, 0.55, 0.55}
\definecolor{caribbeangreen}{rgb}{0.0, 0.8, 0.6}
\definecolor{flamingopink}{rgb}{0.99, 0.56, 0.67}
\definecolor{jazzberryjam}{rgb}{0.65, 0.04, 0.37}
\definecolor{mediumpersianblue}{rgb}{0.0, 0.4, 0.65}
\definecolor{coolblack}{rgb}{0.0, 0.18, 0.39}
\definecolor{bleudefrance}{rgb}{0.19, 0.55, 0.91}
\definecolor{ao}{rgb}{0.0, 0.0, 1.0}
\definecolor{babyblueeyes}{rgb}{0.63, 0.79, 0.95}
\definecolor{darkwarmgray}{rgb}{0.2,0,0}
\definecolor{brightpink}{rgb}{1.0, 0.0, 0.5}
\definecolor{iy}{rgb}{0.0, 0.36, 0.05}
\definecolor{nbcolor}{rgb}{1.0, 0.33, 0.64}

\newcommand{\squishlist}{
 \begin{list}{$\circ$}
  { \setlength{\itemsep}{0pt}
     \setlength{\parsep}{0pt}
     \setlength{\topsep}{0pt}
     \setlength{\partopsep}{0pt}
     \setlength{\leftmargin}{1em}
     \setlength{\labelwidth}{1em}
     \setlength{\labelsep}{0.5em} } }

\newcommand{\squishsublist}{
\begin{list}{$\rightarrow$}
 { \setlength{\itemsep}{0pt}
    \setlength{\parsep}{0pt}
    \setlength{\topsep}{-10em}
    \setlength{\partopsep}{-3pt}
    \setlength{\leftmargin}{1em}
    \setlength{\labelwidth}{1em}
    \setlength{\labelsep}{0.5em} } }

\newcommand{\squishend}{
  \end{list}  }

\newcommand{\head}[1]{\noindent\textbf{#1.}}

\newcounter{take}
\ifdraft
\paperwidth=\dimexpr \paperwidth + 4cm\relax
\oddsidemargin=\dimexpr\oddsidemargin + 2cm\relax
\evensidemargin=\dimexpr\evensidemargin + 2cm\relax
\marginparwidth=\dimexpr \marginparwidth + 2cm\relax
\fi
\newcommand{\gf}[2]{\ifnum#1=\value{version}\textcolor{red}{#2}\else{#2}\fi}
\newcommand{\agy}[2]{\ifnum#1=\value{version}\textcolor{blue}{#2}\else{#2}\fi}
\newcommand{\atb}[2]{\ifnum#1=\value{version}\textcolor{orange}{#2}\else{#2}\fi}
\newcommand{\yct}[2]{\ifnum#1=\value{version}\textcolor{yt}{#2}\else{#2}\fi}
\newcommand{\ous}[2]{\ifnum#1=\value{version}\textcolor{ouscolor}{#2}\else{#2}\fi}
\newcommand{\iey}[2]{\ifnum#1=\value{version}\textcolor{iy}{#2}\else{#2}\fi}
\newcommand{\nb}[2]{\ifnum#1=\value{version}\textcolor{nbcolor}{#2}\else{#2}\fi}
\newcommand{\hluo}[2]{\ifnum#1=\value{version}\textcolor{moegi}{#2}\else{#2}\fi}
\newcommand{\om}[2]{\ifnum#1=\value{version}\textcolor{gfored}{#2}\else#2\fi}

\newcommand{\agytodo}[2]{\ifnum#1=\value{version}\todo[size=\scriptsize, linecolor=orange, bordercolor=orange, backgroundcolor=white]{\textcolor{blue}{TODO:~#2}}\else{}\fi}
\newcommand{\ycttodo}[2]{\ifnum#1=\value{version}\todo[size=\scriptsize, linecolor=orange, bordercolor=orange, backgroundcolor=white]{\textcolor{yt}{TODO:~#2}}\else{}\fi}
\newcommand{\ieytodo}[2]{\ifnum#1=\value{version}\todo[size=\scriptsize, linecolor=orange, bordercolor=orange, backgroundcolor=white]{\textcolor{iey}{TODO:~#2}}\else{}\fi}

\newcommand{\agycomment}[2]{\ifnum#1=\value{version}\todo[size=\scriptsize, linecolor=orange, bordercolor=orange, backgroundcolor=white]{\textcolor{blue}{Giray: #2}}\else{}\fi}
\newcommand{\atbcomment}[2]{\ifnum#1=\value{version}\todo[size=\scriptsize, linecolor=orange, bordercolor=orange, backgroundcolor=white]{\textcolor{orange}{Atb: #2}}\else{}\fi}
\newcommand{\yctcomment}[2]{\ifnum#1=\value{version}\todo[size=\scriptsize, linecolor=orange, bordercolor=orange, backgroundcolor=white]{\textcolor{yt}{Yahya: #2}}\else{}\fi}
\newcommand{\ouscomment}[2]{\ifnum#1=\value{version}\todo[size=\scriptsize, linecolor=orange, bordercolor=orange, backgroundcolor=white]{\textcolor{ouscolor}{Oguzhan: #2}}\else{}\fi}
\newcommand{\gfcomment}[2]{\ifnum#1=\value{version}\todo[size=\scriptsize, linecolor=orange, bordercolor=orange, backgroundcolor=white]{\textcolor{purple}{Geraldo: #2}}\else{}\fi}
\newcommand{\ieycomment}[2]{\ifnum#1=\value{version}\todo[size=\scriptsize, linecolor=orange, bordercolor=orange, backgroundcolor=white]{\textcolor{iy}{Ismail: #2}}\else{}\fi}
\newcommand{\omcomment}[2]{\ifnum#1=\value{version}\todo[size=\scriptsize, linecolor=orange, bordercolor=orange, backgroundcolor=white]{\textcolor{gfored}{Onur: #2}}\else{}\fi}
\newcommand{\nbcomment}[2]{\ifnum#1=\value{version}\todo[size=\scriptsize, linecolor=orange, bordercolor=orange, backgroundcolor=white]{\textcolor{nbcolor}{Nisa: #2}}\else{}\fi}
\newcommand{\hluocomment}[2]{\ifnum#1=\value{version}\todo[size=\scriptsize, linecolor=orange, bordercolor=orange, backgroundcolor=white]{\textcolor{moegi}{Haocong: #2}}\else{}\fi}
\newcommand{\versionedparam}[2]{\ifnum#1=\value{version}{#2}\else{#2}\fi}
\providecommand{\param}[1]{\versionedparam{\value{version}}{#1}}

\newcommand{\X}[0]{Chronus}

\newcommand{\secref}[1]{§\ref{#1}}

\newcommand{\tabref}[1]{Table~\ref{#1}}

\newcommand{\figref}[1]{Fig.~\ref{#1}}
\newcommand{\figsref}[1]{Figs.~\ref{#1}}

\newcommand{\revtag}[1]{}

\newcommand{\rhmemisolationrefs}[0]{\cite{fournaris2017exploiting, poddebniak2018attacking, tatar2018throwhammer, carre2018openssl, barenghi2018software, zhang2018triggering, bhattacharya2018advanced, google-project-zero, kim2014flipping, rowhammergithub, seaborn2015exploiting, van2016drammer, gruss2016rowhammer, razavi2016flip, pessl2016drama, xiao2016one, bosman2016dedup, bhattacharya2016curious, burleson2016invited, qiao2016new, brasser2017can, jang2017sgx, aga2017good, mutlu2017rowhammer, tatar2018defeating, gruss2018another, lipp2018nethammer, van2018guardion, frigo2018grand, cojocar2019eccploit,  ji2019pinpoint, mutlu2019rowhammer, hong2019terminal, kwong2020rambleed, frigo2020trrespass, cojocar2020rowhammer, weissman2020jackhammer, zhang2020pthammer, yao2020deephammer, deridder2021smash, hassan2021utrr, jattke2022blacksmith, tol2022toward, kogler2022half, orosa2022spyhammer, zhang2022implicit, liu2022generating, cohen2022hammerscope, zheng2022trojvit, fahr2022frodo, tobah2022spechammer, rakin2022deepsteal, park2016statistical, park2016experiments,lim2017active, ryu2017overcoming, yun2018study, yang2019trap, walker2021ondramrowhammer, kim2020revisiting, orosa2021deeper, yaglikci2022understanding, khan2018analysis, agarwal2018rowhammer, li2014write, ni2018write, genssler2022reliability, mutlu2023fundamentally}}

\newcommand{\understandingRowHammerAllCitations}[0]{\cite{redeker2002investigation, kim2014flipping, park2014active, park2016statistical, yang2016suppression, park2016experiments,lim2017active, ryu2017overcoming, yang2017scanning, lim2018study, yun2018study, yang2019trap, gautam2019row, walker2021ondramrowhammer, kim2020revisiting, orosa2021deeper, jiang2021quantifying, orosa2022spyhammer, cohen2022hammerscope, yaglikci2022understanding, khan2018analysis, agarwal2018rowhammer, li2014write, ni2018write, genssler2022reliability, mutlu2023fundamentally, he2023whistleblower, baeg2022estimation, frigo2020trrespass, mutlu2017rowhammer, mutlu2018rowhammer, mutlu2019rowhammer, olgun2023hbm, olgun2023drambender, zhou2023double, luo2023rowpress, lang2023blaster, baek2025marionette}}

\newcommand{\mitigatingRowHammerAllCitations}[0]{\cite{AppleRefInc, hp2015rowhammer, lenovo2015rowhammer, greenfield2012throttling, kim2014flipping, kim2014architectural, jedec2017ddr4, aichinger2015ddr, aweke2016anvil, bains-merged, bains2015row, bains2016distributed, bains2016row, gomez2016dummy, yang2016suppression, son2017making, seyedzadeh2018cbt, irazoqui2016mascat, ryu2017overcoming, yang2017scanning, you2019mrloc, lee2019twice, park2020graphene, yaglikci2021security, yaglikci2021blockhammer, frigo2020trrespass, kang2020cattwo, hassan2021utrr, qureshi2022hydra, saileshwar2022randomized, brasser2017can, konoth2018zebram, van2018guardion, vig2018rapid, hassan2019crow, gautam2019row, kim2022mithril, lee2021cryoguard, marazzi2022protrr, zhang2022softtrr, joardar2022learning, juffinger2023csi, yaglikci2022hira, saxena2022aqua, enomoto2022efficient, manzhosov2022revisiting, ajorpaz2022evax, naseredini2022alarm, joardar2022machine, hassan2022acase, zhang2020leveraging,loughlin2021stop, devaux2021method, han2021surround, fakhrzadehgan2022safeguard, saroiu2022price, saroiu2022configure, loughlin2022moesiprime, zhou2022ltpim, hong2023dsac, mutlu2023fundamentally, marazzi2023rega, didio2023copyonflip, sharma2022areview, woo2023scalable, park2022rowhammer, wi2023shadow, kim2023ddr5, gude2023defending, guha2022criticality, france2022modeling, france2022reducing, bennett2021panopticon, arikan2022processor, tomita2022extracting, saxena2023pt, zhou2023dnndefender}}

\newcommand{\mcBasedRowHammerMitigations}[0]{\cite{kim2014flipping, saxena2022aqua, park2020graphene, kim2022mithril, you2019mrloc, kang2020cattwo, yaglikci2022hira, hassan2022acase, zhou2022ltpim, son2017making, saileshwar2022randomized, ryu2017overcoming, saroiu2022price, han2021surround, saroiu2022configure, lee2019twice, yaglikci2021blockhammer, greenfield2012throttling, devaux2021method, lee2021cryoguard, joardar2022learning, joardar2022machine, qureshi2022hydra, mutlu2023fundamentally, ajorpaz2022evax, loughlin2022moesiprime, bains2016distributed, france2022modeling, zhang2022softtrr, van2018guardion, loughlin2021stop, kim2014architectural, bains2015row, frigo2020trrespass, sharma2022areview, aichinger2015ddr, rh-hp, brasser2017can, LenovoRefInc, AppleRefInc, gude2023defending, woo2023scalable, didio2023copyonflip, bains2016row, juffinger2023csi, kim2023ddr5, gautam2019row, bains-merged, enomoto2022efficient, fakhrzadehgan2022safeguard, manzhosov2022revisiting, guha2022criticality, hassan2021utrr, wi2023shadow, konoth2018zebram, vig2018rapid, tomita2022extracting, zhou2023dnndefender, irazoqui2016mascat, arikan2022processor, yang2017scanning, zhang2020leveraging, saxena2023pt, aweke2016anvil, france2022reducing, park2022rowhammer}}

\newcommand{\noCiteMcBasedRowHammerMitigations}[0]{\nocite{kim2014flipping, saxena2022aqua, park2020graphene, kim2022mithril, you2019mrloc, kang2020cattwo, yaglikci2022hira, hassan2022acase, zhou2022ltpim, son2017making, saileshwar2022randomized, ryu2017overcoming, saroiu2022price, han2021surround, saroiu2022configure, lee2019twice, yaglikci2021blockhammer, greenfield2012throttling, devaux2021method, lee2021cryoguard, joardar2022learning, joardar2022machine, qureshi2022hydra, mutlu2023fundamentally, ajorpaz2022evax, loughlin2022moesiprime, bains2016distributed, france2022modeling, zhang2022softtrr, van2018guardion, loughlin2021stop, kim2014architectural, bains2015row, frigo2020trrespass, sharma2022areview, aichinger2015ddr, rh-hp, brasser2017can, LenovoRefInc, AppleRefInc, gude2023defending, woo2023scalable, didio2023copyonflip, bains2016row, juffinger2023csi, kim2023ddr5, gautam2019row, bains-merged, enomoto2022efficient, fakhrzadehgan2022safeguard, manzhosov2022revisiting, guha2022criticality, hassan2021utrr, wi2023shadow, konoth2018zebram, vig2018rapid, tomita2022extracting, zhou2023dnndefender, irazoqui2016mascat, arikan2022processor, yang2017scanning, zhang2020leveraging, saxena2023pt, aweke2016anvil, france2022reducing, park2022rowhammer}}

\newcommand{\refreshBasedRowHammerDefenseCitations}[0]{\cite{lee2019twice, seyedzadeh2017counterbased, seyedzadeh2018mitigating, kang2020cattwo, park2020graphene, kim2022mithril, kim2014architectural, bains2015row, bains2016distributed, bains2016row, aweke2016anvil, AppleRefInc, kim2014flipping, son2017making, you2019mrloc, yaglikci2021security, frigo2020trrespass, hassan2021utrr, qureshi2022hydra, devaux2021method, lee2021cryoguard, marazzi2022protrr, zhang2022softtrr, joardar2022learning}}

\newcommand{\integrityBasedMitigationsAllCitations}[0]{\cite{dell1997white, huang2010ivec, saileshwar2018synergy, chen2014memguard, juffinger2023csi, fakhrzadehgan2022safeguard, qureshi2021rethinking, manzhosov2022revisiting}}

\newcommand{\citeHardwareBasedMitigations}[0]{\cite{kim2014flipping, saxena2022aqua, park2020graphene, kim2022mithril, you2019mrloc, kang2020cattwo, yaglikci2022hira, hassan2022acase, zhou2022ltpim, son2017making, saileshwar2022randomized, ryu2017overcoming, saroiu2022price, han2021surround, saroiu2022configure, lee2019twice, yaglikci2021blockhammer, greenfield2012throttling, devaux2021method, lee2021cryoguard, joardar2022learning, joardar2022machine, qureshi2022hydra, seyedzadeh2018cbt, naseredini2022alarm, kim2015architectural, woo2022scalable, seyedzadeh2017cbt, yang2016suppression, gomez2016dummy, bostanci2024comet, olgun2024abacus, yaglikci2024spatial, marazzi2022protrr, bennett2021panopticon, hassan2019crow, mutlu2023fundamentally, wang2021discreet}}

\newcommand{\noCiteHardwareBasedMitigations}[0]{\nocite{kim2014flipping, saxena2022aqua, park2020graphene, kim2022mithril, you2019mrloc, kang2020cattwo, yaglikci2022hira, hassan2022acase, zhou2022ltpim, son2017making, saileshwar2022randomized, ryu2017overcoming, saroiu2022price, han2021surround, saroiu2022configure, lee2019twice, yaglikci2021blockhammer, greenfield2012throttling, devaux2021method, lee2021cryoguard, joardar2022learning, joardar2022machine, qureshi2022hydra, seyedzadeh2018cbt, naseredini2022alarm, kim2015architectural, woo2022scalable, seyedzadeh2017cbt, yang2016suppression, gomez2016dummy, bostanci2024comet, olgun2024abacus, yaglikci2024spatial, marazzi2022protrr, bennett2021panopticon, hassan2019crow, mutlu2023fundamentally, wang2021discreet}}

\makeatletter
\def\bstctlcite{\@ifnextchar[{\@bstctlcite}{\@bstctlcite[@auxout]}}
\def\@bstctlcite[#1]#2{\@bsphack
 \@for\@citeb:=#2\do{%
   \edef\@citeb{\expandafter\@firstofone\@citeb}%
   \if@filesw\immediate\write\csname #1\endcsname{\string\citation{\@citeb}}\fi}%
 \@esphack}
\makeatother

\newcounter{observation}
\newcounter{corollary}
\newcounter{takeaway}
\newcounter{claim}
\newcounter{proof}

\ifrev
    \definecolor{darkblue}{rgb}{0.0, 0.0, 0.55}

    \newcommand{\cql}[2]{#2\todo[size=\small,color=jazzberryjam]{\textbf{\textrm{\textcolor{white}{#1}}}}}
    \newcommand{\iql}[2]{#2\todo[size=\small,color=babyblueeyes]{\textbf{\textrm{\textcolor{black}{#1}}}}}

\else

    \newcommand{\cql}[1]{}
    \newcommand{\iql}[1]{}
\fi

\title{\huge\X{}: Understanding and Securing the Cutting-Edge Industry Solutions to DRAM Read Disturbance}

\pagenumberstrue
\newcounter{passversion}
\pagenumberstrue
\fancypagestyle{firstpage} {
    \fancyhead{}
    
    \fancyhead[C]{
        \parbox[][12mm][t]{13.5cm}{\hpcayear{} IEEE International Symposium on High-Performance Computer Architecture (HPCA)} 
        \includegraphics[width=12mm,height=12mm]{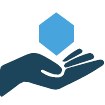}
        \includegraphics[width=12mm,height=12mm]{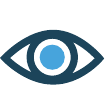}
        \includegraphics[width=12mm,height=12mm]{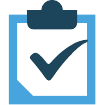}
    }
    \fancyfoot[C]{\thepage}
}

\makeatletter
\def\bstctlcite{\@ifnextchar[{\@bstctlcite}{\@bstctlcite[@auxout]}}
\def\@bstctlcite[#1]#2{\@bsphack
  \@for\@citeb:=#2\do{%
    \edef\@citeb{\expandafter\@firstofone\@citeb}%
    \if@filesw\immediate\write\csname #1\endcsname{\string\citation{\@citeb}}\fi}%
  \@esphack}
\makeatother 

\begin{document}
\maketitle

\thispagestyle{firstpage}
\pagestyle{plain}

\ifcameraready
    \setcounter{version}{99}
\else
    \setcounter{version}{99}
\fi

\begin{abstract}

Read disturbance in modern DRAM is an important \agy{0}{robustness (security, safety, and reliability)} problem, where repeatedly accessing (hammering) a row of DRAM cells (DRAM row) induces bitflips in other physically nearby DRAM rows.
Shrinking technology node size exacerbates \agy{0}{DRAM} read disturbance over generations.
To help mitigate read disturbance, the latest DDR5 \agy{0}{specifications (as of April 2024)} introduced a new RowHammer mitigation framework, called \gls{prac}. \gls{prac}
1)~enables the DRAM chip to accurately track row activations by allocating an activation counter per row and
2)~provides the DRAM chip with \om{2}{the} necessary time window to perform RowHammer-preventive refreshes by introducing a new back-off signal.
Unfortunately, no prior work rigorously studies \gls{prac}'s security guarantees and overheads.
\agy{0}{In this paper, we}
1)~present the first rigorous security, performance, energy, and cost analyses of \gls{prac} and
2)~propose \X{}, \ous{0}{a new mechanism that} addresses \gls{prac}'s two major weaknesses.

Our analysis shows that \gls{prac}'s system performance overhead on benign applications is non-negligible for \ous{0}{modern} DRAM chips and prohibitively large for future DRAM chips that are more vulnerable to read disturbance.
We identify \param{two} weaknesses of \gls{prac} that cause these overheads.
First, \gls{prac} increases critical DRAM access latency parameters due to the additional time required to increment activation counters.
Second, \gls{prac} performs a constant number of preventive refreshes at a time,
making it vulnerable to an adversarial access pattern, known as the wave attack, and consequently requiring it to be configured for significantly smaller activation thresholds.

To address \gls{prac}'s two weaknesses, we propose a new on-DRAM-die RowHammer mitigation mechanism, \X{}. \X{} 
1)~updates row activation counters concurrently \ous{5}{while} serving accesses by separating counters from the data and
2)~prevents the wave attack by dynamically controlling the number of preventive refreshes performed.
Our performance analysis shows that \X{}'s system performance overhead is near-zero for modern DRAM chips and very low for future DRAM chips.
\X{} outperforms \param{three} variants of \gls{prac} and \param{three} other state-of-the-art \om{2}{read disturbance} solutions.
We discuss \X{}'s and \gls{prac}'s implications \ous{0}{for future systems} and foreshadow future research directions.
To aid future research, we open-source our \X{} implementation at \url{https://github.com/CMU-SAFARI/Chronus}.\ouscomment{1}{Repository will be ready after artifact evaluation results}

\end{abstract}

\glsresetall{}
\section{Introduction}
\label{sec:intro}

To ensure system robustness \agy{0}{(security, safety, and reliability)}, it is critical to maintain memory isolation: accessing a memory address should \emph{not} cause unintended side-effects on data stored \om{2}{in} other addresses~\cite{kim2014flipping}.
Unfortunately, with aggressive technology scaling, DRAM~\cite{dennard1968fieldeffect}, the prevalent main memory technology, suffers from increased \emph{read disturbance}: accessing (reading) a row of DRAM cells (i.e., a DRAM row) degrades the data integrity of other physically close but \emph{unaccessed} DRAM rows.\noCiteHardwareBasedMitigations{}\noCiteMcBasedRowHammerMitigations{}
RowHammer~\cite{kim2014flipping} is a prime example of DRAM read disturbance, where a row (i.e., victim row) can experience bitflips when at least one nearby row (i.e., aggressor row) is repeatedly activated (i.e., hammered)~\rhmemisolationrefs{} more times than a threshold, called the \gls{nrh}.
\emph{RowPress}~\cite{luo2023rowpress} is another prime example of DRAM read disturbance that amplifies the effect of RowHammer \ous{2}{by keeping an aggressor row open for longer, thereby causing more disturbance with each activation and consequently reducing} \gls{nrh}.

A simple way of mitigating DRAM read disturbance is to preventively refresh potential victim rows before bitflips occur.
Unfortunately, a preventive refresh blocks accesses to thousands of DRAM rows that are in the same DRAM bank for a non-negligible time window \ous{0}{(e.g., 350ns~\cite{jedec2024jesd795c})}, in which the memory controller should \emph{not} issue any other DRAM command \om{2}{to the bank}.
As DRAM chips become more vulnerable to read disturbance with technology node scaling, preventive refreshes can significantly reduce system performance~\refreshBasedRowHammerDefenseCitations{}. Therefore, it is important to accurately identify when a preventive refresh is needed and perform it \om{2}{in a timely manner}.

To provide DRAM chips with the necessary flexibility to perform preventive refreshes in a timely manner, recent DRAM standards (e.g., DDR5~\cite{jedec2020jesd795, jedec2024jesd795c}) introduce 1) a command called \emph{\gls{rfm}}\cite{jedec2020jesd795} and 2) a \ous{0}{framework} called \emph{\gls{prac}}\cite{jedec2024jesd795c}.
\gls{rfm} is a command that provides the DRAM chip with a time window (e.g., \param{\SI{195}{\nano\second}}~\cite{jedec2024jesd795c}) to perform preventive refreshes.
Specifications before 2024 (e.g., \om{2}{earlier} DDR5~\cite{jedec2020jesd795}) advise the memory controller to issue \gls{rfm} when the number of row activations in a bank or a logical memory region exceeds a threshold (e.g., \param{32}~\cite{jedec2024jesd795c}), \om{2}{a mechanism we call} \emph{periodic RFM (PRFM)}.

A recent update \om{2}{(as of April 2024)} of the JEDEC DDR5 specification~\cite{saroiu2024ddr5, jedec2024jesd795c} introduces a new on-DRAM-die read disturbance mitigation \ous{0}{framework} called \gls{prac}.
\gls{prac} has two key features.
First, \gls{prac} maintains an activation counter per DRAM row~\cite{kim2014flipping} to accurately identify when a preventive refresh is needed.
\gls{prac} increments a DRAM row's activation counter while the row is being closed, which increases the latency of closing a row, i.e., the \gls{trp} and \gls{trc} timing parameters.
Second, \gls{prac} proposes a new \emph{back-off} signal to convey the need for preventive refreshes from the DRAM chip to the memory controller, similar to what prior works propose~\cite{bennett2021panopticon, devaux2021method, yaglikci2021security, hassan2022acase, kim2022mithril, hassan2024self}\omcomment{2}{also cite SMD MICRO'24, update reference blocks everywhere}\ouscomment{2}{ACK. Added SMD. TODO: Create and revise citation blocks}.
The DRAM chip asserts this back-off signal when a DRAM row's activation count reaches a critical value.
Within a predefined time window (e.g., \param{\SI{180}{\nano\second}}~\cite{jedec2024jesd795c}) after receiving the back-off signal, the memory controller has to issue an \gls{rfm} command so that the DRAM chip can perform the necessary preventive refresh operations.
\gls{prac} aims to
1) avoid read disturbance bitflips by performing necessary preventive refreshes in a timely manner and
2) minimize unnecessary preventive refreshes by accurately tracking each row's activation count.
Unfortunately, \emph{no} prior work rigorously investigates \gls{prac}'s security, performance, energy, and \ous{0}{storage} cost\nbcomment{0}{what cost? if you mean area/storage it is better to mention here. cost on its own doesn't mean much. it could mean \$s even.} \agy{0}{implications} for modern and future systems. 

\agy{0}{In this paper, we} 1)~present the first rigorous security, performance, energy, and \ous{0}{storage} cost\nbcomment{0}{:(} analyses of \gls{prac}, \agy{0}{which identifies \gls{prac}'s two major weaknesses} and
2)~propose a new mechanism, \X{}, which addresses \agy{0}{those} weaknesses.

\head{\gls{prac} Analysis}
We analyze \gls{prac} in \param{three} steps.
First, we define a security-oriented adversarial access pattern that achieves the highest possible activation count in systems protected by \gls{prac}.
Second, we conduct a security analysis by evaluating the highest possible activation count that a DRAM row can reach under different configurations of \gls{prac}.
Our security analysis shows that \gls{prac} can be configured for secure operation against an \gls{nrh} value of \param{20} or higher.
Third, we conduct a system performance analysis using cycle-accurate simulations with an open-source simulator, Ramulator 2.0~\cite{luo2023ramulator2, ramulator2github}.
Our system performance results across 60 different four-core multiprogrammed workload mixes show that:
At modern \gls{nrh} values higher than \param{1K}, \gls{prac}'s overheads are mainly dominated by the increased critical DRAM timing parameters where \gls{prac} incurs high average (maximum) \param{\ous{10}{5.8}}\% (\param{\ous{10}{8.9}}\%) system performance and \param{\ous{10}{10.7\%}} (\param{\ous{10}{13.5\%}}) DRAM energy overheads.
For the lowest secure \gls{nrh} value of \param{20}, these overheads increase to \param{\ous{10}{78.5\%}} (\param{\ous{10}{90.7\%}}) and \param{\ous{10}{6.6x}} (\param{\ous{10}{7.1x}}), respectively.\footnote{Our analysis rigorously sweeps \gls{nrh} to extremely low values (e.g., 20). \agy{2}{This is} because a mitigation mechanism that securely scales to low \om{2}{\gls{nrh} values} with low performance and energy overheads provides two benefits:
1)~activation counters maintain fewer bits, resulting in smaller hardware complexity and
2)~read disturbance profiling becomes significantly shorter as \om{2}{each} DRAM row can be tested for a \om{2}{smaller} hammer count (e.g., hammering the row 64 times takes $16\times$ shorter than hammering the row 1024 times).}\ouscomment{10}{Perf Diff\\
old -> new\\
$\nrh{}$: avg (max) -> avg (max)\\
1K: 9.7 (13.4) -> 5.8 (8.9)\\
20: 81.2 (91.9) -> 78.5 (90.7)}\ouscomment{10}{Energy Diff\\
old -> new\\
$\nrh{}$: avg (max) -> avg (max)\\
1K: 18.4 (22.9) -> 10.7 (13.5)\\
20: 7.9x (8.7x) -> 6.6x (7.1x)}

We attribute these large overheads to \param{two} key weaknesses \om{2}{in \gls{prac}}.
\ous{0}{First, \gls{prac} increases critical DRAM \ous{6}{timing} parameters due to the additional time required to increment activation counters.}
\ous{0}{Second, \gls{prac} performs a \ous{2}{fixed} number of preventive refreshes at a time and enforces a \emph{delay period} during which preventive refreshes \om{2}{\emph{cannot}} be requested, making it vulnerable to an adversarial \om{2}{access} pattern known as the \emph{wave attack}.}

\head{\agy{0}{\X{}}}
To address PRAC’s two \om{2}{major} weaknesses, we propose \X{}.
\X{} 1)~updates row activation counters \om{2}{\emph{concurrently}} \ous{5}{while} serving accesses by \om{2}{physically} separating counters from the data and
2)~prevents the wave attack by dynamically controlling the number of preventive refreshes performed \ous{2}{by the memory controller} and removing the delay period \ous{2}{after the memory controller performs preventive refreshes}.\footnote{In Greek mythology, Cronus is renowned for separating heaven and earth~\cite{graf1993greek}, and Chronos for controlling time~\cite{graf1993greek}. We chose their blended name, Chronus, which resembles our proposal: separating counters (similar to Cronus separating worlds) and dynamically controlling the number of preventive refreshes (similar to Chronos controlling time).}

\head{Key Results}
We compare \X{} to 1)~\param{three} variants of PRAC, as the state-of-the-art industry solutions, and 2)~\param{three} other state-of-the-art \nb{0}{academic} proposals, i.e., Graphene~\cite{park2020graphene}, Hydra~\cite{qureshi2022hydra}, and PARA~\cite{kim2014flipping}.\omcomment{2}{Why do we not compare others like abacus, for example?}\ouscomment{2}{Maria was implementing modern mechanisms like Abacus and Comet. I contacted her and will push for more mechanisms.}
Our performance analysis shows that \X{} outperforms \emph{all} six evaluated mechanisms across all studied workloads and \gls{nrh} values.
\X{} incurs near-zero system performance (\ous{6}{$<$\param{0.1\%}} on average) and relatively low (compared to \gls{prac}) DRAM energy (\param{10.3}\% on average) overheads for \ous{2}{modern} DRAM chips \om{2}{($\nrh{} =$ 1K)} and low system performance (\ous{6}{\param{8.3\%}} on average) and DRAM energy (\param{17.9\%} on average) overheads for future DRAM chips \om{2}{($\nrh{} =$ 20)}.

We make the following contributions:
\begin{itemize}
    [noitemsep,topsep=0pt,parsep=0pt,partopsep=0pt,labelindent=0pt,itemindent=0pt,leftmargin=*]
    \item We present the first security analysis of \om{2}{industry's state-of-the-art read disturbance solution, \gls{prac}}, and provide robust configurations against its worst-case access pattern.
    \item We rigorously evaluate the performance, energy, and cost implications of \gls{prac}'s different configurations for modern and future DRAM chips. Our results show that \gls{prac} incurs non-negligible overheads \nb{0}{due} to two \agy{0}{major} weaknesses: 1) increased critical DRAM timing parameters, and 2) wave attack vulnerability due to \om{2}{allowing only} a \ous{2}{fixed} number of preventive refreshes.
    \item We propose \X{}, which addresses the weaknesses of \gls{prac} through \ous{0}{1) updating row activation counters concurrently \ous{5}{while} serving accesses by separating counters from the data and 2) preventing wave attacks by dynamically controlling the number of preventive refreshes as needed}. We show that \X{}'s system performance overhead is near-zero (very low) for modern (future) DRAM chips.
    \item We compare \X{} against three variants of \gls{prac} and three state-of-the-art mitigation mechanisms. Our results show that 1)~\X{} outperforms all evaluated mitigation mechanisms at all evaluated \gls{nrh} values and 2)~\gls{prac} \ous{0}{variants} underperform against \param{three} of the \param{four} (including \X{}) mitigation mechanisms for modern DRAM chips with relatively high (i.e., $\geq$1K) \gls{nrh} values.
    \item To aid future research in a transparent manner, we open-source our implementations \om{2}{with Ramulator 2.0~\cite{chronusgithub}}.
\end{itemize}
\section{Background}
\label{sec:background}
\subsection{DRAM Organization and Operation}
\label{sec:dram_organization}

\begin{figure}[ht]
    \centering
    \includegraphics[width=\linewidth]{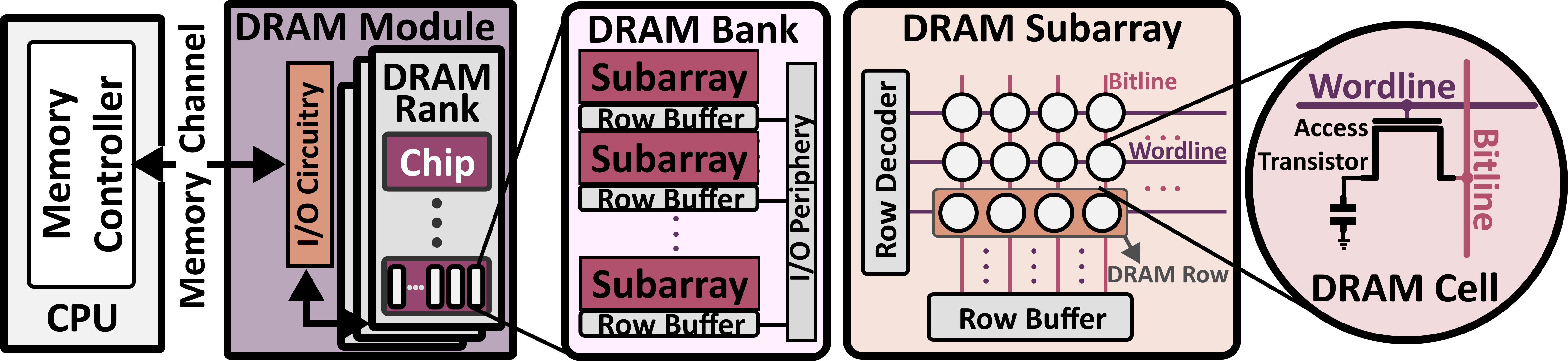}
    \caption{DRAM organization}
    \label{fig:dram_organization}
\end{figure}

\head{Organization}
\figref{fig:dram_organization} shows the hierarchical organization of a modern DRAM-based main memory. The memory controller connects to a DRAM module over a memory channel. A module contains one or multiple DRAM ranks that time-share the memory channel.
A rank consists of multiple DRAM chips.
Each DRAM chip has multiple DRAM banks each of which contains multiple subarrays. A DRAM bank is organized as a two-dimensional array of DRAM cells, where a row of cells is called a DRAM row. A DRAM cell consists of 1) a storage capacitor, which stores one bit of information in the form of electrical charge, and 2) an access transistor.

\head{Operation}
To access a DRAM row, the memory controller issues a set of commands to DRAM over the memory channel. The memory controller sends an $ACT$ command to activate a DRAM row, which asserts the corresponding wordline and loads the row data into the row buffer. Then, the memory controller can issue \emph{RD}/\emph{WR} \om{3}{commands} to read from/write into the DRAM row.
\om{3}{An access} to the row \om{3}{that is already in the row buffer causes} a row hit.
To access a different row, the memory controller must \ous{4}{first close the opened row and prepare the bank for a new activation} by issuing a $PRE$ command.
Therefore, accessing a different row \om{3}{from the one in the row buffer} causes a row conflict \om{3}{in the row buffer}.

DRAM cells are inherently leaky and lose their charge over time due to charge leakage in the access transistor and the storage capacitor.
\om{3}{To} maintain data integrity, the memory controller periodically refreshes each row in a time interval called \gls{trefw} ($32 ms$ for DDR5~\cite{jedec2020ddr5} and $64 ms$ for DDR4~\cite{jedec2017ddr4}).
To ensure all rows are refreshed every \gls{trefw}, the memory controller issues $REF$ commands with a time interval called \gls{trefi} ($3.9 \mu s$ for DDR5~\cite{jedec2020ddr5} and $7.8 \mu s$ for DDR4~\cite{jedec2017ddr4}).

\head{Timing Parameters}
To ensure correct operation, the memory controller must obey specific timing parameters while accessing DRAM.
In addition to \gls{trefw} and \gls{trefi}, we explain \param{three} timing parameters related to the rest of the paper:
i) the minimum time interval between two consecutive ACT commands targeting the same bank ($t_{RC}$),
ii) the minimum time needed to issue a PRE command following an ACT command ($t_{RAS}$), and
iii) the minimum time needed to issue an ACT command following a PRE command ($t_{RP}$).
\om{3}{Detailed explanations of these parameters can be found in \cite{kim2012acase, lee2013tiered, lee2015adaptive}}.

\subsection{DRAM Read Disturbance}
As DRAM manufacturing technology node size shrinks, interference across rows increases, \om{3}{exacerbating} circuit-level read disturbance mechanisms.
Two prime examples of read disturbance mechanisms are RowHammer~\cite{kim2014flipping} and RowPress~\cite{luo2023rowpress}, where repeatedly activating \ous{3}{(i.e., hammering)} a DRAM row (i.e., aggressor row) or keeping the aggressor row active for a long time \ous{3}{(i.e., pressing)} induces bitflips in physically nearby rows (i.e., victim rows), respectively.
\ous{3}{To induce read disturbance bitflips, 1) an aggressor row needs to be hammered more than a threshold value called \gls{nrh}~\cite{kim2014flipping} or 2) an aggressor needs to be pressed for long enough~\cite{kim2020revisiting, orosa2021deeper, yaglikci2022understanding, luo2023rowpress}}.\omcomment{3}{define and use hammer and press. Background is similar to PACRAM and same mistakes are repeated. Can we make them less similar?}\ouscomment{3}{Acknowledged. This background should be from our DRAMSec version and BreakHammer but I will revisit the background and improve it.}
\om{3}{Various} characterization studies~\understandingRowHammerAllCitations{} show that as DRAM \om{3}{technology} scaling continues to smaller \om{4}{technology} node \om{3}{sizes}, DRAM chips are \om{3}{becoming} more vulnerable to \ous{3}{read disturbance} (i.e., newer chips have lower \gls{nrh} values).
For example, one can induce RowHammer bitflips \ous{4}{by activating two aggressors that are physically adjacent to a victim row (i.e., double-sided RowHammer) 4.8K times each} in the chips manufactured in 2020 while a row needs to be hammered \ous{4}{69.2K times in older chips manufactured in 2013~\cite{kim2020revisiting}}.\omcomment{5}{Please find out if this issue exists elsewhere.}\ouscomment{5}{Acknowledged. The following papers seem fine: DRAMSec version, BreakHammer, PACRAM, and VRD.}
To make matters worse, RowPress~\cite{luo2023rowpress} reduces \gls{nrh} by \om{3}{1-2 orders} of magnitude. 

\head{DRAM Read Disturbance Mitigation Mechanisms}
Many prior works propose mitigation techniques~\mitigatingRowHammerAllCitations{} to protect DRAM chips against RowHammer bitflips leveraging different approaches.
These mechanisms perform two tasks: 1)~execute a trigger algorithm and 2)~perform preventive actions.
The trigger algorithm observes the memory access patterns and triggers a preventive action based on the result of a probabilistic or a deterministic process. The preventive action is one of  1)~preventively refreshing victim row~\refreshBasedRowHammerDefenseCitations{},
2) dynamically remapping aggressor rows~\cite{saileshwar2022randomized, saxena2022aqua, wi2023shadow, woo2023scalable}, and
3)~throttling unsafe accesses~\cite{greenfield2012throttling, yaglikci2021blockhammer}.
Existing RowHammer mitigation mechanisms can also prevent RowPress bitflips when their trigger algorithms are configured to be more aggressive, which is practically equivalent to configuring them for sub-1K \gls{nrh} values~\cite{luo2023rowpress}.
Unfortunately, existing RowHammer mitigation mechanisms incur prohibitively large performance overheads at low \gls{nrh} values (i.e., sub-1K) \om{3}{because they more aggressively perform} preventive actions.
Given that DRAM read disturbance worsens with shrinking technology node size, \gls{nrh} values are expected to reduce even more~\cite{yaglikci2024spatial, kim2014flipping, orosa2021deeper}.
Therefore, reducing the performance overhead of existing RowHammer mitigation mechanisms is critical.

\head{\gls{prac}}
Various prior works discuss the use of per-row activation counters to detect how many times each row in DRAM is activated within a refresh interval~\cite{kim2014flipping,kim2014architectural,bennett2021panopticon,kim2023ddr5,yaglikci2021security}.
A recent update (as of April 2024) of the JEDEC DDR5 specification~\cite{saroiu2024ddr5, jedec2024jesd795c} introduces a similar on-DRAM-die read disturbance mitigation mechanism called \gls{prac} (explained in \secref{sec:briefsummary}), which aims to ensure robust operation at low overhead by preventively refreshing victim rows when necessary.
Although \gls{prac} is a promising DRAM specification advancement, \emph{no} prior work rigorously analyzes \gls{prac}'s impact on security, performance, energy, and cost for modern and future systems.\footnote{\om{3}{An earlier version of this paper was presented at DRAMSec 2024~\cite{canpolat2024understanding}}.}
\section{A Brief Summary of RFM and PRAC}
\label{sec:briefsummary}

This section briefly explains the \gls{rfm} command, \gls{prac} mechanism, and assumptions we use for our evaluations.

\head{RFM Command}
\gls{rfm} is a DRAM command that provides the DRAM chip with a time window (e.g.,~\param{\SI{195}{\nano\second}}~\cite{jedec2024jesd795c}) so that the DRAM chip \om{3}{can} preventively refresh potential victim rows.
The DRAM chip is responsible for identifying and preventively refreshing potential victim rows, and the memory controller is responsible for issuing \gls{rfm} commands.

\head{\gls{prac} Overview}
\gls{prac}~\cite{jedec2024jesd795c}, \om{3}{as described in the JEDEC DDR5 standard} \om{4}{updated in April 2024}~\cite{jedec2024jesd795c}, implements an activation counter for each DRAM row, and thus accurately measures the activation counts of \emph{all} rows.
When a row's activation count reaches a threshold, the DRAM chip asserts a back-off signal which forces the memory controller to issue an RFM command.
The DRAM chip preventively refreshes potential victim rows upon receiving an RFM command. 

\head{Assumptions About \gls{prac}}
We make two assumptions:
1)~\gls{prac} always refreshes victims of the row with the maximum activation count during each \gls{rfm} command\ouscomment{4}{Some PRAC implementations (Moin's MOAT) only refresh rows that exceed a threshold. Doing so accelerates the wave attack.}\omcomment{4}{Do they show the negative effect on the wave attack?}\ouscomment{4}{I triple-checked all analyses (both our's and moin's) and MOAT's approach does not affect the wave attack (because wave attack starts from $N_{BO}$ activations, which is always higher than "MOAT-style" refresh threshold). We only need to assume that PRAC refreshes the row with max activations. If we still have the ``why is this a good assumption'' question on the currently remaining part: I am not sure if we can mathematically analyze the security if we do not know if PRAC correctly refreshes victims of a row with maximal activation count.}
and 2)~physically-adjacent DRAM rows can experience bitflips when a DRAM row is activated \ous{0}{at least \gls{nrh} times}.

\head{\gls{prac}'s Operation and Parameters}
\gls{prac} increments a row's activation count \ous{3}{during a precharge command}.
\ous{2}{When a bank receives a precharge command, the bank internally reads, modifies, and writes the open row's counter before de-asserting the wordline.
As such, \gls{prac}'s counter update} affects several DRAM timing parameters \ous{2}{around row accesses}.
Table~\ref{tab:practiming}\ouscomment{4}{TODO: fix table overextending to the right} summarizes DRAM timing parameter changes \om{4}{when \gls{prac} is enabled} for the DRAM 3200AN~\cite{jedec2024jesd795c} speed bin.

\begin{table}[ht]
    \centering
    \footnotesize
    \vspace{1em}
    \caption{DRAM Timing Parameter Changes \om{4}{with \gls{prac}}}
    \begin{tabular}{l|l||@{\hspace{2pt}}c@{\hspace{3pt}}c}
    \makecell[b]{Parameter} & \makecell[bl]{Description} & \makecell[b]{\om{3}{DDR5} \\ \om{3}{without \gls{prac}}} & \makecell[b]{\om{4}{DDR5} \\ \om{4}{with \gls{prac}}} \\
    \hline
    \hline
    $t_{RAS}$ & \makecell[l]{minimum time for a $PRE$ \\ after an $ACT$ to the same bank} & \SI{32}{\nano\second} & \SI{16}{\nano\second} \\ 
    \hline
    $t_{RP}$ & \makecell[l]{minimum time for an $ACT$ \\ after a $PRE$ to the same bank} & \SI{15}{\nano\second} & \SI{36}{\nano\second} \\ 
    \hline
    $t_{RC}$ & \makecell[l]{minimum time for two $ACT$s \\ to the same bank} & \SI{47}{\nano\second} & \SI{52}{\nano\second} \\ 
    \hline
    $t_{RTP}$ & \makecell[l]{minimum time for a $PRE$ \\ after a $RD$ to the same bank} & \SI{7.5}{\nano\second} & \SI{5}{\nano\second} \\ 
    \hline
    $t_{WR}$ & \makecell[l]{minimum time for a $PRE$ \\ after a $WR$ to the same bank} & \SI{30}{\nano\second} & \SI{10}{\nano\second} \\ 
    \hline
    \end{tabular}
    \label{tab:practiming}
\end{table}

\ous{3}{We make two observations from Table~\ref{tab:practiming}.
First, the $\trp{}$ and $\trc{}$ timing parameters increase because of the additional time needed to update the counter before the wordline is de-asserted.
Second, the $\tras{}$, $\trtp{}$, and $\twr{}$ timing parameters likely reduce to account for the additional time the row stays open as the counter is updated.
We note that \gls{prac} timing parameters can also be better utilizing the aggressive guardbands existing in modern DRAM chips as shown in~\cite{lee2015adaptive, chang2016understanding, chang2017understanding, chang2017understandingphd, kim2018solar, yaglikci2022understanding, mathew2017using, tugrul2025understanding} to reduce one or more of the \om{4}{latter three} timing parameters.}


The DRAM chip asserts the back-off signal when a row's activation count reaches a fraction of \gls{nrh}, denoted as \gls{aboth}, where the fraction can be configured to either 70\%, 80\%, 90\%, or 100\%~\cite{jedec2024jesd795c}.
The memory controller receives the back-off signal shortly after (e.g., \param{$\approx$\SI{5}{\nano\second}}~\cite{jedec2024jesd795c}) issuing a \gls{pre} command.
The memory controller and the DRAM chip go through three phases when the back-off signal is asserted.
First, during \gls{taboact}~\cite{jedec2024jesd795c}, the memory controller has a limited time window (e.g., \param{\SI{180}{\nano\second}}~\cite{jedec2024jesd795c}) to serve requests after receiving the back-off signal.
A DRAM row can receive up to \gls{taboact}/\gls{trc} activations in this window.
Second, during the \emph{recovery period}~\cite{jedec2024jesd795c}, the memory controller issues a number of \gls{rfm} commands, which we denote as $\bonrefs{}$ (e.g., 1, 2 or 4~\cite{jedec2024jesd795c}).
An \gls{rfm} command can further increment the activation count of a row before its potential victims are refreshed.
Third, during the \emph{delay period} or \gls{tbodelay}~\cite{jedec2024jesd795c}, the DRAM chip \emph{cannot} reassert the back-off signal until it receives a number of activate commands, which we denote as $\bonacts{}$  (e.g., 1, 2 or 4~\cite{jedec2024jesd795c}).\footnote{Current DDR5 specification~\cite{jedec2024jesd795c} notes that $\bonrefs$ and $\bonacts$ always have the same value. To comprehensively assess PRAC's security guarantees, we also consider different values in our security analysis (\secref{sec:configurationandsecurity}).}
\om{3}{Taking into account} these three phases, \secref{sec:configurationandsecurity} calculates the highest achievable activation count to any row in a \gls{prac}-protected system.

\head{Determining Rows to Refresh During an RFM Command}
\ous{3}{\gls{prac} maintains a large number of counters in each DRAM bank}.
As such, it is \emph{not} \ous{4}{practical} for the DRAM bank to search for the row with the maximum activation count during an RFM command.
To track the rows with the highest activation counts, our \gls{prac} implementation uses a relatively small \ous{2}{per-bank} table \ous{2}{(e.g., 4 \om{3}{entries} \ous{3}{store enough rows for each \gls{rfm} command to refresh during the recovery period}) that} we call \om{3}{the} \emph{Aggressor Tracking Table} (ATT).
The table starts empty (i.e., all entries are invalid).
When a row is precharged, the table is updated \ous{3}{with the precharged row's counter value} if
1) the precharged row exists in the table,
2) an entry in the table is invalid, or
3) the precharged row's counter value exceeds the table entry with the \emph{lowest} activation count.
When \ous{2}{a DRAM bank receives} an RFM command, the bank invalidates and refreshes the potential victims of the table entry with the \emph{maximum} activation count.

\head{\gls{rfm} and \gls{prac} Implementations}
We analyze \param{three} different \gls{rfm} and \gls{prac} implementations:
1) \emph{\gls{prfm}}, where the memory controller issues an \gls{rfm} command \emph{periodically} when the total number of activations to a bank reaches a predefined threshold called \ous{3}{\gls{rfmth}}\ouscomment{3}{defined $\rfmth{}$ here} with \emph{no} back-off signal from the DRAM chip, as described in early DDR5 standards~\cite{jedec2020jesd795};
2) \emph{\gls{prac}-N}, where the memory controller issues N back-to-back \gls{rfm} commands \emph{only} after receiving a back-off signal from the DRAM chip, as described in the latest JEDEC DDR5 standard~\cite{saroiu2024ddr5, jedec2024jesd795c};
3) \emph{\gls{prac}+\gls{prfm}}, where the memory controller issues an \gls{rfm} command when $i$) the total number of activations to a bank reaches \om{3}{$\rfmth{} = 75$} or $ii$) it receives a back-off signal from the DRAM chip.\footnote{The \gls{rfmth} of 75 \om{4}{is} provided \om{4}{in} the example \gls{prac}+\gls{prfm} configuration in the latest (as of April 2024) JEDEC DDR5 standard~\cite{jedec2024jesd795c}.}\omcomment{3}{Is RFMth defined in the footnote?}
\om{3}{As shown in \secref{sec:configurationandsecurity}}, \gls{prac}-N implementations are \emph{not} secure at \gls{nrh} values lower than \param{20}.
Therefore, combining \gls{prac} and \gls{prfm} enables security at lower \gls{nrh} values at the cost of potentially refreshing the victims of aggressor rows whose activation counts are \emph{not} close to \gls{nrh}.
\section{Adversarial Access Pattern: The Wave Attack}
\label{sec:adversarial}

\ous{6}{We describe} our threat model and the adversarial access pattern, known as \emph{the wave attack}~\cite{yaglikci2021security, devaux2021method} or \emph{the feinting attack}~\cite{marazzi2022protrr}.

\head{Threat Model}
We assume that the attacker
1)~knows the physical layout of DRAM rows (as in~\cite{yaglikci2021blockhammer}),
2)~accurately detects when a row is internally refreshed (preventively or periodically, as in U-TRR~\cite{hassan2021utrr}), and
3)~precisely times \emph{all} DRAM commands except \gls{ref} and \gls{rfm} commands (as in~\cite{yaglikci2021blockhammer, hassan2021utrr}). 

\begin{figure}[h]
\centering
\includegraphics[width=\linewidth]{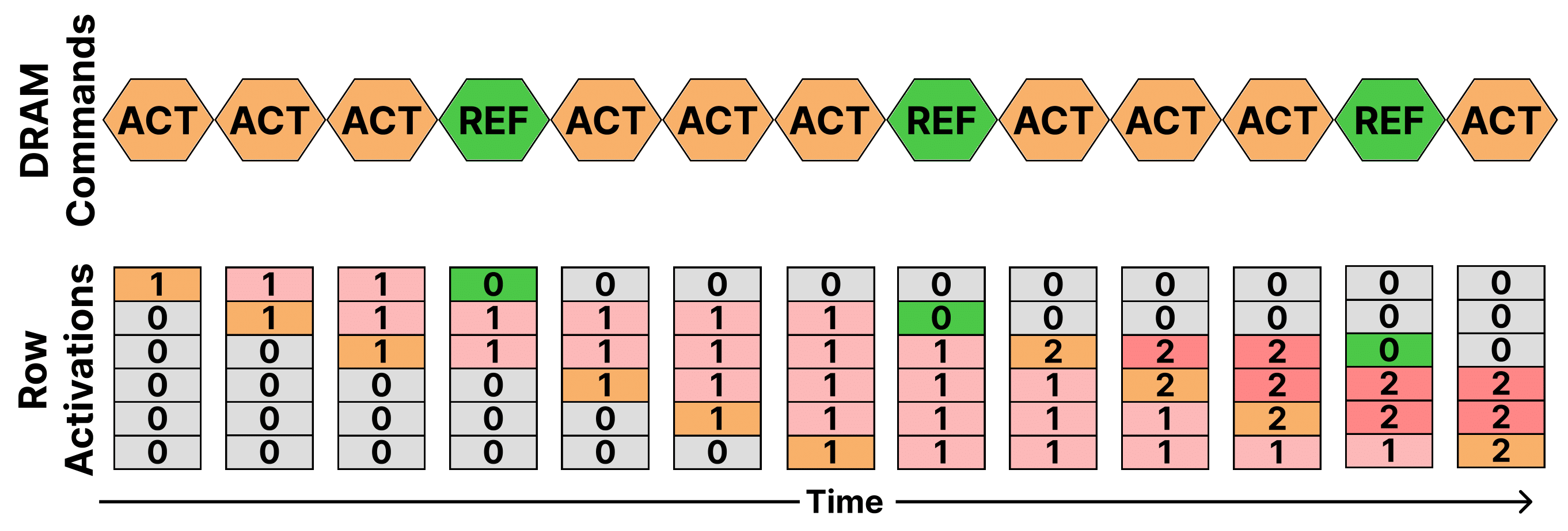}
\caption{Wave attack buildup visualization}
\label{fig:waveattack}
\end{figure}

\head{Overview}
The wave attack aims to achieve the highest activation count for a given row in an \gls{rfm} and \gls{prac}-protected DRAM chip by overwhelming the mitigation mechanism using a number of decoy rows.
\figref{fig:waveattack} visualizes the buildup of a wave attack against a periodic read disturbance mitigation \nb{0}{mechanism} (e.g., \gls{prfm}) that refreshes the potential victims of an aggressor for every three row activations.\omcomment{4}{Cannot see the figure.}\ouscomment{4}{Replaced with a PNG. Is it fixed?}

In this access pattern, the attacker hammers several rows in a balanced manner, such that the mitigation mechanism can perform preventive refreshes \emph{only} for a small subset of the hammered rows when a preventive refresh is issued.
When an aggressor row's victims are refreshed, the attacker excludes the aggressor row in the next round of activations.
By doing so, this adversarial access pattern achieves the highest possible activation count for the row whose victims are preventively refreshed last.
\ous{3}{More details and analyses of the attack can be found in~\cite{yaglikci2021security, devaux2021method, marazzi2022protrr}}.
\section{Configuration of PRAC and Security Analysis}
\label{sec:configurationandsecurity}
This section investigates different \gls{rfm} and \gls{prac} configurations and their impact on security under the wave attack.

\head{Key Parameters}\ouscomment{4}{Moved before notation} We assume
a \emph{blast radius} of 2~\cite{kogler2022half},
a \gls{trc} of \SI{52}{\nano\second}~\cite{jedec2024jesd795c},
and a \gls{trfm} of \SI{350}{\nano\second}~\cite{jedec2024jesd795c}, which allows an RFM command to refresh four victim rows of one aggressor row.

\head{Notation} \agy{0}{$R_i$ is} the set of rows that the wave attack hammers in round $i$ \agy{0}{and $|R_{i}|$ is} the \ous{0}{number of rows in} $R_i$.

\head{PRFM}
In round 1, the wave attack \agy{0}{hammers each row in $R_1$ once, causing the memory controller to issue $\lfloor(|R_{1}|/\rfmth{})\rfloor$ \gls{rfm} commands, \ous{0}{each refreshing} \ous{0}{four} victims of one aggressor row.}
In round 2, the wave attack \agy{0}{hammers} the non-mitigated rows $R_2$, where $|R_2| = |R_1| - \lfloor ({|R_1|}/{\rfmth{}}) \rfloor$.
By repeating this calculation $i$ times, Equation~\ref{eqn:rfmrec} evaluates the number of non-mitigated rows at an arbitrary round $i$ ($|R_{i}|$). 

\vspace{-8pt}
\begin{equation}
\label{eqn:rfmrec}
|R_{i}| = |R_{1}| - 
\Bigl\lfloor
\frac{
\sum_{k=1}^{i-1} |R_{k}|
}{
RFM_{th}
}
\Bigr\rfloor
\end{equation}

\vspace{2pt}
To cause bitflips, the wave attack must make sure that 1) at least one aggressor row is not mitigated (i.e., its victims are \emph{not} refreshed by an RFM command until the end of round $N_{RH}$, i.e., $|R_{N_{RH}}| > 0$\nb{0}{)}, and 2) the time taken by the attacker's row activations and RFM preventive refreshes do \emph{not} exceed \gls{trefw}, i.e., aggressor's victims are \emph{not} periodically refreshed before being activated \gls{nrh} times.
We rigorously sweep the wave attack's parameters and identify the maximum hammer count of an aggressor row before its victims are refreshed.

\head{\gls{prac}-N}
We adapt our \gls{prfm} wave attack analysis to \gls{prac}-N by leveraging two key insights.
First, \gls{prac}-N mechanism will \emph{not} preventively refresh any row until a row's activation count reaches \glsfirst{aboth} \ous{4}{(i.e., the back-off threshold)}.\omcomment{4}{Is there a meaning to \gls{aboth}? i.e.,...}\ouscomment{4}{It is the back-off threshold. It is defined in PRAC's operation and parameters but added a reminder here to help reader follow easier.}
We prepare rows in $R_1$ such that each row is already hammered \gls{aboth}-1 times.
Doing so, the number of rounds necessary to induce a bitflip is reduced by \gls{aboth}-1.
Second, at least one row's activation counter remains above \gls{aboth} across \emph{all} rounds after initialization until the end of the wave attack.
This causes \gls{prac}-N to assert the back-off signal as frequently as possible, i.e., with a time period containing a recovery period ($\bonrefs\times\trfm$), a delay period (\gls{tbodelay}), and a window of normal traffic (\gls{taboact})~\cite{jedec2024jesd795c}.
Leveraging these insights, we update Equation~\ref{eqn:rfmrec} to derive Equation~\ref{eqn:pracrec}.

\vspace{-8pt}
\begin{equation}
\label{eqn:pracrec}
|R_{i}| = |R_{1}| - \bonrefs \times
\Bigl\lfloor
\frac{
\sum_{k=1}^{i-1} |R_{k}|
}{
(\bonacts + (\taboact / \trc))
}
\Bigr\rfloor
\end{equation}

\vspace{2pt}
For a \gls{prac}-N system to be secure,
an attacker should \emph{not} be able to obtain $|R_{\nrh{}-\aboth{}}| > 0$ within \gls{trefw} for any $R_1$.
We analyze the maximum hammer count of an aggressor row before its victims are refreshed in a \gls{prac}-N-protected system 
for a wide set of \gls{aboth} and $|R_{1}|$ configurations. 

\begin{figure}[h]
\centering
\includegraphics[width=\linewidth]{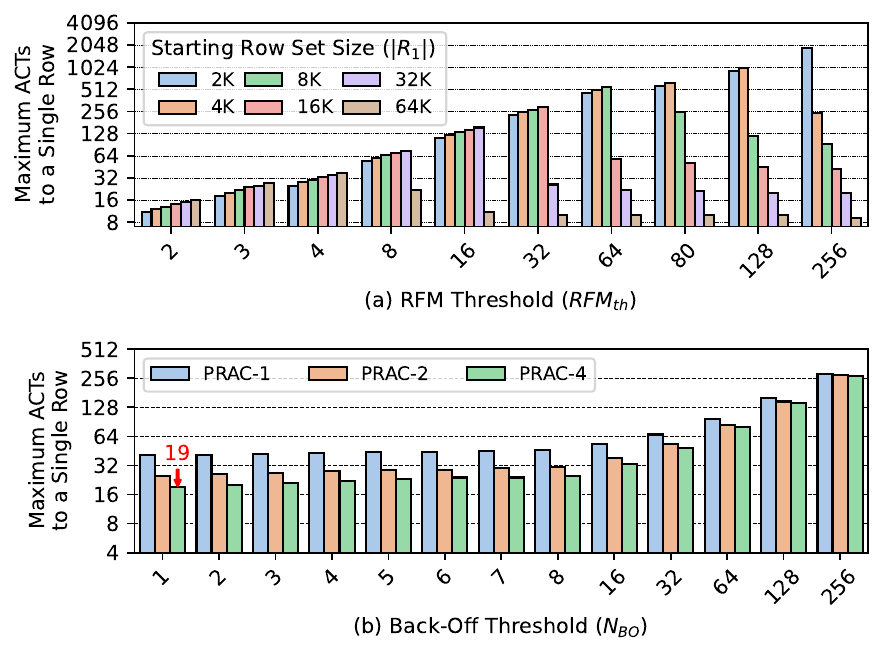}
\caption{Maximum activations to a row allowed by (a) \gls{prfm} and (b) \gls{prac}-N}
\label{fig:rfmpracanalysis}
\end{figure}

\head{Configuration Sweep}
\figref{fig:rfmpracanalysis} shows the maximum activation count an aggressor row can reach before its victims are refreshed (y-axis) for \gls{prfm} and \gls{prac}-N in \figsref{fig:rfmpracanalysis}a and~\ref{fig:rfmpracanalysis}b, respectively.
\figref{fig:rfmpracanalysis}a shows the bank activation threshold to issue an RFM command ($\rfmth{}$) on the x-axis and starting \om{4}{(i.e., round 1)} row set size ($|R_1|$) color-coded. \figref{fig:rfmpracanalysis}b shows the back-off threshold ($\aboth{}$)\omcomment{4}{NBO should have been defined and explained earlier in this section}\ouscomment{4}{Fixed with your comment above.} on the x-axis and $\bonrefs$ color-coded.
\ous{4}{In \figref{fig:rfmpracanalysis}b, each bar depicts the \om{5}{\emph{worst-case}} starting row set size (i.e., $|R_1|$ that yields the highest activation count)}.

From \figref{fig:rfmpracanalysis}a, we observe that to prevent bitflips for very low \gls{nrh} values (e.g., 32 on the y axis), \gls{rfmth} should be configured to very low values (e.g., $<$4), as only such \gls{rfmth} values result in activation counts less than \gls{nrh} for all $|R_1|$ values.
From \agy{0}{\figref{fig:rfmpracanalysis}b}, we observe that \gls{prac}-N provides security at \gls{nrh} values as low as \param{20} (because a row can receive \param{19} activations as annotated) when configured to 1) trigger a back-off as frequently as possible ($N_{BO}=1$) and 2) issue \param{four} RFMs
in the recovery period (i.e., \gls{prac}-4).
For the remaining sections, we configure \gls{prfm} and \gls{prac} using these secure thresholds.\footnote{In this study, we assume that we can accurately determine \gls{nrh}. \gf{0}{However, that} is a difficult problem, as determining \gls{nrh} for every row has many challenges and a time-consuming process~\cite{kim2014flipping, orosa2021deeper, luo2023rowpress, kim2020revisiting, saroiu2022configure, olgun2023understanding, olgun2024read, zhou2023threshold, olgun2025variable}).}
We also assume that \gls{prac} \emph{borrows} time from periodic refreshes to transparently refresh the potential victims of one aggressor row every other periodic refresh command (\ous{5}{an operation we call \emph{borrowed refresh}}, as shown to be possible in modern DRAM chips~\cite{hassan2021utrr}), thereby reducing the number of back-offs needed.
We do \emph{not} consider periodic refreshes in our security analysis because the memory controller can potentially delay \ous{4}{a periodic refresh up to 4 times~\cite{jedec2024jesd795c}.\omcomment{5}{Did anyone publish a system level attack that exploits this fresh postponing?}\ouscomment{5}{Not that I am aware of. Also, this was reduced from 7 or 8 in DDR4 (?) to 4 in DDR5. I think this is a ``soft'' mitigation against RowPress. MINT (Moin's MICRO paper) discusses delayed refreshes as a problem though.}
These delays result in a maximum time between two consecutive $REF$ commands of $5*\trefi{}$ \ous{7}{(i.e., \SI{19500}{\nano\second} or 414 row activations for a single DDR5 bank with $\trc{} =$~\SI{47}{\nano\second})}\omcomment{7}{If 19500 was correct keep it as it was. Also, why e.g. for 414 activations?}\ouscomment{7}{It is correct (5 * 3900ns). I wrote e.g. because tRC could vary (also added what tRC I assumed and switched back to i.e.)}.
This large window where the DRAM cells \emph{cannot} be protected with a periodic refresh} significantly reduces the security of mechanisms that rely on borrowed refreshes.\omcomment{5}{Borrowed refresh is undefined? Also very odd terminology. Fix.}\ouscomment{5}{Acknowledged. Defined above.}
\section{Overhead Analysis \om{4}{of \gls{prac}}}
\label{sec:sensitivity}

We evaluate \gls{prac}'s performance overheads for existing and future DRAM chips, by sweeping \gls{nrh} from \param{1K} down to \param{20} (lowest secure \gls{nrh} value for \gls{prac}).
To evaluate performance, we conduct cycle-level simulations using Ramulator 2.0~\cite{luo2023ramulator2, ramulator2github} \om{4}{(which builds on the original Ramulator~\cite{kim2016ramulator, ramulatorgithub})}.
We extend Ramulator 2.0 with the implementations of \gls{prac}, \gls{rfm}, and the back-off signal.
We evaluate system performance using the weighted speedup metric~\cite{eyerman2008systemlevel, snavely2000symbiotic}.

\tabref{table:system_configuration} shows our system configuration.
We assume a realistic quad-core system, connected to a dual-rank memory with \param{eight} bank groups, each containing \param{four} banks (\param{64} banks total).
The memory controller employs the FR-FCFS memory scheduler\cite{frfcfs, zuravleff1997controller} with a Cap on Column-Over-Row Reordering (FR-FCFS+Cap)~\cite{mutlu2007stall} of \param{four}.
We extend the memory controller to delay requests that \emph{cannot} be served within \gls{taboact}.

\newcolumntype{C}[1]{>{\let\newline\\\arraybackslash\hspace{0pt}}m{#1}}
\begin{table}[ht]
\scriptsize
\centering
\caption{{Simulated} System Configuration}
\begin{tabular}{l|C{5.8cm}}
 \hline
 \textbf{Processor} & {\SI{4.2}{\giga\hertz}, 4-core, 4-wide issue, {128-entry} instr. window} \\ \hline
 \textbf{Last-Level Cache} & {64-byte} cache line, 8-way {set-associative, \SI{8}{\mega\byte}} \\ \hline
 \textbf{Memory Controller} & {64-entry read/write request queues; Scheduling policy: FR-FCFS+Cap of \param{4}~\cite{mutlu2007stall}}; Address mapping: MOP~\cite{kaseridis2011minimalistic} \\ \hline
 \textbf{Main Memory} & DDR5 DRAM~\cite{ramulator2github}, 1 channel, 2 ranks, 8 bank groups, 4 banks/bank group, {64K} rows/bank\\ \hline
 \end{tabular}
 \label{table:system_configuration}
\end{table}

\head{Workloads}
We evaluate applications from five benchmark suites: SPEC CPU2006~\cite{spec2006}, SPEC CPU2017~\cite{spec2017}, TPC~\cite{tpc}, MediaBench~\cite{fritts2009media}, and YCSB~\cite{ycsb}. We group all applications into three memory-intensity groups based on their row buffer misses per kilo instructions (RBMPKIs), similar to prior works~\cite{olgun2024abacus, bostanci2024comet}. These groups are High (H), Medium (M), and Low (L) for the lowest MPKI values of \param{10, 2, and 0}, respectively. Then, we create \param{60} workload mixes with \param{10} of each HHHH, MMMM, LLLL, HHMM, MMLL, and LLHH combination types. We simulate each workload until all cores execute \param{100}M instructions \om{4}{each}.      

\subsection{Performance Evaluation}

\begin{figure}[h]
\centering
\includegraphics[width=\linewidth]{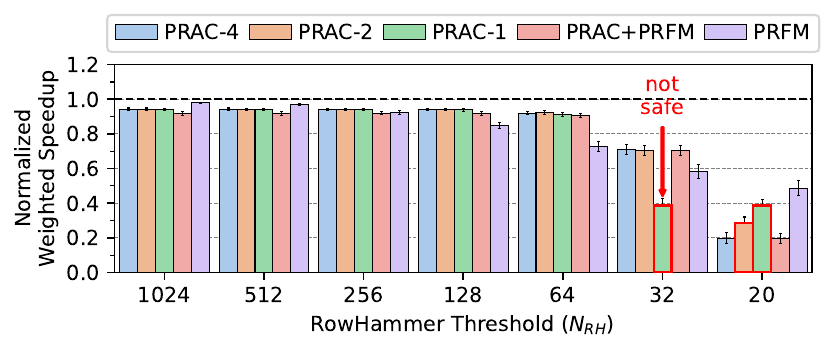}
\caption{Performance impact of evaluated \gls{prac} and \gls{rfm} configurations on 60 benign four-core workloads}
\label{fig:sensitivity_performance}
\end{figure}

\figref{fig:sensitivity_performance} presents the performance overheads of the evaluated read disturbance mitigation mechanisms as \gls{nrh} decreases.
Axes respectively show the \gls{nrh} values (x axis) and system performance (y axis) in terms of weighted speedup~\cite{eyerman2008systemlevel, snavely2000symbiotic} normalized to a baseline with \emph{no} read disturbance mitigation (higher y value is better).
Different bars identify different read disturbance mitigation mechanisms and the red edge color indicates \om{5}{unsafe (i.e., read disturbance vulnerable)} configurations.
We make \param{six} observations from \figref{fig:sensitivity_performance}.

\head{1. Reducing $\bm{\nrh{}}$ increases performance overheads}
As \gls{nrh} decreases, performance overheads of all studied mitigation configurations increase, as expected, due to the more frequent mitigating actions (i.e., preventive refreshes) performed.

\head{2. \gls{prac} \agy{0}{has non-negligible performance overheads}}
At \ous{6}{$\nrh{} =$~1K}, \gls{prac}-4 incurs an average (maximum) system performance overhead of \param{\ous{10}{5.8\%}} (\param{\ous{10}{8.9\%}})\ouscomment{10}{Diff\\
old -> new\\
$\nrh{}$: avg (max) -> avg (max)\\
1K: 9.7 (13.4) -> 5.8 (8.9)\\}
across all workloads.
We attribute \gls{prac}'s non-negligible overhead at relatively high \gls{nrh} values to \gls{prac}'s increased DRAM timing parameters~\cite{jedec2024jesd795c} (\secref{sec:briefsummary}) as \gls{prac}-4 sends only 6 back-offs (1.9$\times$10$^{-8}$ back-offs per million cycles) on average across all workloads.

\head{3. \gls{prac} overheads slightly increase until $\bm{\nrh{} =}$~64}
When \gls{nrh} decreases from \param{1K} to \param{64}, \gls{prac}-4's system performance overheads slightly increase (i.e., $\approx$\param{2\%}).
We attribute the steady system performance as \gls{nrh} decreases to \gls{prac} accurately tracking aggressor row activations and refreshing rows by \emph{borrowing} time from periodic refreshes.

\head{4. \gls{prac} becomes prohibitively expensive at $\bm{\nrh{} \leq}$~64}
Between \gls{nrh} values of \param{64} and \param{20}, \gls{prac}-4's average (maximum) system performance overheads across all workloads significantly increase from \param{\ous{10}{8.1\%}} (\param{\ous{10}{11.8\%}}) to \param{\ous{10}{78.5\%}} (\param{\ous{10}{90.7\%}}).\ouscomment{10}{Diff\\
old -> new\\
$\nrh{}$: avg (max) -> avg (max)\\
64: 11.7 (16.0) -> 8.1 (11.8)\\
20: 81.2 (91.9) -> 78.5 (90.7)\\}
We attribute the significant increase in system performance overhead to \gls{prac} performing more frequent preventive refreshes.
For example, with \gls{prac}-4 at \gls{nrh} values of \param{64} and \param{20}, the four-core benign workload of \emph{523.xalancbmk}, \emph{435.gromacs}, \emph{459.GemsFDTD}, and \emph{434.zeusmp} trigger \ous{10}{13.4 and 125.5}\ouscomment{10}{Diff:
old -> new\\
11.7 and 124.6 -> 13.4 and 125.5}
\ous{4}{back-offs} per million cycles, resulting in \param{\ous{10}{15.5\%}} and \param{\ous{10}{87.4\%}}\ouscomment{10}{Diff:
old -> new\\
18.6 and 88.0 -> 15.5 and 87.4}
system performance overhead, respectively.

\head{5. \gls{prfm} performs poorly}
When \gls{nrh} decreases from \param{1K} to \param{20}, the average (maximum) system performance overhead of \gls{prfm} increases from \param{2.1}\% (\param{3.7}\%) to \param{51.8}\% (\param{68.7}\%).
We attribute this significant overhead increase to \gls{prfm}'s configuration against the wave attack drastically increasing the frequency of preventive refreshes as \gls{nrh} decreases, similar to \gls{prac}.

\head{6. \gls{prac}+\gls{prfm} performs poorly}
Pairing \gls{prac}-4 with \gls{prfm} increases \gls{prac}'s system performance overhead by an average (maximum) of \param{\ous{10}{21.0\%}} (\param{\ous{10}{27.9\%}})\ouscomment{10}{Diff\\
old -> new\\
13.6 (40.2) -> 21.0 (27.9)\\}
across all $\nrh{}$ values.
This is because 1)~\gls{prac}'s secure configurations (\secref{sec:configurationandsecurity}) already preventively refresh all rows before they reach a critical level and
2)~pairing \gls{prac} with \gls{prfm} causes performance degradation due to unnecessary preventive refreshes.

We conclude that
1)~\gls{prac}'s increased DRAM timing parameters incur significant overheads even under infrequent preventive refreshes for modern DRAM chips (i.e., \gls{nrh} $=$ 1K),
2)~\gls{prac}'s system performance overheads slightly increase as \gls{nrh} decreases until \param{32}, where \gls{prac} starts performing significantly worse,
3)~\gls{prfm} incurs significant system performance loss with decreasing \gls{nrh}, and
4)~pairing \gls{prac} with \gls{prfm} provides no system performance advantage.

\subsection{PRAC's Two Major Outstanding Problems}

We identify \param{two} major outstanding problems \om{4}{with \gls{prac}}.
First, \gls{prac} counters are incremented \om{4}{while} a DRAM row is \om{4}{closed}.
Doing so increases \om{4}{some} timing parameters and degrades system performance.
Second, \gls{prac} allows a \ous{2}{fixed} number of preventive refreshes with each back-off and forces the memory controller to send the same number of activations as refreshed rows before new preventive refreshes can be requested.
The possible number of row activations exceeds the number of preventive refreshes performed with \gls{prac}'s preventive actions.
Thus, \gls{prac} is vulnerable to the wave attack~\cite{yaglikci2021security, devaux2021method}, which significantly reduces the necessary thresholds to mitigate read disturbance and makes mounting memory performance attacks~\cite{mutlu2007stall} easier (\secref{sec:evaluation_dos}).

\addtocounter{figure}{1}
\begin{figure*}[hb!]
\centering
\includegraphics[width=\linewidth]{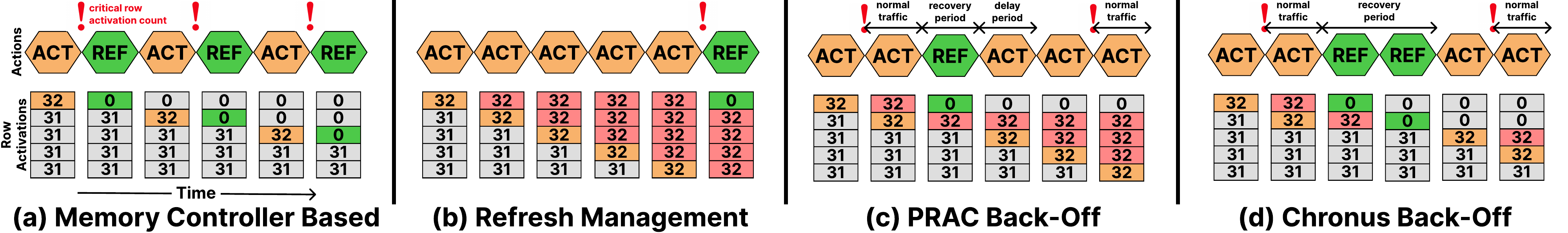}
\caption{Wave attack visualization for four different classes of read disturbance mitigation mechanisms}
\label{fig:waveattackcheck}
\end{figure*}
\addtocounter{figure}{-2}

\section{\X{}: Alleviating Counter Access Latencies and Back-Off Delays}
\label{sec:mechanism}

We propose \X{}, a new mechanism that addresses \gls{prac}'s \param{two} \ous{2}{major} weaknesses.
\ous{5}{\X{} implements two key components to address these weaknesses: \emph{Concurrent Counter Update} (CCU) and \emph{Chronus Back-Off}}.
First, \ous{5}{CCU} updates row activation counters concurrently \ous{5}{while} serving accesses.
\X{} achieves this by \om{4}{physically} separating counters from the data stored in the DRAM array.
As such, \X{} prevents the increase in DRAM timing parameters \om{4}{due to \gls{prac}}.
Second, \ous{5}{Chronus Back-Off} extends \gls{prac}'s back-off policy such that \X{} performs as many preventive refreshes as needed when a back-off is \ous{2}{triggered} as opposed to \gls{prac}'s policy of \emph{only} allowing a \ous{2}{fixed} number of preventive refreshes and enforcing a delay between consecutive back-offs to serve memory accesses.
By doing so, \X{} prevents an attacker from maliciously using the back-off delay for a wave attack.

\subsection{\ous{5}{CCU: Concurrent Counter Update}}
The latest JEDEC DDR5 DRAM specification \ous{2}{(as of April 2024~\cite{jedec2024jesd795c})} \om{4}{allows} manufacturers to extend DRAM rows with counter bits.
These row activation counters are incremented while the row is being closed, which increases critical \ous{2}{DRAM} timing parameters (\secref{sec:briefsummary}) and degrades system performance even at relatively high \ous{4}{read disturbance} thresholds, e.g., $\nrh{} =$ 1K, as we show in \secref{sec:sensitivity}.
These performance overheads can be eliminated by updating counters \om{4}{\emph{concurrently \ous{5}{while} serving accesses}}, \ous{5}{i.e., Concurrent Counter Update (CCU)}.\ouscomment{5}{CCU defined here}
\om{4}{To achieve this}, we propose \om{4}{physically} separating counters \om{4}{from data} by leveraging subarray-level parallelism~\cite{kim2012acase, chang2014improving, yaglikci2022hira, yuksel2024functionally}.\omcomment{6}{Revise to clearly say what we add and what methodology we use to evaluate their overhead. Point me to where it is said.}

\begin{figure}[h]
\centering
\includegraphics[width=0.7\linewidth]{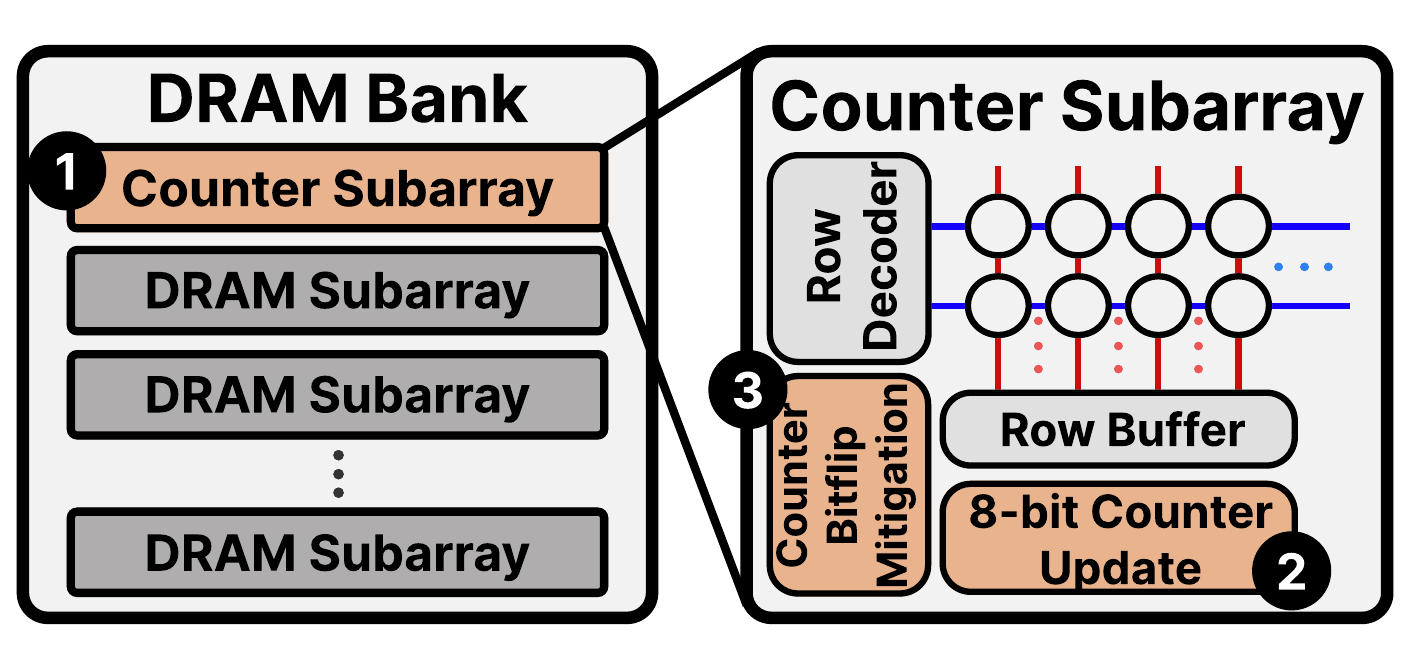}
\caption{\om{4}{Overview of \X{}' \om{6}{Concurrent Counter Update technique}}}
\label{fig:chronushighlevel}
\end{figure}

\head{High Level Explanation}
\figref{fig:chronushighlevel} depicts a high-level overview of \om{4}{\X{}'} separate counters.
We propose leveraging subarray-level parallelism by storing row activation counters of a DRAM bank in a sufficiently small (e.g., \om{4}{\param{64}-row}) additional subarray, which we call \om{4}{the} \emph{counter subarray}~\circled{1}, within each bank.
The counter subarray 1) \om{4}{stores} the row activation counts \om{4}{of all data rows in the bank,} \ous{6}{2) implements custom circuitry to update~\circled{2}} (read -- increment by 1 -- write back) an activation count when a row is activated, \ous{6}{and 3) employs a mechanism to prevent bitflips within the counter subarray~\circled{3}}.\ouscomment{6}{Revised here to \emph{clearly} say what we add at a high level. Each detailed section provides overheads and the methodology on how the overhead is calculated (I also attached comments to where their overheads are discussed)}
As the counter subarray can be accessed concurrently with \om{5}{\emph{regular subarrays}} that store data~\cite{kim2012acase, chang2014improving, yaglikci2022hira, yuksel2024functionally, yuksel2024simultaneous}, the latency of reading, updating, and writing a counter can be hidden by the latency of opening, accessing, and closing the corresponding row.\footnote{We propose using subarray-level parallelism~\cite{kim2012acase} only between the counter subarray and regular subarrays. \X{} can also be combined with \om{6}{exploiting} subarray-level parallelism between regular subarrays~\cite{kim2012acase}. We leave the exploration of the benefits and challenges of such design to future work.}

\addtocounter{figure}{1}

\head{Storing the Counters \circled{1}}
\X{} leverages the high density of the DRAM array by storing the activation counters in a set of DRAM rows, located in the counter subarray.
Assuming a realistic implementation, where\ouscomment{6}{Counter subarray storage overhead calculation}
1)~each row has an \param{8-bit} row activation count,\footnote{\X{}'s performance and energy overheads are reasonably low ($<$\param{0.1\%}) for an \gls{nrh} of \param{256} (\secref{sec:evaluation}). Therefore, we assume a counter width of \param{8} bits.}
2)~a DRAM bank contains \param{128K} DRAM rows~\cite{jedec2024jesd795c}, and
3)~each DRAM row contains \param{16Kbit} in a DRAM chip~\cite{jedec2024jesd795c},
\X{} maintains \param{128KB} (128K rows $\times$ 8-bits) of activation count metadata for a DRAM bank of \param{256MB} (128K rows $\times$ 16Kbit), which fits into \param{64} DRAM rows (128KB $/$ 16Kbit).
As such, the counter subarray incurs \om{4}{only} a \param{0.05\%} of capacity overhead.

\head{Updating the Counters \circled{2}}
When a DRAM row is activated, \X{} updates the corresponding activation counter in \param{five} steps.
First, \X{} activates the row that contains the corresponding activation count in the counter subarray. 
Second, \X{} reads the column that contains the corresponding activation count. 
Third, \X{} reads the activation count by parsing the corresponding bits in the read column.
Fourth, \X{} updates the activation count.
Fifth, \X{} writes the updated counter back.
To read/write the counter, \X{} parses three sets of bits from the externally provided row address and identifies the row, column, and byte addresses used in the counter subarray.
To achieve low hardware complexity, \X{} uses custom circuitry to decrement the counter for a row by one \ous{5}{(we call this circuit \emph{the decrementer circuit})} when the row is activated and triggers a \ous{6}{back-off} when the counter reaches zero (e.g., every \param{256} activations).
For \gls{nrh} values smaller than \param{256}, \X{} compares the counter against \param{$256 - \nrh{}$} and triggers \ous{6}{a back-off} accordingly. 

\ous{4}{We implement the decrementer circuit using NAND, NOR, NOT, and MUX gates (gates that are already implemented in the local sense amplifiers).
Our design consists of \param{21} gates and can be implemented with \param{96} transistors.
We evaluate the critical path delay of the decrementer circuit to be \SI{0.627}{\nano\second}\ouscomment{6}{decrementer hardware complexity},\footnote{We use \om{4}{the} Synopsys Design Compiler~\cite{synopsys} with a Global Foundaries 22nm technology~\cite{carter201622nm}, configured to take into account the performance degradation of implementing logic using the DRAM process.
We set the output load of the decrementer circuit to be 1000fF and \om{4}{apply} a latency penalty factor of 22.91\% \om{5}{used in prior work}~\cite{chen1996Assessing}.}
which is significantly smaller than the row cycle latency (\gls{trc}) of \SI{47}{\nano\second} \om{4}{and hence can be completed \om{5}{concurrently} with an access to another subarray}.
Appendix~\ref{apx:decrementer} provides a detailed implementation of the decrementer circuit}.

\head{Avoiding Bitflips in the Counter Subarray~\circled{3}}\ouscomment{6}{Moved ``bitflip mitigation'' up (such that headings follow 1, 2, and 3 in the figure)}
\ous{2}{The counter subarray consists of DRAM cells and thus it is also vulnerable to read disturbance bitflips}.
To avoid bitflips in the counter subarray, we recommend implementing one of three effective countermeasures:
1)~tracking the activation counts of rows in the counter subarray in a separate SRAM array and refreshing them when necessary, similar to the Silver Bullet technique~\cite{devaux2021method, yaglikci2021security},
2)~refreshing potential victim rows in the counter subarray frequently in parallel to accessing other rows by latching the fetched row in a redundant array of sense amplifiers, similar to REGA~\cite{marazzi2023rega}, and
3)~allocating guard rows in between consecutive rows in the counter subarray, similar to GuardION~\cite{van2018guardion} and ZebRAM~\cite{razavi2016flip}.
Although these can be costly solutions for protecting all rows, their costs can be reasonable when limited to the counter subarray \ous{2}{(equivalent to protecting only \param{0.05\%} of a DRAM bank)}.\ouscomment{6}{Bitflip mitigation hardware complexity}
Our CACTI~\cite{cacti} and mathematical evaluations show that \ous{4}{applying any of these solutions} incur near-zero ($<$0.1\%) area overhead per DRAM bank.\omcomment{4}{What about REGA?}\ouscomment{4}{Using the methodology in the REGA paper, the overhead is 0.0157\% (again extremely low because we only protect the counter subarray)}

\head{Resetting the Counters}
\ous{4}{As DRAM cells are periodically refreshed to retain their data, DRAM read disturbance mechanisms periodically reset activation counters to reduce the number of preventive actions taken~\cite{park2020graphene, olgun2024abacus, lee2019twice, bostanci2024comet, qureshi2022hydra}.
Doing so also prevents an adversarial access pattern that can hog a significant amount of DRAM bandwidth availability due to preventive actions.
This adversarial pattern\nbcomment{7}{I am not aware of any works actually describing this potential attack.}\omcomment{7}{Not even in submission? If so, ok.}\ouscomment{7}{Not even in submission. (verified by Nisa)}
1) \om{5}{hammers} many rows close to the preventive action threshold and
2) \om{5}{triggers} preventive actions on each row quickly.\footnote{\ous{4}{An attacker can also hog a significant amount of DRAM bandwidth availability with different adversarial access patterns at future \gls{nrh} values, as shown by prior work~\cite{canpolat2024breakhammer}. We perform a similar study by analyzing \X{} and \gls{prac}'s vulnerability to memory performance attacks~\cite{mutlu2007stall} in \secref{sec:evaluation_dos}.}}
To reduce the number of \ous{5}{preventive actions (i.e., back-off)} and to mitigate the adversarial access pattern, \X{} borrows time from periodic refreshes (similarly to \gls{prac}, as we describe in \secref{sec:configurationandsecurity}).
As such, \X{} periodically \ous{5}{refreshes the victims of \emph{a recently accessed row with a relatively high activation count}}, thereby 1) resetting the counter of the row before it is activated enough times to trigger a back-off and 2) preventing the build-up of an adversarial access pattern that could hog DRAM bandwidth availability}.

\head{DRAM Energy Overhead of the Counter Subarray}
\om{5}{A} row activation in the counter subarray \om{4}{(which is concurrent with an activation \om{5}{of} a row in a regular subarray)} does \emph{not} significantly increase DRAM power consumption, as shown by two independent prior studies~\cite{yuksel2024simultaneous,mathew2017using}.
First, real DRAM chip measurements with multiple row activations~\cite{yuksel2024simultaneous} show that additional row activations introduce marginal \om{4}{additional} power consumption, e.g., simultaneously activating 32 rows consumes 21.19\% less power than a periodic refresh operation.
Second, a separate prior work~\cite{mathew2017using} \ous{4}{shows} that power consumption largely comes from driving the peripheral circuitry instead of asserting wordlines.
For our evaluations, we conduct SPICE~\cite{nagel1973spice} simulations using prior state-of-the-art open-source DRAM subarray models~\cite{hassan2019crow} to estimate Chronus' power and energy overheads.
Our SPICE simulations show that the row activation \om{5}{in the counter subarray and the} counter update increase the DRAM energy consumption of a DRAM row access (i.e., opening and closing a row) by \param{19.07}\%.

\subsection{\ous{5}{\X{} Back-Off}}
The wave attack is only possible if an attacker can perform row activations \om{4}{more quickly} than the \ous{2}{read disturbance} mitigation mechanism refreshes \ous{2}{victims}.
\figref{fig:waveattackcheck} visualizes the wave attack buildup for \param{four} classes of mitigation mechanisms: (a) Memory Controller Based~\mcBasedRowHammerMitigations{}, (b) Refresh Management~\cite{jedec2024jesd795c}, (c) \gls{prac} Back-Off~\cite{jedec2024jesd795c}, and (d) \X{} Back-Off (this work).

The memory controller based mechanisms control the flow of traffic to DRAM.
Therefore, \ous{0}{these mechanisms can} stop \ous{0}{serving} DRAM \ous{0}{accesses} and \ous{0}{freely} refresh all potential victims of aggressor rows that reach \ous{2}{a critical activation count} (e.g., \param{32} in \figref{fig:waveattackcheck}).
Refresh Management, \agy{0}{as described in} the JEDEC standard~\cite{jedec2024jesd795c}, requires \gls{rfm} commands to be issued periodically based on the number of activations \ous{0}{to a bank}.
\agy{0}{Therefore, the} activations between periodic preventive refreshes \ous{2}{enable a wave attack}.
\gls{prac} Back-Off suffers from \param{three} limitations that both reduce performance and \ous{2}{enable a wave attack}:
$L1$) The memory controller \emph{can} issue activations (with a maximum \ous{0}{number} of \SI{180}{\nano\second}/$\trc{}$) \ous{0}{after a back-off is triggered (during the window of normal traffic)},
$L2$) the memory controller inefficiently sends a \ous{2}{fixed} number of preventive refresh commands ($\bonrefs{}$) with each back-off \ous{2}{(regardless of the number of preventive refreshes needed to refresh all potential victims)}, and
$L3$) the DRAM module \emph{cannot} trigger another back-off \ous{0}{(delay period)} until the DRAM module receives a certain number of activations ($\bonacts{}$).
\om{4}{In contrast}, \X{} Back-Off \om{4}{securely} \emph{prevents} \ous{2}{a wave attack} (and performance loss due to unnecessary preventive refreshes) by solving \param{two} key weaknesses ($L2$ and $L3$)\footnote{We do \emph{not} propose \ous{5}{reducing or removing} the window of normal traffic \ous{5}{(L1) because such a change} requires 1) the \texttt{alert\_n} to be a fast signal and 2) the memory controller to react to a back-off more quickly, thereby potentially \ous{5}{increasing memory scheduling complexity}. We leave a rigorous analysis of \om{5}{using} a more intelligent DRAM interface and protocol (such as~\cite{hassan2024self}) \om{5}{together with \X{}} to future work.} of \gls{prac} Back-Off.
\ous{4}{First}, \X{} keeps the back-off signal asserted until the DRAM module \ous{4}{refreshes the last aggressor row with an activation count that exceeds \gls{aboth}.
While the back-off signal is asserted, the memory controller continues to send preventive refreshes.
Second}, \X{} does \emph{not} enforce a delay period.
Doing so prevents unnecessary refreshes by allowing \X{} to dynamically adjust the number of preventive refreshes.

\head{Challenges of Enabling Chronus Back-Off}
We discuss \param{two} design \ous{2}{considerations} \ous{6}{to enable} Chronus Back-Off:
1) latency of the back-off signal and
2) re-asserting the back-off signal without a delay period.
First, the latency of the back-off signal is low enough to provide dynamic control over the number of RFM commands because an RFM command has a relatively high duration (e.g., \SI{350}{\nano\second}~\cite{jedec2024jesd795c}).
This high duration is enough for the memory controller to acknowledge the de-asserted back-off signal and stop sending RFM commands;
The JEDEC DDR5 standard~\cite{jedec2024jesd795c} already utilizes the pin used by back-off signals (i.e., \texttt{alert\_n} pin) at a lower latency in two places:
1) \emph{Write CRC Error Handling}~\cite{jedec2024jesd795c} requires \texttt{alert\_n} to be asserted for 12 to 20 command clock cycles with a potential delay of \SI{3}{\nano\second} to \SI{13}{\nano\second} and
2) \emph{PRAC Back-Off}~\cite{jedec2024jesd795c} requires the memory controller to acknowledge \texttt{alert\_n} with a latency of \SI{180}{\nano\second} (i.e., duration of normal traffic). 
Second, \emph{PRAC Back-Off} already \emph{allows} the \texttt{alert\_n} to be asserted during the delay period \om{4}{due to} other causes (e.g., Write DQ CRC errors~\cite{jedec2024jesd795c}).
We conclude that there are \emph{no} fundamental limitations in the JEDEC DDR5 standard~\cite{jedec2024jesd795c} against enabling Chronus Back-Offs.

\section{Security Analysis \om{4}{of \X{}}}

We analyze the security of \X{} against read disturbance bitflips.
\ous{5}{We base our analysis on \X{}'s \param{three} key properties:
$P1$) \X{} accurately tracks the activation count of all rows (as it employs per row activation counters)\om{6}{;}
$P2$) \X{} can trigger a back-off at any time (as it does \emph{not} enforce a delay period)\om{6}{;} and
$P3$) \X{} back-offs remain in effect until all rows that reach the back-off threshold have been refreshed}.\ouscomment{5}{Properties are defined here}

Let $A(i)$ denote the activation count of row $i$ and $N_{BO}$ denote the back-off threshold.
We assume that a read disturbance bitflip is only possible if a row can be activated at least \gls{nrh} times \ous{0}{before \om{4}{its} victims are preventively refreshed}.
Therefore, a system is secure if and only if \param{$A(i) < \nrh{}$} \ous{0}{holds} for all rows $i$ at all times.
We prove that \X{} is secure against read disturbance bitflips by investigating the system state in four phases:
1) before a back-off is triggered,
2) during the window of normal traffic,
3) during the recovery period, and
4) after the recovery period.

First, before a back-off is triggered, the maximum number of activations \om{4}{to} a single row \om{4}{can be at most} $N_{BO} - 1$ (\om{6}{following from} property $P1$ and \ous{5}{property} $P2$).
Second, \ous{0}{during the window of normal traffic (after a back-off is triggered)}, a single row \ous{0}{can be activated} at most \ous{4}{\param{$A_{normal} = \SI{180}{\nano\second}$/$\trc{}$}}\ouscomment{4}{Anormal calculation is defined here} times before the first preventive refresh is issued.
Therefore, the total number of activations \om{4}{to} a row $i$ before the \ous{0}{memory controller issues the} first preventive refresh is $A(i) = N_{BO} + A_{normal}$ \ous{0}{(because $N_{BO}$th activation triggers the back-off)}.
Third, during the recovery period, \X{} preventively refreshes the potential victims and resets the activation counts of all rows $i$ where \param{$A(i) \geq N_{BO}$} (\om{6}{following from} \ous{5}{property $P1$} and \ous{5}{property $P3$}).\omcomment{6}{Sounds odd. By P3 meaning?}\ouscomment{6}{I was just referring to the properties that ensure the sentence. For example, \X{} tracks all activation correctly (P1) and refreshes any row reaching the back-off threshold during the recovery period (P3). Therefore, each back-off refreshes all rows $i$ where \param{$A(i) \geq N_{BO}$}.}\omcomment{6}{Is this all described?}\ouscomment{6}{I think properties themselves are clear. We could be more verbose within the text but it would mostly be redundant to what the properties say.}
Fourth, after the recovery period, a row \ous{0}{has} at most $N_{BO} - 1$ activations \ous{0}{(which is the same as step one)}.

These four phases show that \X{} performs preventive refreshes in a way that \param{$A(i) \leq N_{BO} + A_{normal}$} for all $i$ at all times.
Therefore, \X{} is secure against read disturbance bitflips \ous{0}{in configurations where} \param{$N_{BO} < \nrh{} - A_{normal}$}.

\head{Determining a Secure Aggressor Tracking Table Size}
An attacker can try to overwhelm the \emph{\om{4}{Aggressor} Tracking Table} (ATT, see \secref{sec:briefsummary}) by forcing many rows to reach $N_{BO}$ activations.
Our proof shows that an attacker can perform at most $A_{normal}$ additional activations after triggering a back-off.
The attacker can maximize the number of rows that require a refresh in three steps.
First, the attacker activates $A_{normal} + 1$ rows $N_{BO} - 1$ times each \emph{without} triggering a back-off.
Second, the attacker activates one of the rows for the $N_{BO}$th time, which triggers a back-off.
Third, the attacker activates the other $A_{normal}$ rows that did \emph{not} trigger the back-off for the $N_{BO}$th time during the window of normal traffic.
By doing so, an attacker can force \emph{at most} $A_{normal} + 1$ rows to reach $N_{BO}$ activations.
Therefore, ATT should be able to hold at least $A_{normal} + 1$ entries \ous{4}{(e.g., $\lfloor{} \SI{180}{\nano\second}/\trc{} \rfloor{} + 1 = 4$ entries for a $\trc{}$ of \SI{47}{\nano\second})}.\omcomment{4}{How is this calculation done?}\ouscomment{4}{changed to match the Anormal definition above}
\begin{figure*}[b]
\centering
\includegraphics[width=\linewidth]{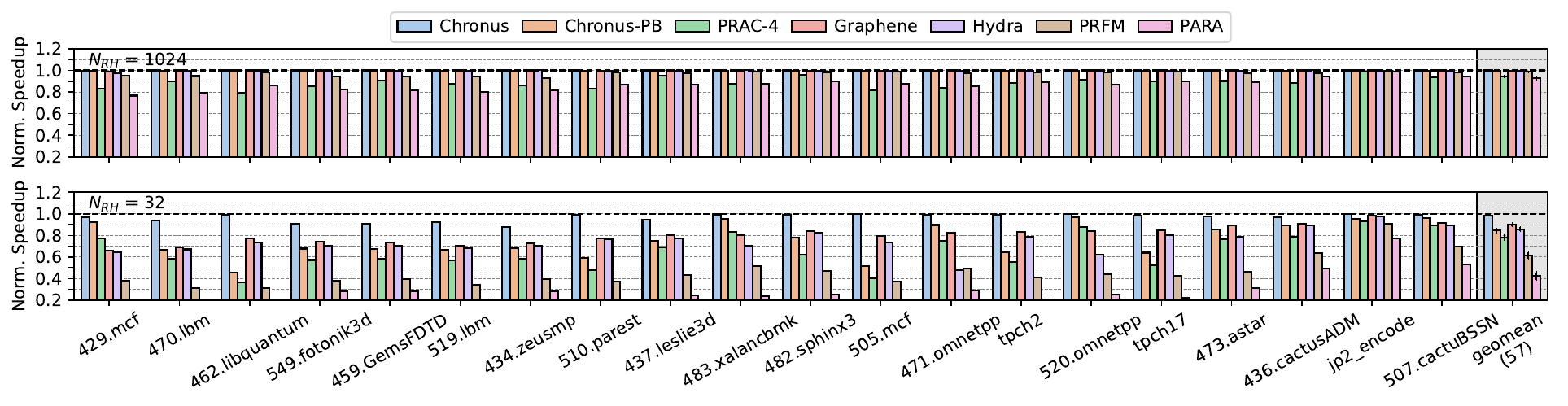}
\caption{Performance impact of evaluated read disturbance mitigation mechanisms on benign single-core workloads}
\label{fig:benign_singlecore}
\end{figure*}

\section{Experimental Methodology}
\label{sec:methodology}

We evaluate \X{}'s overheads on performance and DRAM energy consumption for existing and future DRAM chips, by sweeping \gls{nrh} from \param{1K} down to \param{20}.
\ous{0}{We consider \param{two} \X{} \om{6}{variants} to separately assess the effects of \om{6}{Concurrent Counter Update (CCU) and Chronus Back-Off:}
1) \om{6}{\emph{\X{}}}, our complete proposal that implements \om{7}{both} CCU \emph{and} Chronus Back-Off and
2) \X{} with \gls{prac} Back-Off (\om{6}{\emph{\X{}-PB}}), which implements \om{7}{only CCU and}, \om{4}{like \gls{prac}}, \emph{always} sends \param{four} preventive refreshes \emph{with} a delay period (i.e., \gls{prac}-4 with CCU).}

\head{\ous{6}{Comparison Points}} We compare \X{} to \gls{prac}~\cite{jedec2024jesd795c}, \gls{prfm}~\cite{jedec2024jesd795c}, and \param{three} \ous{2}{state-of-the-art} read disturbance mitigation mechanisms:
1) \om{6}{\emph{Graphene}}~\cite{park2020graphene}, a \om{4}{deterministic} mechanism that \ous{4}{uses the Misra-Gries frequent item counting algorithm~\cite{misra1982finding} and maintains frequently accessed row counters} completely within the \om{4}{memory controller}; 
2) \om{6}{\emph{PARA}}~\cite{kim2014flipping}, a \om{4}{probabilistic} \om{5}{memory-controller-based} mechanism that does \emph{not} maintain any counters; and
3) \om{6}{\emph{Hydra}}~\cite{qureshi2022hydra}, a \om{4}{deterministic} mechanism that maintains counters in the DRAM chip and caches them in the \om{4}{memory controller}.
\ous{9}{Appendix \ref{apx:abacuscomparison}} compares \X{} to ABACuS, a storage-optimized deterministic mechanism that maintains counters in the memory controller.\footnote{We evaluate ABACuS separately because \ous{9}{ABACuS performs} best with ABACuS's own DRAM address mapping configuration (as described in~\cite{olgun2024abacus}).}
To evaluate performance and DRAM energy consumption, we conduct cycle-level simulations using Ramulator 2.0~\cite{ramulator2github, luo2023ramulator2}, integrated with DRAMPower~\cite{drampower}.
We extend Ramulator 2.0~\cite{ramulator2github, luo2023ramulator2} with the implementations of \X{} \om{4}{and open source all of our code and scripts~\cite{chronusgithub}}.
We use the same system configuration and workloads in~\secref{sec:sensitivity}.

\section{Experimental Evaluation}
\label{sec:evaluation}

\head{\ous{2}{Single-core Performance}}
\figref{fig:benign_singlecore} presents the performance overheads of the evaluated read disturbance mitigation mechanisms for \om{4}{different} single-core applications at \gls{nrh} values of \param{1K} (top) and \param{32} (bottom).
Axes respectively show the single-core applications (x axis) and system performance (y axis) in terms of weighted speedup normalized to a baseline with \emph{no} read disturbance mitigation.
Different bars identify different read disturbance mitigation mechanisms.
\om{5}{\emph{geomean}} depicts the geometric mean of each mechanism across \param{57} single-core applications \ous{5}{and error bars show the standard error of the mean}~\cite{altman2005standard}.

We make \param{two} observations from \figref{fig:benign_singlecore}.
First, at \ous{6}{$\nrh{} =$~1K}, \X{} incurs the \om{5}{lowest} system performance overhead ($<$\param{0.1}\% on average) across all evaluated mitigation mechanisms.
We attribute the low system performance overhead of \X{} at relatively high \gls{nrh} values (i.e., $\nrh{} \geq$ 1K) to \ous{5}{the CCU mechanism of \X{} because \X{}-PB \om{6}{(which implements CCU but \emph{not} Chronus Back-Off)} performs similarly well at these thresholds}.
Second, \X{} outperforms all mitigation mechanisms even at very low \gls{nrh} values.
At \ous{6}{$\nrh{} =$~32}, \X{} induces \om{4}{\emph{only}} \param{6.8}\% average system performance overhead.
In \om{4}{contrast}, Graphene, Hydra, and \gls{prac}-4 induce \param{28.1}\%, \param{30.6}\%, and \param{\ous{10}{42.5\%}} average system performance overhead, respectively.
We attribute the low system performance overhead of \X{} at relatively low \gls{nrh} values (i.e., $\nrh{} \leq$ 32) to \ous{4}{securely preventing a wave attack and thereby having a less aggressive back-off threshold because \X{}-PB and \gls{prac}, \ous{5}{mechanisms that are vulnerable to a wave attack due to using \gls{prac} Back-Off,} both induce high system performance \om{6}{overheads} (respectively \param{32.2}\% and \param{\ous{10}{42.5\%}})}\ouscomment{10}{Diff:\\
old -> new:\\
46.5 -> 42.5}
for these same thresholds.

We conclude that
1) \X{} incurs \ous{5}{the lowest} system performance overhead to single-core applications at \gls{nrh} values \om{4}{similar to current DRAM chips} (e.g., $\nrh{} =$ \param{1K}) and
2) \X{} \ous{5}{is the best performing mechanism with $<$\param{6.8}\% system performance overhead at} future \gls{nrh} values (e.g., $\nrh{} <$ \param{64}).

\begin{figure}[h]
\centering
\includegraphics[width=\linewidth]{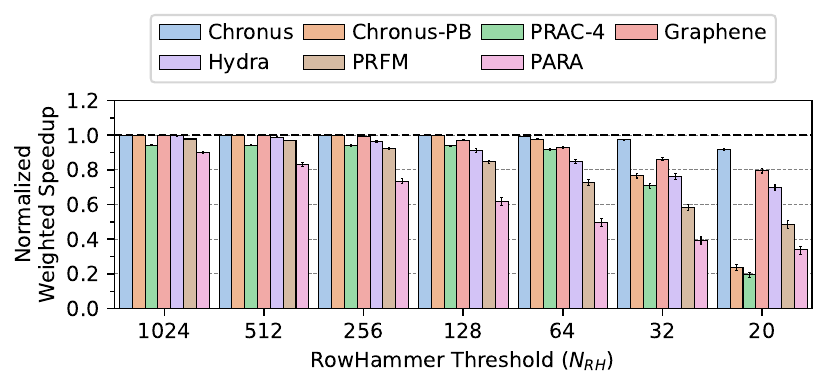}
\caption{Performance impact of evaluated read disturbance mitigation mechanisms on 60 benign four-core workloads}
\label{fig:benign_scaling}
\end{figure}

\head{\ous{2}{Multi-core System Performance}}
\figref{fig:benign_scaling} presents the performance overheads of the evaluated read disturbance mitigation mechanisms across \param{60} benign \param{four}-core workloads \om{4}{for \gls{nrh} values from \param{1K} to \param{20}}.
x and y axes respectively show the \gls{nrh} values and system performance in terms of weighted speedup normalized to a baseline with \emph{no} read disturbance mitigation (higher y value is better).
\ous{6}{Each colored bar depicts the mean system performance of a mechanism across 60 four-core workloads and error bars show the standard error of the mean across 60 four-core workloads}.

We make \ous{5}{\param{four}} observations from \figref{fig:benign_scaling}.
\ous{5}{First}, across all evaluated \gls{nrh} values, \X{} outperforms all evaluated read disturbance mitigation mechanisms.
\ous{5}{Second, across all \ous{6}{evaluated} mechanisms, \X{}'s system performance scales the best as the \iey{6}{concurrent\ous{6}{ly} running} application count in a workload increases.
For example, as the \ous{6}{concurrently running} application count increases from \param{one} (57 single-core workloads total \om{6}{in \figref{fig:benign_singlecore}}) to \param{four} (60 four-core workloads total \om{6}{in \figref{fig:benign_scaling}}) at \ous{6}{$\nrh{} =$~32},
\X{}'s average system performance improves respectively over \gls{prac}-4 by \param{\ous{10}{18.0\%}},\ouscomment{10}{Diff:\\
old -> new\\
23.5 -> 18.0}
Graphene by \param{14.8}\%, Hydra by \param{5.0}\%, and PARA by \param{64.4}\%.}
\ous{5}{Third}, as \gls{nrh} decreases, the system performance overhead of all \ous{6}{evaluated} mitigation mechanisms increases.
\ous{5}{Fourth}, \gls{prac}-4 underperforms against all \ous{6}{evaluated} mitigation mechanisms \om{6}{at $\nrh{} =$~20}.
\ous{5}{This is because \gls{prac}-4 inefficiently performs 4 preventive refreshes with each back-off and requires conservative configuration} against a potential wave attack.
In contrast, \X{}'s performance does \emph{not} get drastically \ous{6}{(i.e., $\geq$10\% overhead)} exacerbated \ous{6}{at $\nrh{} =$~20 for two reasons: \X{}}
\ous{6}{1)} dynamically adjusts the number of preventive refreshes as necessary and
\ous{6}{2)} can be less aggressively configured due to \om{7}{its mechanism} securely preventing a potential wave attack.

\ous{6}{We conclude that
1) \X{} incurs low system performance overhead on evaluated multi-core workloads at \gls{nrh} values similar to current DRAM chips (\ous{6}{e.g.}, $\nrh{} =$~\param{1K}) and
2) \X{} scales the best to future \gls{nrh} values (e.g., $\nrh{} <$~\param{64}) with $<$\param{8.3}\% system performance overhead}.

\begin{figure}[h]
\centering
\includegraphics[width=\linewidth]{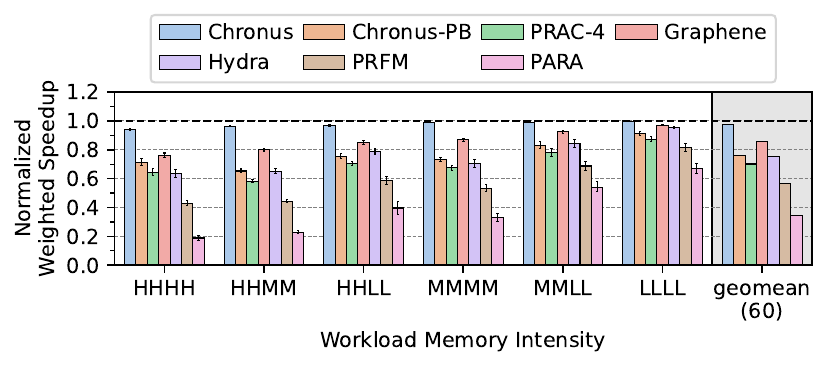}
\caption{Performance impact of evaluated read disturbance mitigation mechanisms on \om{5}{\param{six}} workload types of varying \om{5}{memory} intensities}
\label{fig:benign_workloads}
\end{figure}

\head{\ous{2}{Sensitivity to Workload Memory Intensity}}
We further study the system performance overhead \ous{4}{of each mechanism at different memory intensity levels}.
\figref{fig:benign_workloads} presents the performance overheads of the evaluated read disturbance mitigation mechanisms for \ous{0}{\param{four}-core} workload types of varying intensities at \ous{6}{$\nrh{} =$~32}.
Axes respectively show the workload types (x axis) and system performance (y axis) in terms of weighted speedup normalized to a baseline with \emph{no} read disturbance mitigation (higher y value is better).
Each letter in the workload type identifies the memory intensity of an application as High (H), Medium (M), or Low (L).
\ous{6}{In each workload type, colored bars show the mean system performance of a mechanism across 10 workloads and error bars show the standard error of the mean across 10 workloads}.
\om{5}{\emph{geomean}} depicts the geometric mean of each mechanism across \param{60} four-core workloads.

We make \param{two} observations from \figref{fig:benign_workloads}.
First, \X{} outperforms all evaluated mitigation mechanisms across all workload intensity types.
Second, the system performance overhead of evaluated mitigation mechanisms increases with the memory intensity of the workloads.
\ous{6}{We conclude that \X{}'s system performance stays the best with varying memory intensity}.

\begin{figure}[h]
\centering
\includegraphics[width=\linewidth]{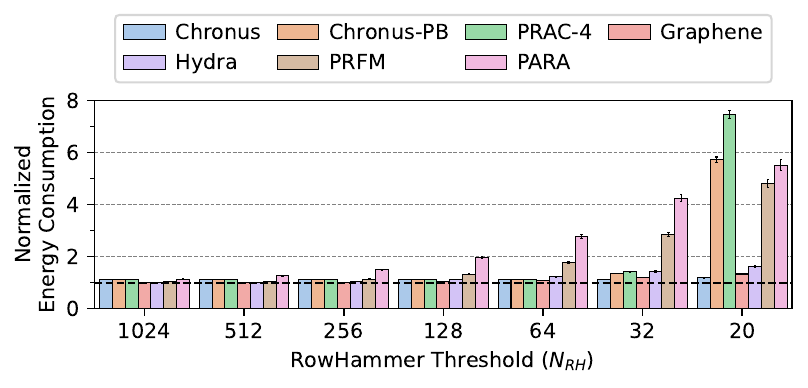}
\caption{Energy impact of evaluated read disturbance mitigation mechanisms on \param{60} benign four-core workloads}
\label{fig:benign_energy}
\end{figure}

\head{\ous{2}{DRAM Energy}}
We study the DRAM energy overheads of read disturbance mechanisms.
\figref{fig:benign_energy} presents the energy consumption of the evaluated read disturbance mitigation mechanisms (y axis) \ous{4}{normalized to a baseline with \emph{no} read disturbance mitigation mechanism} as \gls{nrh} (x axis) decreases.
\ous{6}{Each colored bar depicts the mean energy consumption of a mechanism across 60 four-core workloads and error bars show the standard error of the mean across 60 four-core workloads}.\omcomment{6}{Where is the mean?}\ouscomment{6}{Each bar is a mean across 60 four-core workloads}

We make \ous{4}{\param{seven}} observations from \figref{fig:benign_energy}.
First, as \gls{nrh} decreases, the energy \om{7}{overheads} of all \ous{6}{evaluated} mitigation mechanisms increases.
Second, at \ous{6}{$\nrh{} =$~1K}, \X{} increases the average (maximum) energy \ous{6}{consumption} by \param{10.3}\% (\param{10.7}\%) \ous{4}{over a baseline with \emph{no} read disturbance mechanism}.
At this threshold, \om{11}{\X{} has \param{\ous{10}{3.5}}\% lower energy overhead than \gls{prac}-4}.\ouscomment{10}{Diff:
old -> new\\
44.4 -> 3.5}
This is because \X{}'s \ous{5}{CCU} \om{5}{mechanism} \ous{4}{introduces less energy overhead than} \gls{prac}'s energy overheads, \om{5}{which are high due to higher} DRAM timing parameters and \om{5}{consequently slower} \ous{4}{execution}.
Third, when \gls{nrh} decreases from \param{1K} to \param{20}, \X{}'s average (maximum) \ous{6}{energy overhead} increases from \param{10.3}\% (\param{10.7}\%) to \param{17.9}\% (\param{20.7}\%).
\ous{4}{Fourth}, \X{} \om{7}{leads to lower energy consumption than} all \ous{6}{evaluated} mitigation mechanisms \ous{6}{at $\nrh{} \leq$~64}.
\ous{4}{Fifth}, when \gls{nrh} decreases from \param{1K} to \param{20}, \gls{prfm}'s average energy \om{7}{overhead} increases from \ous{6}{$<$1.1x to 4.8x}.
\ous{4}{Sixth}, when \gls{nrh} decreases from \param{1K} to \param{20}, \gls{prac}-4's average energy \om{7}{overhead} significantly increases from \ous{10}{1.1x} to \ous{10}{6.6x}.\ouscomment{10}{Diff:
old -> new\\
1.2x and 7.9x -> 1.1x and 6.6x}
We attribute the high \om{7}{increases} in energy consumption \om{7}{overheads} of \gls{prfm} and \gls{prac}-4 as \gls{nrh} decreases to
1) their conservative preventive refresh thresholds against \ous{6}{a potential} wave attack and
2) benign applications triggering many preventive refreshes.
\ous{4}{Seventh}, as \gls{nrh} decreases from \param{1K} to \param{20}, average energy overheads of Graphene and Hydra increase \om{5}{greatly}, from $<$\param{0.1}\% and \param{0.3}\% to \param{33.0}\% and \param{62.0}\%, respectively.

We conclude that
1) \X{} consumes \om{5}{lower} DRAM energy compared to \gls{prac} at \gls{nrh} values \ous{4}{similar to current DRAM chips} (e.g., $\nrh{} =$~\param{1K}) and
2) \X{} \om{5}{is the lowest energy-overhead solution at} \ous{5}{future \gls{nrh} values (e.g., $\nrh{} <$~64)}.

\begin{figure}[h]
\centering
\includegraphics[width=\linewidth]{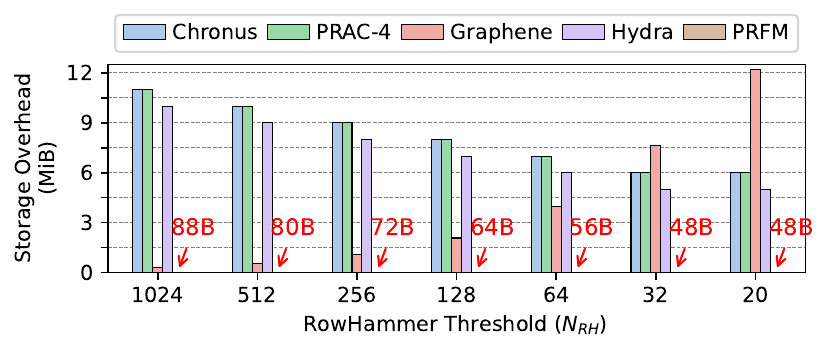}
\caption{Storage used by \ous{0}{\X{} (DRAM), Graphene (CAM), Hydra (DRAM+SRAM), PRAC (DRAM), and PRFM (SRAM)} as a function of RowHammer threshold for a DRAM module with \param{64} banks and \param{128K} rows per bank}
\label{fig:benign_storage}
\end{figure}

\head{\ous{2}{Storage}}
\label{subsec:storageeval}
\figref{fig:benign_storage} shows the storage overhead \ous{0}{in MiB (y axis)} of the evaluated read disturbance mitigation mechanisms as \gls{nrh} decreases \ous{0}{(x axis)}.
\ous{0}{We evaluate the storage usage of \X{} (DRAM), Graphene (CAM), Hydra (DRAM+SRAM), PRAC (DRAM), and PRFM (SRAM) as a function of \gls{nrh} for a DRAM module with \param{64} banks and \param{128K} rows per bank.}\footnote{\ous{4}{PARA~\cite{kim2014flipping} is \emph{stateless} (i.e., does \emph{not} track or store row activation history) and only requires a random number generator. Therefore, we do \emph{not} include PARA in our storage overhead evaluation.}}\ouscomment{5}{This footnote causes the line (and the previous paragraphs) to get placed weirdly and I can't find a way to quickly fix now. I will hunt it if it remains this way in the ``beautification'' process.}

\ous{6}{We make \param{four} observations from \figref{fig:benign_storage}}.
First, as \gls{nrh} decreases from \param{1K} to \param{20}, \X{}'s storage overhead in DRAM \ous{0}{is the same as \gls{prac}'s as they both employ a counter per DRAM row}.
Second, as \gls{nrh} decreases from \param{1K} to \param{20}, Graphene's storage overhead in CPU increases significantly (by \param{50.3}x) due to the need to track many more rows.
Third, as \gls{nrh} decreases from \param{1K} to \param{20}, \ous{0}{\X{}} and Hydra's storage overheads in DRAM reduce by \param{45.5}\% and \param{50.0}\%, respectively.
We note that while Hydra's cache structure in CPU does \emph{not} change, the overall cache size reduces with \gls{nrh} (by \param{43.9}\% from \param{1K} to \param{20}) as smaller cache entries are sufficient to track activations.
Fourth, \gls{prfm} incurs the least storage overhead in the CPU among the evaluated mitigation techniques as it requires \om{4}{only} one counter per bank.

We conclude that \X{}, \gls{prac}, \gls{prfm}, and Hydra incur low storage overheads and scale well with decreasing \gls{nrh} values as they either $i$) keep counters in DRAM where a large amount of storage is available at high density or $ii$) require only a small set of counters.

\section{Performance Degradation Attack}
\label{sec:evaluation_dos}

An attacker can take advantage of \ous{0}{\X{} and} \gls{prac} to mount memory performance (or denial of memory service) attacks~\cite{mutlu2007memory} by triggering many \ous{0}{back-offs}.
This section presents
1) the \ous{5}{\emph{worst-case} access pattern that causes the maximum theoretical DRAM bandwidth consumption in a} \X{} and \gls{prac}-protected system and
2) simulation results.\footnote{\ous{9}{We prove the worst-case property of the adversarial pattern in Appendix~\ref{apx:adversarialproof}}.}

\head{\gls{prac} Theoretical Analysis}
We calculate the maximum possible fraction of \om{4}{execution} time that preventive actions take in a \gls{prac}-protected system.
\ous{0}{Triggering a \gls{prac}} back-off signal takes $\aboth\times\trc$, which causes $\bonrefs$ RFM commands, blocking the bank for a time window of $\bonrefs\times\trfm$.\footnote{Attacking rows \om{5}{concurrently in different banks} does \emph{not} increase attack efficiency because the memory controller issues \om{4}{all-bank RFM commands (\emph{RFMab})} after a back-off, thereby refreshing the potential victim \om{7}{rows} of all concurrent aggressor \om{7}{rows} \emph{without} additional overhead.}
Therefore, an attacker can block a DRAM bank for $(\bonrefs\times\trfm)/(\bonrefs\times\trfm+\aboth\times\trc)$ of time.
We configure $\aboth$, $\bonrefs$, $\trfm$, and $\trc$ as \param{1}, \param{4}, \param{\SI{350}{\nano\second}}, and \param{\SI{52}{\nano\second}} \om{7}{for} \ous{6}{$\nrh{} =$~20}, based on the \param{DDR5-3200AN} \ous{0}{speed bin} in the JEDEC standard~\cite{jedec2024jesd795c}.
We observe that an attacker can \emph{theoretically} consume \param{94\%} of DRAM throughput by triggering \gls{prac} back-offs.

\head{\X{} Theoretical Analysis}
\ous{0}{We calculate the maximum possible fraction of \ous{4}{execution} time that preventive actions take in a \X{}-protected system.
Triggering a \X{} back-off signal takes $\aboth\times\trc$, which causes \param{one}\footnote{An attacker should \emph{not} trigger more than one \gls{rfm} command per back-off as it takes $N_{BO}\times\trc$ per additional \gls{rfm} command (time to increase another rows activation count), significantly decreasing attack efficiency.} RFM command, blocking the bank for a time window of $\trfm$.
Therefore, an attacker can block a DRAM bank for $(\trfm)/(\trfm+\aboth\times\trc)$ of time.
We configure $\aboth$, $\trfm$, and $\trc$ as \param{16}, \param{\SI{350}{\nano\second}}, and \param{\SI{47}{\nano\second}} \om{7}{for} \ous{6}{$\nrh{} =$~20}, based on the \param{DDR5-3200AN} \ous{0}{speed bin} in the JEDEC standard~\cite{jedec2024jesd795c}.
We observe that an attacker can \emph{theoretically} consume \agy{0}{up to} \param{32\%} of DRAM throughput by triggering \X{} back-offs, which is significantly smaller \om{4}{than} \gls{prac}'s \emph{theoretical} overheads (\param{94}\%).}

\head{Simulation}
To understand the system performance degradation an attacker could cause by hogging the available DRAM throughput with preventive refreshes, we simulate \param{60} \param{four}-core workload mixes of varying memory intensities where one core maliciously hammers \param{8} rows in each of \param{4} banks.\footnote{We experimentally found these values to yield the highest performance overhead for \X{} and \gls{prac} in our system configuration.}

\ous{0}{Our system performance (weighted speedup~\cite{eyerman2008systemlevel, snavely2000symbiotic}) and maximum slowdown \om{4}{of} a single application~\cite{kim2010thread} results for \gls{nrh} values of 128 and 20 show that
1) \gls{prac}-4 reduces system performance on average (maximum) by \ous{10}{\param{15.2}\% (\param{26.3}\%) and \param{84.1}\% (\param{93.5}\%)}\ouscomment{10}{Diff:\\
$\nrh{}$: old -> new\\
128: 18.4 (29.0) -> 15.2 (26.3)\\
20: 86.8 (94.6) -> 84.1 (93.5)}
with a maximum slowdown of \ous{10}{\param{63.3}\% and \param{97.5}\%},\ouscomment{10}{Diff:\\
$\nrh{}$: old -> new\\
128: 64.5 -> 63.3\\
20: 97.7 -> 97.5}
respectively and
2) \X{} reduces system performance on average (maximum) by \param{13.3}\% (\param{24.9}\%) and \param{25.9}\% (\param{39.3}\%) with a maximum slowdown of \param{62.3}\% and \param{69.9}\%, respectively.}
\ous{0}{We conclude that \X{} significantly reduces the memory performance attack vulnerability compared to \gls{prac}.}
\section{Future Research Directions}
\label{sec:implications}

Although \X{} addresses major weaknesses of \gls{prac}, \om{5}{we believe there is still significant room for improvement in read disturbance solutions (including \X{} and \gls{prac})}.
\ous{5}{In this section, we discuss \param{five} directions to improve read disturbance solutions}.

A first direction is to determine read disturbance thresholds more accurately.
Read disturbance mitigation mechanisms provide security under the assumption that a safe \gls{nrh} value is known.
However, determining a safe \gls{nrh} value to cover all rows is not easy due to the need to identify the lowest \gls{nrh} in the presence of \gls{nrh} variation across \om{7}{data patterns~\cite{kim2014flipping, kim2020revisiting, luo2023rowpress, yaglikci2022understanding, olgun2025variable}, temperature~\cite{kim2020revisiting, orosa2022spyhammer}, access patterns~\cite{luo2023rowpress, orosa2021deeper, kogler2022half, luo2024experimental}, voltage~\cite{yaglikci2022understanding}, physical row locations~\cite{orosa2021deeper, yaglikci2024spatial}, and time~\cite{olgun2025variable}}, as shown by multiple works~\cite{kim2014flipping, orosa2021deeper, luo2023rowpress, kim2020revisiting, saroiu2022configure, olgun2023understanding, zhou2023threshold, olgun2025variable, tugrul2025understanding, luo2024experimental, orosa2022spyhammer, yaglikci2022understanding}.
A second direction to explore is to implement \X{} and \gls{prac} counters more efficiently.
Future work can improve the efficiency with
$i$) counter update policies against different read disturbance attacks (e.g., RowPress~\cite{luo2023rowpress})\footnote{One way to mitigate RowPress is to configure RowHammer solutions at lower thresholds (e.g., $<$500)~\cite{luo2023rowpress}. We already show that \X{} outperforms state-of-the-art \om{5}{RowPress} solutions~\cite{park2020graphene,qureshi2022hydra,kim2014flipping,jedec2024jesd795c} at such low thresholds.} and
$ii$) different architectures that improve area and DRAM energy overheads of both \X{} and \gls{prac}.
A third direction is to leverage the significant variation in read disturbance vulnerability across rows to avoid overprotecting the vast majority of the rows~\cite{yaglikci2024spatial,orosa2021deeper}.
For example, Svärd~\cite{yaglikci2024spatial} enhances existing RowHammer solutions to become spatial RowHammer threshold aware.
\X{} could similarly be combined with Svärd to improve system performance by extending the counter subarray to store the necessary meta-data about the \ous{4}{read disturbance vulnerability}\omcomment{4}{Defined?}\ouscomment{4}{revised, the original paper uses "vulnerability bin"} (e.g., very low, low, average, high) of each row.
\X{} already has relatively low performance overheads (e.g., $<$0.1\%) \emph{without} spatial variation awareness even at very low RowHammer thresholds (e.g., $<$100) and we expect \X{}’ performance overheads would \om{5}{further reduce} with spatial variation awareness.
A fourth direction is to defend against malicious attackers that exploit preventive refreshes.
Attackers can trigger increasing amounts of preventive refreshes as \gls{nrh} decreases, allowing a new attack vector to conduct memory performance attacks~\cite{mutlu2007memory}.
Preventing these performance attacks may be possible by accurately detecting and throttling workloads that trigger many preventive refreshes~\cite{yaglikci2021blockhammer, canpolat2024breakhammer}.
\ous{4}{A fifth direction is to explore a more flexible DRAM interface design and protocol \om{6}{(as in~\cite{hassan2024self})}.
Improving the flexibility of the interface and intelligently dividing the work between the memory controller and DRAM by separating \om{5}{the responsibility of handling/controlling} access operations (e.g., reading and writing data) and maintenance operations (e.g., DRAM refresh and read disturbance mitigation, \om{5}{like \X{}}) can significantly improve system performance and energy, \om{6}{as shown by Self-Managing DRAM}~\cite{hassan2024self}}.
\section{Related Work}
\label{sec:relatedwork}

This is the first work that
1) rigorously analyzes the security and performance and
2) solves the major problems of \gls{prac}, a key feature introduced in the latest JEDEC DDR5 DRAM specification~\cite{jedec2024jesd795c} \om{5}{to prevent read disturbance bitflips~\cite{kim2014flipping, luo2023rowpress}}.
\om{4}{An earlier version of this paper}~\cite{canpolat2024understanding} analyzes the security and performance benefits of \gls{prac}, \om{5}{but~\cite{canpolat2024understanding}} does \emph{not} propose or evaluate a mechanism that solves \gls{prac}'s major weaknesses.

We propose \X{}, which addresses the \om{5}{major} weaknesses of \gls{prac} \om{5}{by} 1) \om{5}{reducing the latency of its counter maintenance operations} and 2) preventing adversarial access patterns.
\ous{5}{\secref{sec:evaluation} and \secref{sec:evaluation_dos}} \om{4}{already} qualitatively and quantitatively compare \X{} to \gls{prac} and \ous{4}{three prominent read disturbance} mitigation mechanisms: \ous{4}{Graphene~\cite{park2020graphene}, PARA~\cite{kim2014flipping}, and Hydra~\cite{qureshi2022hydra}.
We demonstrate that \X{} outperforms these evaluated mechanisms for \om{5}{modern} and future DRAM chips.
In this section, we discuss other read disturbance mitigation techniques}.

\noindent
\textbf{Per-Row Activation Tracking.}\omcomment{4}{Start with the decade discussion, then Panopticon, and finally Hydra.}\ouscomment{4}{Acknowledged. Reordered.}
\om{4}{Per}-row activation counters were already discussed close to a decade ago in the original RowHammer paper~\cite{kim2014flipping} and other works~\cite{kim2014architectural,yaglikci2021security,bains2016distributed,bains2016row}.
Building on these \om{5}{works}, our paper implements these counters \emph{without} inducing high performance and hardware complexity overheads to modern and future DRAM chips.
Panopticon~\cite{bennett2021panopticon} proposes implementing separate per-row counters in a \emph{counter mat} and using the \texttt{alert\_n} signal to request time for preventive refreshes.
Panopticon does \emph{not} discuss performance, energy, or hardware complexity overhead.
On the other hand, \X{} replicates an existing DRAM bank structure (i.e., a subarray) and quantifies the performance, energy, and hardware complexity overheads.
Hydra~\cite{qureshi2022hydra}, a memory controller-based mechanism, proposes storing per-row tracking entries in DRAM and caching the entries in the memory controller.
Hydra does \emph{not} perfectly track per-row activations and does \emph{not} propose a way to update in-DRAM counters \ous{6}{concurrently while} serving memory requests.
\om{5}{In \secref{sec:evaluation}, we show that \X{} greatly outperforms Hydra.}\omcomment{5}{Does Hydra need this much discussion?}\ouscomment{5}{Chopped Hydra a bit}

\noindent
\textbf{Other \om{4}{On-DRAM}-\ous{4}{Die} Mitigation Techniques.}
DRAM manufacturers implement read disturbance mitigation techniques, also known as Target Row Refresh (TRR)~\cite{jedec2020jesd795,jedec2017ddr4,hassan2021utrr,frigo2020trrespass}, in commercial DRAM chips.
The specific designs of these techniques are not openly disclosed.
Recent research shows that custom attacks can bypass these mechanisms~\cite{frigo2020trrespass, hassan2021utrr, jattke2022blacksmith, deridder2021smash, van2016drammer, saroiu2022price} and cause read disturbance bitflips.

\noindent
\textbf{Hardware-based Mitigation Techniques.}
Prior works propose hardware-based mitigation techniques~\citeHardwareBasedMitigations{} to prevent read disturbance bitflips.
\om{5}{Some} of these works~\cite{kim2014flipping, you2019mrloc, son2017making, wang2021discreet,yaglikci2022hira,saroiu2022configure,kim2021hammerfilter} \om{5}{propose} probabilistic preventive refresh mechanisms to mitigate read disturbance at low area cost.
\om{5}{These} mechanisms do not provide deterministic read disturbance prevention and thus \om{5}{cause} high system performance overhead as \gls{nrh} decreases \om{5}{(as they need to generate many more preventive refreshes)}.
\om{5}{Three} prior works~\cite{joardar2022learning, joardar2022machine, naseredini2022alarm} propose machine-learning-based mechanisms.
\ous{6}{These mechanisms are \emph{not} fully secure \ous{7}{because they have} increasing bitflip probability \om{8}{with lower} read disturbance thresholds~\cite{joardar2022learning, joardar2022machine} or \om{7}{they require} error-correction codes to correct a small number of bitflips~\cite{naseredini2022alarm}}.
Another group of prior works~\cite{seyedzadeh2017cbt, seyedzadeh2018cbt, kang2020cattwo, lee2019twice, saileshwar2022randomized, saxena2022aqua, kim2022mithril, marazzi2022protrr, park2020graphene, woo2022scalable, olgun2024abacus} propose using the Misra-Gries frequent item counting algorithm~\cite{misra1982finding}.
\ous{4}{Misra-Gries-based mechanisms use} a large number of counters implemented with \ous{5}{content-addressable memory} for low \gls{nrh} values~\cite{olgun2024abacus, bostanci2024comet}, thereby inducing high \om{5}{hardware} area overheads \om{5}{(as we showed for Graphene~\cite{park2020graphene} in \secref{sec:evaluation})}.
In contrast, \X{} provides \om{6}{\emph{deterministic}} security guarantees \om{5}{at} low \gls{nrh} values \om{5}{with} \om{6}{\emph{low system performance and hardware overheads}}.

\noindent
\textbf{Software-based Mitigation Techniques.}
Several software-based read disturbance mitigation techniques~\cite{konoth2018zebram, van2018guardion, brasser2017can, bock2019riprh, aweke2016anvil, zhang2022softtrr, enomoto2022efficient} propose to avoid hardware-level modifications. However, these works cannot monitor \textit{all} memory requests and thus, {many} of them are shown to be defeated by recent attacks~\cite{qiao2016new, gruss2016rowhammer, gruss2018another, cojocar2019eccploit, zhang2019telehammer, kwong2020rambleed, zhang2020pthammer}.

\noindent
\textbf{Integrity-based Mitigation Techniques.}
Another set of mitigation techniques~\integrityBasedMitigationsAllCitations{} implements integrity check mechanisms that identify and correct potential bitflips.
However, it is either impossible or too costly to address all read disturbance bitflips with these mechanisms.

\section{Conclusion}
\label{sec:conclusion}

We presented the first rigorous security, performance, energy, and cost analyses of \gls{prac} and proposed a new mechanism, \X{}, which addresses \gls{prac}'s two major weaknesses.
Our analyses show that \gls{prac} increases the critical DRAM access latency parameters due to the additional time required to increment activation counters and performs a \ous{4}{fixed} number of preventive refreshes at a time, making it vulnerable to an adversarial access pattern, known as the \emph{the wave attack} or \emph{the feinting attack}.
These two weaknesses of \gls{prac} cause significant performance \om{5}{overheads} at current and future DRAM chips.
Our mechanism, \X{}, \om{6}{solves} \gls{prac}'s two \om{5}{major} problems \om{6}{by} \ous{5}{1)} updating row activation counters concurrently \ous{5}{while} serving accesses by physically separating counters from the data and
\ous{5}{2)} dynamically controlling the number of preventive refreshes performed.
Our evaluation shows that \X{} outperforms \ous{5}{\gls{prac} and \gls{prfm}, the \om{6}{state-of-the-art} industry solutions to read disturbance}, and \param{three} other state-of-the-art \om{5}{read disturbance mitigation} proposals in terms of \om{6}{both} system performance and DRAM energy, \om{6}{especially for future DRAM chips with higher read disturbance vulnerability}.
\om{6}{We believe \X{} provides a robust and efficient solution to read disturbance \ous{6}{for current and future DRAM chips at low area, performance, and energy costs}.
We hope that future research continues to improve read disturbance solutions with new ideas, interfaces, and even more efficient techniques.}



\section*{Acknowledgments} {
This paper is a significantly extended version of an earlier work presented at DRAMSec 2024~\cite{canpolat2024understanding}.
This version of the paper is an updated version of our original \om{14}{HPCA 2025} conference publication~\cite{canpolat2025chronus} that fixes a bug reported by~\cite{kim2025per} regarding \gls{prac}'s DRAM timing constraints, as described in Appendix~\ref{apx:errata}.
We thank the anonymous reviewers of DRAMSec 2024 and HPCA 2025 (both main submission and artifact evaluation) for the encouraging feedback.
We thank the SAFARI Research Group members for valuable feedback and the stimulating scientific and intellectual environment.
We acknowledge the generous gift funding provided by our industrial partners (especially Google, Huawei, Intel, Microsoft, VMware), which has been instrumental in enabling the research we have been conducting on read disturbance in DRAM since 2011~\cite{mutlu2023retrospective}.
This work was also in part supported by \om{3}{a} Google Security and Privacy Research Award, the Microsoft Swiss Joint Research Center, \ous{3}{and the ETH Future Computing Laboratory (EFCL)}\ouscomment{3}{Oguz Hoca requested to include EFCL}.}\ouscomment{7}{Dennard patent was cited but bumped its citation up (2 now). Changed patent application citations to the final patent numbers and fixed broken citations.}

\balance
\bibliographystyle{unsrt}
\bibliography{refs}

\begin{thebibliography}{100}

\bibitem{kim2014flipping}
Y.~{Kim}, R.~{Daly}, J.~{Kim}, C.~{Fallin}, J.~H. {Lee}, D.~{Lee}, C.~{Wilkerson}, K.~{Lai}, and O.~{Mutlu}.
\newblock {Flipping Bits in Memory Without Accessing Them: An Experimental Study of DRAM Disturbance Errors}.
\newblock In {\em ISCA}, 2014.

\bibitem{dennard1968fieldeffect}
Robert~H. Dennard.
\newblock {Field-Effect Transistor Memory}.
\newblock U.S. Patent 3,387,286, 1968.

\bibitem{saxena2022aqua}
Anish Saxena, Gururaj Saileshwar, Prashant~J. Nair, and Moinuddin Qureshi.
\newblock {AQUA: Scalable Rowhammer Mitigation by Quarantining Aggressor Rows at Runtime}.
\newblock In {\em MICRO}, 2022.

\bibitem{park2020graphene}
Yeonhong Park, Woosuk Kwon, Eojin Lee, Tae~Jun Ham, Jung~Ho Ahn, and Jae~W Lee.
\newblock {Graphene: Strong yet Lightweight Row Hammer Protection}.
\newblock In {\em MICRO}, 2020.

\bibitem{kim2022mithril}
Michael~Jaemin Kim, Jaehyun Park, Yeonhong Park, Wanju Doh, Namhoon Kim, Tae~Jun Ham, Jae~W Lee, and Jung~Ho Ahn.
\newblock {Mithril: Cooperative Row Hammer Protection on Commodity DRAM Leveraging Managed Refresh}.
\newblock In {\em HPCA}, 2022.

\bibitem{you2019mrloc}
Jung~Min You and Joon-Sung Yang.
\newblock {MRLoc: Mitigating Row-Hammering Based on Memory Locality}.
\newblock In {\em DAC}, 2019.

\bibitem{kang2020cattwo}
Ingab Kang, Eojin Lee, and Jung~Ho Ahn.
\newblock {CAT-TWO: Counter-Based Adaptive Tree, Time Window Optimized for {DRAM} Row-Hammer Prevention}.
\newblock {\em {IEEE} Access}, 2020.

\bibitem{yaglikci2022hira}
A.~Giray Ya{\u{g}}lik{\c{c}}i, Ataberk Olgun, Minesh Patel, Haocong Luo, Hasan Hassan, Lois Orosa, O{\u{g}}uz Ergin, and Onur Mutlu.
\newblock {HiRA: Hidden Row Activation for Reducing Refresh Latency of Off-the-Shelf DRAM Chips}.
\newblock In {\em MICRO}, 2022.

\bibitem{hassan2022acase}
Hasan Hassan, Ataberk Olgun, A.~Giray Ya\u{g}l{\i}k\c{c}{\i}, Haocong Luo, and Onur Mutlu.
\newblock {A Case for Self-Managing DRAM Chips: Improving Performance, Efficiency, Reliability, and Security via Autonomous in-DRAM Maintenance Operations}.
\newblock arXiv:2207.13358 [cs.AR], 2022.

\bibitem{zhou2022ltpim}
Ranyang Zhou, Sepehr Tabrizchi, Arman Roohi, and Shaahin Angizi.
\newblock {LT-PIM: An LUT-Based Processing-in-DRAM Architecture with RowHammer Self-Tracking}.
\newblock {\em CAL}, 2022.

\bibitem{son2017making}
Mungyu Son, Hyunsun Park, Junwhan Ahn, and Sungjoo Yoo.
\newblock {Making DRAM Stronger Against Row Hammering}.
\newblock In {\em DAC}, 2017.

\bibitem{saileshwar2022randomized}
Gururaj Saileshwar, Bolin Wang, Moinuddin Qureshi, and Prashant~J Nair.
\newblock {Randomized Row-Swap: Mitigating Row Hammer by Breaking Spatial Correlation Between Aggressor and Victim Rows}.
\newblock In {\em ASPLOS}, 2022.

\bibitem{ryu2017overcoming}
Seong-Wan Ryu, Kyungkyu Min, Jungho Shin, Heimi Kwon, Donghoon Nam, Taekyung Oh, Tae-Su Jang, Minsoo Yoo, Yongtaik Kim, and Sungjoo Hong.
\newblock {Overcoming the Reliability Limitation in the Ultimately Scaled DRAM using Silicon Migration Technique by Hydrogen Annealing}.
\newblock In {\em IEDM}, 2017.

\bibitem{saroiu2022price}
Stefan Saroiu, Alec Wolman, and Lucian Cojocar.
\newblock {The Price of Secrecy: How Hiding Internal DRAM Topologies Hurts Rowhammer Defenses}.
\newblock In {\em IRPS}, 2022.

\bibitem{han2021surround}
Jin Han, Jungsik Kim, Dafna Beery, K~Deniz Bozdag, Peter Cuevas, Amitay Levi, Irwin Tain, Khai Tran, Andrew~J Walker, Senthil~Vadakupudhu Palayam, et~al.
\newblock {Surround Gate Transistor With Epitaxially Grown Si Pillar and Simulation Study on Soft Error and Rowhammer Tolerance for DRAM}.
\newblock {\em TED}, 2021.

\bibitem{saroiu2022configure}
Stefan Saroiu and Alec Wolman.
\newblock {How to Configure Row-Sampling-Based Rowhammer Defenses}.
\newblock DRAMSec, 2022.

\bibitem{lee2019twice}
Eojin Lee, Ingab Kang, Sukhan Lee, G~{Edward Suh}, and Jung {Ho Ahn}.
\newblock {TWiCe: Preventing Row-Hammering by Exploiting Time Window Counters}.
\newblock In {\em ISCA}, 2019.

\bibitem{yaglikci2021blockhammer}
A.~Giray Ya{\u{g}}l{\i}k{\c{c}}{\i}, Minesh Patel, Jeremie~S. Kim, Roknoddin Azizibarzoki, Ataberk Olgun, Lois Orosa, Hasan Hassan, Jisung Park, Konstantinos Kanellopoullos, Taha Shahroodi, Saugata Ghose, and Onur Mutlu.
\newblock {BlockHammer: Preventing RowHammer at Low Cost by Blacklisting Rapidly-Accessed DRAM Rows}.
\newblock In {\em HPCA}, 2021.

\bibitem{greenfield2012throttling}
Zvika Greenfield and Tomer Levy.
\newblock {Throttling Support for Row-Hammer Counters}, 2016.

\bibitem{devaux2021method}
Fabrice Devaux and Renaud Ayrignac.
\newblock {Method and Circuit for Protecting a DRAM Memory Device from the Row Hammer Effect}.
\newblock U.S. Patent 10,885,966, 2021.

\bibitem{lee2021cryoguard}
Gyu-Hyeon Lee, Seongmin Na, Ilkwon Byun, Dongmoon Min, and Jangwoo Kim.
\newblock {CryoGuard: A Near Refresh-Free Robust DRAM Design for Cryogenic Computing}.
\newblock In {\em ISCA}, 2021.

\bibitem{joardar2022learning}
Biresh~Kumar Joardar, Tyler~K Bletsch, and Krishnendu Chakrabarty.
\newblock {Learning to Mitigate RowHammer Attacks}.
\newblock In {\em DATE}, 2022.

\bibitem{joardar2022machine}
Biresh~Kumar Joardar, Tyler~K. Bletsch, and Krishnendu Chakrabarty.
\newblock {Machine Learning-Based Rowhammer Mitigation}.
\newblock {\em TCAD}, 2022.

\bibitem{qureshi2022hydra}
Moinuddin Qureshi, Aditya Rohan, Gururaj Saileshwar, and Prashant~J Nair.
\newblock {Hydra: Enabling Low-Overhead Mitigation of Row-Hammer at Ultra-Low Thresholds via Hybrid Tracking}.
\newblock In {\em ISCA}, 2022.

\bibitem{seyedzadeh2018cbt}
S.~M. {Seyedzadeh}, A.~K. {Jones}, and R.~{Melhem}.
\newblock {Mitigating Wordline Crosstalk Using Adaptive Trees of Counters}.
\newblock In {\em ISCA}, 2018.

\bibitem{naseredini2022alarm}
Amir Naseredini, Martin Berger, Matteo Sammartino, and Shale Xiong.
\newblock {ALARM: Active LeArning of Rowhammer Mitigations}.
\newblock \url{https://users.sussex.ac.uk/~mfb21/rh-draft.pdf}, 2022.

\bibitem{kim2015architectural}
Dae-Hyun Kim, Prashant~J Nair, and Moinuddin~K Qureshi.
\newblock {Architectural Support for Mitigating Row Hammering in DRAM Memories}.
\newblock {\em CAL}, 2015.

\bibitem{woo2022scalable}
Jeonghyun Woo, Gururaj Saileshwar, and Prashant~J. Nair.
\newblock {Scalable and Secure Row-Swap: Efficient and Safe Row Hammer Mitigation in Memory Systems}.
\newblock arXiv:2212.12613 [cs.CR], 2022.

\bibitem{seyedzadeh2017cbt}
Seyed~Mohammad Seyedzadeh, Alex~K. Jones, and Rami Melhem.
\newblock {Counter-Based Tree Structure for Row Hammering Mitigation in DRAM}.
\newblock {\em CAL}, 2017.

\bibitem{yang2016suppression}
Chia Yang, Chen~Kang Wei, Yu~Jing Chang, Tieh~Chiang Wu, Hsiu~Pin Chen, and Chao~Sung Lai.
\newblock {Suppression of RowHammer Effect by Doping Profile Modification in Saddle-Fin Array Devices for Sub-30-nm DRAM Technology}.
\newblock {\em TDMR}, 2016.

\bibitem{gomez2016dummy}
H.~{Gomez}, A.~{Amaya}, and E.~{Roa}.
\newblock {{DRAM} Row-Hammer Attack Reduction Using Dummy Cells}.
\newblock In {\em NORCAS}, 2016.

\bibitem{bostanci2024comet}
F~Nisa Bostanci, ISmail~Emir Y{\"u}ksel, Ataberk Olgun, Konstantinos Kanellopoulos, Yahya~Can Tu{\u{g}}rul, A~Giray Ya{\u{g}}li{\c{c}}i, Mohammad Sadrosadati, and Onur Mutlu.
\newblock {CoMeT: Count-Min-Sketch-Based Row Tracking to Mitigate RowHammer at Low Cost}.
\newblock In {\em HPCA}, 2024.

\bibitem{olgun2024abacus}
Ataberk Olgun, Yahya~Can Tugrul, Nisa Bostanci, Ismail~Emir Yuksel, Haocong Luo, Steve Rhyner, Abdullah~Giray Yaglikci, Geraldo~F Oliveira, and Onur Mutlu.
\newblock {ABACuS: All-Bank Activation Counters for Scalable and Low Overhead RowHammer Mitigation}.
\newblock In {\em USENIX Security}, 2024.

\bibitem{yaglikci2024spatial}
A.~Giray Ya{\u{g}}l{\i}k{c}{\i}, Yahya~Can Tu{\u{g}}rul, Geraldo~F De~Oliviera, Ismail~Emir Yüksel, Ataberk Olgun, Haocong Luo, and Onur Mutlu.
\newblock {Spatial Variation-Aware Read Disturbance Defenses: Experimental Analysis of Real DRAM Chips and Implications on Future Solutions}.
\newblock In {\em HPCA}, 2024.

\bibitem{marazzi2022protrr}
Michele Marazzi, Patrick Jattke, Flavien Solt, and Kaveh Razavi.
\newblock {ProTRR}: {Principled} yet {Optimal} {In-DRAM} {Target Row Refresh}.
\newblock In {\em {S\&P}}, 2022.

\bibitem{bennett2021panopticon}
Tanj Bennett, Stefan Saroiu, Alec Wolman, and Lucian Cojocar.
\newblock {Panopticon: A Complete In-DRAM Rowhammer Mitigation}.
\newblock DRAMSec, 2021.

\bibitem{hassan2019crow}
H.~{Hassan}, M.~{Patel}, J.~S. {Kim}, A.~G. {Ya\u{g}l{\i}k\c{c}{\i}}, N.~{Vijaykumar}, N.~{Mansouri Ghiasi}, S.~{Ghose}, and O.~{Mutlu}.
\newblock {CROW: A Low-Cost Substrate for Improving DRAM Performance, Energy Efficiency, and Reliability}.
\newblock In {\em ISCA}, 2019.

\bibitem{mutlu2023fundamentally}
Onur Mutlu, Ataberk Olgun, and A.~Giray Yaglikci.
\newblock {Fundamentally Understanding and Solving RowHammer}.
\newblock In {\em ASP-DAC}, 2023.

\bibitem{wang2021discreet}
Yicheng Wang, Yang Liu, Peiyun Wu, and Zhao Zhang.
\newblock {Discreet-PARA: Rowhammer Defense with Low Cost and High Efficiency}.
\newblock In {\em ICCD}, 2021.

\bibitem{ajorpaz2022evax}
Samira~Mirbagher Ajorpaz, Daniel Moghimi, Jeffrey~Neal Collins, Gilles Pokam, Nael Abu-Ghazaleh, and Dean Tullsen.
\newblock {EVAX: Towards a Practical, Pro-Active \& Adaptive Architecture for High Performance \& Security}.
\newblock In {\em MICRO}, 2022.

\bibitem{loughlin2022moesiprime}
Kevin Loughlin, Stefan Saroiu, Alec Wolman, Yatin~A. Manerkar, and Baris Kasikci.
\newblock {MOESI-Prime: Preventing Coherence-Induced Hammering in Commodity Workloads}.
\newblock In {\em ISCA}, 2022.

\bibitem{bains2016distributed}
K.S. Bains and J.B. Halbert.
\newblock {Distributed Row Hammer Tracking}.
\newblock U.S. Patent 9,299,400, April~3 2016.

\bibitem{france2022modeling}
Lo{\"\i}c France, Florent Bruguier, Maria Mushtaq, David Novo, and Pascal Benoit.
\newblock {Modeling Rowhammer in the gem5 Simulator}.
\newblock In {\em CHES}, 2022.

\bibitem{zhang2022softtrr}
Zhi Zhang, Yueqiang Cheng, Minghua Wang, Wei He, Wenhao Wang, Surya Nepal, Yansong Gao, Kang Li, Zhe Wang, and Chenggang Wu.
\newblock {SoftTRR: Protect Page Tables against Rowhammer Attacks using Software-Only Target Row Refresh}.
\newblock In {\em USENIX ATC}, 2022.

\bibitem{van2018guardion}
Victor {van der Veen}, Martina Lindorfer, Yanick Fratantonio, Harikrishnan~Padmanabha Pillai, Giovanni Vigna, Christopher Kruegel, Herbert Bos, and Kaveh Razavi.
\newblock {GuardION: Practical Mitigation of DMA-Based Rowhammer Attacks on ARM}.
\newblock In {\em {DIMVA}}, 2018.

\bibitem{loughlin2021stop}
Kevin Loughlin, Stefan Saroiu, Alec Wolman, and Baris Kasikci.
\newblock {Stop! Hammer Time: Rethinking Our Approach to Rowhammer Mitigations}.
\newblock In {\em HotOS}, 2021.

\bibitem{kim2014architectural}
Dae-Hyun Kim, Prashant~J Nair, and Moinuddin~K Qureshi.
\newblock {Architectural Support for Mitigating Row Hammering in DRAM Memories}.
\newblock {\em CAL}, 2014.

\bibitem{bains2015row}
Kuljit Bains, John Halbert, Christopher Mozak, Theodore Schoenborn, and Zvika Greenfield.
\newblock {Row Hammer Refresh Command}.
\newblock US Patent 9,117,544, 2015.

\bibitem{frigo2020trrespass}
Pietro Frigo, Emanuele Vannacci, Hasan Hassan, Victor {van der Veen}, Onur Mutlu, Cristiano Giuffrida, Herbert Bos, and Kaveh Razavi.
\newblock {TRRespass: Exploiting the Many Sides of Target Row Refresh}.
\newblock In {\em {S\&P}}, 2020.

\bibitem{sharma2022areview}
Sonia Sharma, Debdeep Sanyal, Arpit Mukhopadhyay, and Ramij~Hasan Shaik.
\newblock {A Review on Study of Defects of DRAM-RowHammer and Its Mitigation}.
\newblock {\em Journal For Basic Sciences}, 2022.

\bibitem{aichinger2015ddr}
Barbara Aichinger.
\newblock {DDR Memory Errors Caused by Row Hammer}.
\newblock In {\em HPEC}, 2015.

\bibitem{rh-hp}
{Hewlett-Packard Enterprise}.
\newblock {HP Moonshot Component Pack V2015.05.0}, 2015.

\bibitem{brasser2017can}
Ferdinand Brasser, Lucas Davi, David Gens, Christopher Liebchen, and Ahmad-Reza Sadeghi.
\newblock {Can't Touch This: Software-Only Mitigation Against Rowhammer Attacks Targeting Kernel Memory}.
\newblock In {\em USENIX Security}, 2017.

\bibitem{LenovoRefInc}
{Lenovo Group Ltd.}
\newblock {Row Hammer Privilege Escalation}.
\newblock \url{https://support.lenovo.com/us/en/product_security/row_hammer}, 2015.

\bibitem{AppleRefInc}
{Apple Inc.}
\newblock {About the Security Content of Mac EFI Security Update 2015-001}.
\newblock \url{https://support.apple.com/en-us/HT204934}.
\newblock {June 2015}.

\bibitem{gude2023defending}
C~Gude~Ramarao, K~Tejesh Kumar, G~Ujjinappa, and B~Vasu~Deva Naidu.
\newblock {Defending SoCs with FPGAs from Rowhammer Attacks}.
\newblock {\em Material Science}, 2023.

\bibitem{woo2023scalable}
Jeonghyun Woo, Gururaj Saileshwar, and Prashant~J Nair.
\newblock {Scalable and Secure Row-Swap: Efficient and Safe Row Hammer Mitigation in Memory Systems}.
\newblock In {\em HPCA}, 2023.

\bibitem{didio2023copyonflip}
Andrea Di~Dio, Koen Koning, Herbert Bos, and Cristiano Giuffrida.
\newblock {Copy-on-Flip: Hardening ECC Memory Against Rowhammer Attacks}.
\newblock In {\em NDSS}, 2023.

\bibitem{bains2016row}
Kuljit~S Bains and John~B Halbert.
\newblock {Row Hammer Monitoring Based on Stored Row Hammer Threshold Value}.
\newblock US Patent 10,083,737, 2016.

\bibitem{juffinger2023csi}
Jonas Juffinger, Lukas Lamster, Andreas Kogler, Maria Eichlseder, Moritz Lipp, and Daniel Gruss.
\newblock {CSI: Rowhammer-Cryptographic Security and Integrity against Rowhammer}.
\newblock In {\em SP}, 2023.

\bibitem{kim2023ddr5}
Woongrae Kim, Chulmoon Jung, Seongnyuh Yoo, Duckhwa Hong, Jeongjin Hwang, Jungmin Yoon, Ohyong Jung, Joonwoo Choi, Sanga Hyun, Mankeun Kang, et~al.
\newblock {A 1.1V 16Gb DDR5 DRAM with Probabilistic-Aggressor Tracking, Refresh-Management Functionality, Per-Row Hammer Tracking, a Multi-Step Precharge, and Core-Bias Modulation for Security and Reliability Enhancement}.
\newblock In {\em ISSCC}, 2023.

\bibitem{gautam2019row}
SK~Gautam, SK~Manhas, Arvind Kumar, Mahendra Pakala, and Ellie Yieh.
\newblock {Row Hammering Mitigation Using Metal Nanowire in Saddle Fin DRAM}.
\newblock {\em IEEE TED}, 2019.

\bibitem{bains-merged}
K.~Bains et~al.
\newblock {Row Hammer Refresh Command}.
\newblock U.S. Patents: 9,117,544 9,236,110 10,210,925, 2015.

\bibitem{enomoto2022efficient}
Shuhei Enomoto, Hiroki Kuzuno, and Hiroshi Yamada.
\newblock {Efficient Protection Mechanism for CPU Cache Flush Instruction Based Attacks}.
\newblock {\em IEICE Trans. Inf. Syst.}, 2022.

\bibitem{fakhrzadehgan2022safeguard}
Ali Fakhrzadehgan, Yale~N. Patt, Prashant~J. Nair, and Moinuddin~K. Qureshi.
\newblock {SafeGuard: Reducing the Security Risk from Row-Hammer via Low-Cost Integrity Protection}.
\newblock In {\em HPCA}, 2022.

\bibitem{manzhosov2022revisiting}
Evgeny Manzhosov, Adam Hastings, Meghna Pancholi, Ryan Piersma, Mohamed Tarek~Ibn Ziad, and Simha Sethumadhavan.
\newblock {Revisiting Residue Codes for Modern Memories}.
\newblock In {\em MICRO}, 2022.

\bibitem{guha2022criticality}
Krishnendu Guha and Amlan Chakrabarti.
\newblock {Criticality Based Reliability from Rowhammer Attacks in Multi-User-Multi-FPGA Platform}.
\newblock In {\em VLSID}, 2022.

\bibitem{hassan2021utrr}
Hasan Hassan, Yahya~Can Tugrul, Jeremie~S. Kim, Victor van~der Veen, Kaveh Razavi, and Onur Mutlu.
\newblock {Uncovering In-DRAM RowHammer Protection Mechanisms: A New Methodology, Custom RowHammer Patterns, and Implications}.
\newblock In {\em MICRO}, 2021.

\bibitem{wi2023shadow}
Minbok Wi, Jaehyun Park, Seoyoung Ko, Michael~Jaemin Kim, Nam~Sung Kim, Eojin Lee, and Jung~Ho Ahn.
\newblock {SHADOW: Preventing Row Hammer in DRAM with Intra-Subarray Row Shuffling}.
\newblock In {\em HPCA}, 2023.

\bibitem{konoth2018zebram}
Radhesh~Krishnan Konoth, Marco Oliverio, Andrei Tatar, Dennis Andriesse, Herbert Bos, Cristiano Giuffrida, and Kaveh Razavi.
\newblock {ZebRAM: Comprehensive and Compatible Software Protection Against Rowhammer Attacks}.
\newblock In {\em OSDI}, 2018.

\bibitem{vig2018rapid}
Saru Vig, Sarani Bhattacharya, Debdeep Mukhopadhyay, and Siew-Kei Lam.
\newblock {Rapid Detection of Rowhammer Attacks Using Dynamic Skewed Hash Tree}.
\newblock In {\em HASP}, 2018.

\bibitem{tomita2022extracting}
Chihiro Tomita, Makoto Takita, Kazuhide Fukushima, Yuto Nakano, Yoshiaki Shiraishi, and Masakatu Morii.
\newblock {Extracting the Secrets of OpenSSL with RAMBleed}.
\newblock {\em Sensors}, 2022.

\bibitem{zhou2023dnndefender}
Ranyang Zhou, Sabbir Ahmed, Adnan~Siraj Rakin, and Shaahin Angizi.
\newblock {DNN-Defender: An In-DRAM Deep Neural Network Defense Mechanism for Adversarial Weight Attack}.
\newblock arXiv:2305.08034 [cs.CR], 2023.

\bibitem{irazoqui2016mascat}
Gorka Irazoqui, Thomas Eisenbarth, and Berk Sunar.
\newblock {MASCAT: Stopping Microarchitectural Attacks Before Execution}.
\newblock {\em IACR Cryptology}, 2016.

\bibitem{arikan2022processor}
Kerem Ar{\i}kan, Alessandro Palumbo, Luca Cassano, Pedro Reviriego, Salvatore Pontarelli, Giuseppe Bianchi, O{\u{g}}uz Ergin, and Marco Ottavi.
\newblock {Processor Security: Detecting Microarchitectural Attacks via Count-Min Sketches}.
\newblock {\em VLSI}, 2022.

\bibitem{yang2017scanning}
Chia-Ming Yang, Chen-Kang Wei, Hsiu-Pin Chen, Jian-Shing Luo, Yu~Jing Chang, Tieh-Chiang Wu, and Chao-Sung Lai.
\newblock {Scanning Spreading Resistance Microscopy for Doping Profile in Saddle-Fin Devices}.
\newblock {\em TNANO}, 2017.

\bibitem{zhang2020leveraging}
Zhenkai Zhang, Zihao Zhan, Daniel Balasubramanian, Bo~Li, Peter Volgyesi, and Xenofon Koutsoukos.
\newblock {Leveraging EM Side-Channel Information to Detect Rowhammer Attacks}.
\newblock In {\em SP}, 2020.

\bibitem{saxena2023pt}
Anish Saxena, Gururaj Saileshwar, Jonas Juffinger, Andreas Kogler, Daniel Gruss, and Moinuddin Qureshi.
\newblock {PT-Guard: Integrity-Protected Page Tables to Defend Against Breakthrough Rowhammer Attacks}.
\newblock In {\em DSN}, 2023.

\bibitem{aweke2016anvil}
Zelalem~Birhanu Aweke, Salessawi~Ferede Yitbarek, Rui Qiao, Reetuparna Das, Matthew Hicks, Yossi Oren, and Todd Austin.
\newblock {ANVIL: Software-Based Protection Against Next-Generation Rowhammer Attacks}.
\newblock In {\em ASPLOS}, 2016.

\bibitem{france2022reducing}
Lo{\"\i}c France, Florent Bruguier, David Novo, Maria Mushtaq, and Pascal Benoit.
\newblock {Reducing the Silicon Area Overhead of Counter-Based Rowhammer Mitigations}.
\newblock In {\em 18th CryptArchi Workshop}, 2022.

\bibitem{park2022rowhammer}
Jin~Hyo Park, Su~Yeon Kim, Dong~Young Kim, Geon Kim, Je~Won Park, Sunyong Yoo, Young-Woo Lee, and Myoung~Jin Lee.
\newblock {Row Hammer Reduction Using a Buried Insulator in a Buried Channel Array Transistor}.
\newblock {\em IEEE TED}, 2022.

\bibitem{fournaris2017exploiting}
Apostolos~P Fournaris, Lidia Pocero~Fraile, and Odysseas Koufopavlou.
\newblock {Exploiting Hardware Vulnerabilities to Attack Embedded System Devices: A Survey of Potent Microarchitectural Attacks}.
\newblock {\em Electronics}, 2017.

\bibitem{poddebniak2018attacking}
Damian Poddebniak, Juraj Somorovsky, Sebastian Schinzel, Manfred Lochter, and Paul R{\"o}sler.
\newblock {Attacking Deterministic Signature Schemes Using Fault Attacks}.
\newblock In {\em EuroS\&P}, 2018.

\bibitem{tatar2018throwhammer}
Andrei Tatar, Radhesh~Krishnan Konoth, Elias Athanasopoulos, Cristiano Giuffrida, Herbert Bos, and Kaveh Razavi.
\newblock {Throwhammer: {Rowhammer} {Attacks} Over the {Network} and {Defenses}}.
\newblock In {\em {USENIX} {ATC}}, 2018.

\bibitem{carre2018openssl}
Sebastien Carre, Matthieu Desjardins, Adrien Facon, and Sylvain Guilley.
\newblock {OpenSSL Bellcore's Protection Helps Fault Attack}.
\newblock In {\em DSD}, 2018.

\bibitem{barenghi2018software}
Alessandro Barenghi, Luca Breveglieri, Niccol{\`o} Izzo, and Gerardo Pelosi.
\newblock {Software-Only Reverse Engineering of Physical DRAM Mappings for Rowhammer Attacks}.
\newblock In {\em IVSW}, 2018.

\bibitem{zhang2018triggering}
Zhenkai Zhang, Zihao Zhan, Daniel Balasubramanian, Xenofon Koutsoukos, and Gabor Karsai.
\newblock {Triggering Rowhammer Hardware Faults on ARM: A Revisit}.
\newblock In {\em ASHES}, 2018.

\bibitem{bhattacharya2018advanced}
Sarani Bhattacharya and Debdeep Mukhopadhyay.
\newblock {Advanced Fault Attacks in Software: Exploiting the Rowhammer Bug}.
\newblock {\em Fault Tolerant Architectures for Cryptography and Hardware Security}, 2018.

\bibitem{google-project-zero}
Mark Seaborn and Thomas Dullien.
\newblock {Exploiting the DRAM Rowhammer Bug to Gain Kernel Privileges}.
\newblock \url{http://googleprojectzero.blogspot.com.tr/2015/03/exploiting-dram-rowhammer-bug-to-gain.html}, 2015.

\bibitem{rowhammergithub}
{SAFARI Research Group}.
\newblock {RowHammer --- GitHub Repository}.
\newblock \url{https://github.com/CMU-SAFARI/rowhammer}.

\bibitem{seaborn2015exploiting}
Mark Seaborn and Thomas Dullien.
\newblock {Exploiting the DRAM Rowhammer Bug to Gain Kernel Privileges}.
\newblock {\em Black Hat}, 2015.

\bibitem{van2016drammer}
Victor van~der Veen, Yanick Fratantonio, Martina Lindorfer, Daniel Gruss, Clementine Maurice, Giovanni Vigna, Herbert Bos, Kaveh Razavi, and Cristiano Giuffrida.
\newblock {Drammer: Deterministic Rowhammer Attacks on Mobile Platforms}.
\newblock In {\em CCS}, 2016.

\bibitem{gruss2016rowhammer}
Daniel Gruss, Cl{\'e}mentine Maurice, and Stefan Mangard.
\newblock {Rowhammer.js: A Remote Software-Induced Fault Attack in Javascript}.
\newblock DIMVA, 2016.

\bibitem{razavi2016flip}
Kaveh Razavi, Ben Gras, Erik Bosman, Bart Preneel, Cristiano Giuffrida, and Herbert Bos.
\newblock {Flip Feng Shui: Hammering a Needle in the Software Stack}.
\newblock In {\em USENIX Security}, 2016.

\bibitem{pessl2016drama}
Peter Pessl, Daniel Gruss, Cl{\'e}mentine Maurice, Michael Schwarz, and Stefan Mangard.
\newblock {DRAMA: Exploiting DRAM Addressing for Cross-CPU Attacks}.
\newblock In {\em USENIX Security}, 2016.

\bibitem{xiao2016one}
Yuan Xiao, Xiaokuan Zhang, Yinqian Zhang, and Radu Teodorescu.
\newblock {One Bit Flips, One Cloud Flops: Cross-VM Row Hammer Attacks and Privilege Escalation}.
\newblock In {\em USENIX Security}, 2016.

\bibitem{bosman2016dedup}
Erik Bosman, Kaveh Razavi, Herbert Bos, and Cristiano Giuffrida.
\newblock {Dedup Est Machina: Memory Deduplication as an Advanced Exploitation Vector}.
\newblock In {\em S\&P}, 2016.

\bibitem{bhattacharya2016curious}
Sarani Bhattacharya and Debdeep Mukhopadhyay.
\newblock {Curious Case of Rowhammer: Flipping Secret Exponent Bits Using Timing Analysis}.
\newblock In {\em CHES}, 2016.

\bibitem{burleson2016invited}
Wayne Burleson, Onur Mutlu, and Mohit Tiwari.
\newblock {Invited: Who is the Major Threat to Tomorrow's Security? You, the Hardware Designer}.
\newblock In {\em DAC}, 2016.

\bibitem{qiao2016new}
Rui Qiao and Mark Seaborn.
\newblock {A New Approach for RowHammer Attacks}.
\newblock In {\em HOST}, 2016.

\bibitem{jang2017sgx}
Yeongjin Jang, Jaehyuk Lee, Sangho Lee, and Taesoo Kim.
\newblock {SGX-Bomb: Locking Down the Processor via Rowhammer Attack}.
\newblock In {\em SOSP}, 2017.

\bibitem{aga2017good}
Misiker~Tadesse Aga, Zelalem~Birhanu Aweke, and Todd Austin.
\newblock {When Good Protections Go Bad: Exploiting Anti-DoS Measures to Accelerate Rowhammer Attacks}.
\newblock In {\em HOST}, 2017.

\bibitem{mutlu2017rowhammer}
Onur Mutlu.
\newblock {The RowHammer Problem and Other Issues We May Face as Memory Becomes Denser}.
\newblock In {\em DATE}, 2017.

\bibitem{tatar2018defeating}
Andrei Tatar, Cristiano Giuffrida, Herbert Bos, and Kaveh Razavi.
\newblock {Defeating Software Mitigations Against Rowhammer: A Surgical Precision Hammer}.
\newblock In {\em RAID}, 2018.

\bibitem{gruss2018another}
Daniel Gruss, Moritz Lipp, Michael Schwarz, Daniel Genkin, Jonas Juffinger, Sioli O'Connell, Wolfgang Schoechl, and Yuval Yarom.
\newblock {Another Flip in the Wall of Rowhammer Defenses}.
\newblock In {\em S\&P}, 2018.

\bibitem{lipp2018nethammer}
Moritz Lipp, Misiker~Tadesse Aga, Michael Schwarz, Daniel Gruss, Cl{\'e}mentine Maurice, Lukas Raab, and Lukas Lamster.
\newblock {Nethammer: Inducing Rowhammer Faults Through Network Requests}.
\newblock EuroS\&PW, 2020.

\bibitem{frigo2018grand}
Pietro Frigo, Cristiano Giuffrida, Herbert Bos, and Kaveh Razavi.
\newblock {Grand Pwning Unit: Accelerating Microarchitectural Attacks with the GPU}.
\newblock In {\em S\&P}, 2018.

\bibitem{cojocar2019eccploit}
Lucian Cojocar, Kaveh Razavi, Cristiano Giuffrida, and Herbert Bos.
\newblock {Exploiting Correcting Codes: On the Effectiveness of ECC Memory Against Rowhammer Attacks}.
\newblock In {\em S\&P}, 2019.

\bibitem{ji2019pinpoint}
Sangwoo Ji, Youngjoo Ko, Saeyoung Oh, and Jong Kim.
\newblock {Pinpoint Rowhammer: Suppressing Unwanted Bit Flips on Rowhammer Attacks}.
\newblock In {\em ASIACCS}, 2019.

\bibitem{mutlu2019rowhammer}
Onur Mutlu and Jeremie~S Kim.
\newblock {RowHammer: A Retrospective}.
\newblock {\em TCAD}, 2019.

\bibitem{hong2019terminal}
Sanghyun Hong, Pietro Frigo, Yi\u{g}itcan Kaya, Cristiano Giuffrida, and Tudor Dumitra\c{s}.
\newblock {Terminal Brain Damage: Exposing the Graceless Degradation in Deep Neural Networks Under Hardware Fault Attacks}.
\newblock In {\em USENIX Security}, 2019.

\bibitem{kwong2020rambleed}
Andrew Kwong, Daniel Genkin, Daniel Gruss, and Yuval Yarom.
\newblock {RAMBleed: Reading Bits in Memory Without Accessing Them}.
\newblock In {\em S\&P}, 2020.

\bibitem{cojocar2020rowhammer}
Lucian Cojocar, Jeremie Kim, Minesh Patel, Lillian Tsai, Stefan Saroiu, Alec Wolman, and Onur Mutlu.
\newblock {Are We Susceptible to Rowhammer? An End-to-End Methodology for Cloud Providers}.
\newblock In {\em S\&P}, 2020.

\bibitem{weissman2020jackhammer}
Zane Weissman, Thore Tiemann, Daniel Moghimi, Evan Custodio, Thomas Eisenbarth, and Berk Sunar.
\newblock {JackHammer: Efficient Rowhammer on Heterogeneous FPGA--CPU Platforms}.
\newblock TCHES, 2020.

\bibitem{zhang2020pthammer}
Zhi Zhang, Yueqiang Cheng, Dongxi Liu, Surya Nepal, Zhi Wang, and Yuval Yarom.
\newblock {PTHammer: Cross-User-Kernel-Boundary Rowhammer {T}hrough Implicit Accesses}.
\newblock In {\em MICRO}, 2020.

\bibitem{yao2020deephammer}
Fan Yao, Adnan~Siraj Rakin, and Deliang Fan.
\newblock {Deephammer: Depleting the Intelligence of Deep Neural Networks Through Targeted Chain of Bit Flips}.
\newblock In {\em USENIX Security}, 2020.

\bibitem{deridder2021smash}
Finn de~Ridder, Pietro Frigo, Emanuele Vannacci, Herbert Bos, Cristiano Giuffrida, and Kaveh Razavi.
\newblock {SMASH}: {Synchronized} {Many-Sided} {Rowhammer} {Attacks} from {JavaScript}.
\newblock In {\em {USENIX Security}}, 2021.

\bibitem{jattke2022blacksmith}
Patrick Jattke, Victor van~der Veen, Pietro Frigo, Stijn Gunter, and Kaveh Razavi.
\newblock {Blacksmith: Scalable Rowhammering in the Frequency Domain}.
\newblock In {\em SP}, 2022.

\bibitem{tol2022toward}
M~Caner Tol, Saad Islam, Berk Sunar, and Ziming Zhang.
\newblock {Toward Realistic Backdoor Injection Attacks on DNNs using RowHammer}.
\newblock {arXiv:2110.07683v2 [cs.LG]}, 2022.

\bibitem{kogler2022half}
Andreas Kogler, Jonas Juffinger, Salman Qazi, Yoongu Kim, Moritz Lipp, Nicolas Boichat, Eric Shiu, Mattias Nissler, and Daniel Gruss.
\newblock {Half-Double: Hammering From the Next Row Over}.
\newblock In {\em USENIX Security}, 2022.

\bibitem{orosa2022spyhammer}
Lois Orosa, Ulrich R{\"u}hrmair, A~Giray Yaglikci, Haocong Luo, Ataberk Olgun, Patrick Jattke, Minesh Patel, Jeremie Kim, Kaveh Razavi, and Onur Mutlu.
\newblock {SpyHammer: Using RowHammer to Remotely Spy on Temperature}.
\newblock IEEE Access, 2024.

\bibitem{zhang2022implicit}
Zhi Zhang, Wei He, Yueqiang Cheng, Wenhao Wang, Yansong Gao, Dongxi Liu, Kang Li, Surya Nepal, Anmin Fu, and Yi~Zou.
\newblock {Implicit Hammer: Cross-Privilege-Boundary Rowhammer through Implicit Accesses}.
\newblock {\em TDSC}, 2022.

\bibitem{liu2022generating}
Liang Liu, Yanan Guo, Yueqiang Cheng, Youtao Zhang, and Jun Yang.
\newblock {Generating Robust DNN with Resistance to Bit-Flip Based Adversarial Weight Attack}.
\newblock {\em TC}, 2022.

\bibitem{cohen2022hammerscope}
Yaakov Cohen, Kevin~Sam Tharayil, Arie Haenel, Daniel Genkin, Angelos~D Keromytis, Yossi Oren, and Yuval Yarom.
\newblock {HammerScope: Observing DRAM Power Consumption Using Rowhammer}.
\newblock In {\em CCS}, 2022.

\bibitem{zheng2022trojvit}
Mengxin Zheng, Qian Lou, and Lei Jiang.
\newblock {TrojViT: Trojan Insertion in Vision Transformers}.
\newblock CVPR, 2023.

\bibitem{fahr2022frodo}
Michael Fahr~Jr, Hunter Kippen, Andrew Kwong, Thinh Dang, Jacob Lichtinger, Dana Dachman-Soled, Daniel Genkin, Alexander Nelson, Ray Perlner, Arkady Yerukhimovich, et~al.
\newblock {When Frodo Flips: End-to-End Key Recovery on FrodoKEM via Rowhammer}.
\newblock {\em CCS}, 2022.

\bibitem{tobah2022spechammer}
Youssef Tobah, Andrew Kwong, Ingab Kang, Daniel Genkin, and Kang~G. Shin.
\newblock {SpecHammer: Combining Spectre and Rowhammer for New Speculative Attacks}.
\newblock In {\em SP}, 2022.

\bibitem{rakin2022deepsteal}
Adnan~Siraj Rakin, Md~Hafizul~Islam Chowdhuryy, Fan Yao, and Deliang Fan.
\newblock {DeepSteal: Advanced Model Extractions Leveraging Efficient Weight Stealing in Memories}.
\newblock In {\em SP}, 2022.

\bibitem{park2016statistical}
Kyungbae Park, Donghyuk Yun, and Sanghyeon Baeg.
\newblock {Statistical Distributions of Row-Hammering Induced Failures in DDR3 Components}.
\newblock {\em Microelectronics Reliability}, 2016.

\bibitem{park2016experiments}
Kyungbae Park, Chulseung Lim, Donghyuk Yun, and Sanghyeon Baeg.
\newblock {Experiments and Root Cause Analysis for Active-Precharge Hammering Fault in DDR3 SDRAM under 3xnm Technology}.
\newblock {\em Microelectronics Reliability}, 2016.

\bibitem{lim2017active}
Chulseung Lim, Kyungbae Park, and Sanghyeon Baeg.
\newblock {Active Precharge Hammering to Monitor Displacement Damage Using High-Energy Protons in 3x-nm SDRAM}.
\newblock {\em TNS}, 2017.

\bibitem{yun2018study}
Donghyuk Yun, Myungsang Park, Chulseung Lim, and Sanghyeon Baeg.
\newblock {Study of TID Effects on One Row Hammering using Gamma in DDR4 SDRAMs}.
\newblock In {\em IRPS}, 2018.

\bibitem{yang2019trap}
Thomas Yang and Xi-Wei Lin.
\newblock {Trap-Assisted DRAM Row Hammer Effect}.
\newblock {\em EDL}, 2019.

\bibitem{walker2021ondramrowhammer}
Andrew~J. Walker, Sungkwon Lee, and Dafna Beery.
\newblock {On DRAM RowHammer and the Physics on Insecurity}.
\newblock {\em IEEE TED}, 2021.

\bibitem{kim2020revisiting}
Jeremie~S. Kim, Minesh Patel, Abdullah~Giray Ya\u{g}l{\i}k\c{c}{\i}, Hasan Hassan, Roknoddin Azizi, Lois Orosa, and Onur Mutlu.
\newblock {Revisiting RowHammer: An Experimental Analysis of Modern Devices and Mitigation Techniques}.
\newblock In {\em ISCA}, 2020.

\bibitem{orosa2021deeper}
Lois Orosa, A~Giray Ya{\u{g}}l{\i}k{\c{c}}{\i}, Haocong Luo, Ataberk Olgun, Jisung Park, Hasan Hassan, Minesh Patel, Jeremie~S. Kim, and Onur Mutlu.
\newblock {A Deeper Look into RowHammer's Sensitivities: Experimental Analysis of Real DRAM Chips and Implications on Future Attacks and Defenses}.
\newblock In {\em MICRO}, 2021.

\bibitem{yaglikci2022understanding}
A.~Giray Ya{\u{g}}l{\i}k{\c{c}}{\i}, Haocong Luo, Geraldo~F De~Oliviera, Ataberk Olgun, Minesh Patel, Jisung Park, Hasan Hassan, Jeremie~S Kim, Lois Orosa, and Onur Mutlu.
\newblock {Understanding RowHammer Under Reduced Wordline Voltage: An Experimental Study Using Real DRAM Devices}.
\newblock In {\em DSN}, 2022.

\bibitem{khan2018analysis}
Mohammad Nasim~Imtiaz Khan and Swaroop Ghosh.
\newblock {Analysis of Row Hammer Attack on STTRAM}.
\newblock In {\em ICCD}, 2018.

\bibitem{agarwal2018rowhammer}
S.~Agarwal, H.~Dixit, D.~Datta, M.~Tran, D.~Houssameddine, D.~Shum, and F.~Benistant.
\newblock {Rowhammer for Spin Torque Based Memory: Problem or Not?}
\newblock In {\em INTERMAG}, 2018.

\bibitem{li2014write}
Haitong Li, Hong-Yu Chen, Zhe Chen, Bing Chen, Rui Liu, Gang Qiu, Peng Huang, Feifei Zhang, Zizhen Jiang, Bin Gao, Lifeng Liu, Xiaoyan Liu, Shimeng Yu, H.-S.~Philip Wong, and Jinfeng Kang.
\newblock {Write Disturb Analyses on Half-Selected Cells of Cross-Point RRAM Arrays}.
\newblock In {\em IRPS}, 2014.

\bibitem{ni2018write}
Kai Ni, Xueqing Li, Jeffrey~A. Smith, Matthew Jerry, and Suman Datta.
\newblock {Write Disturb in Ferroelectric FETs and Its Implication for 1T-FeFET AND Memory Arrays}.
\newblock {\em IEEE EDL}, 2018.

\bibitem{genssler2022reliability}
Paul~R. Genssler, Victor~M. van Santen, Jörg Henkel, and Hussam Amrouch.
\newblock {On the Reliability of FeFET On-Chip Memory}.
\newblock {\em TC}, 2022.

\bibitem{luo2023rowpress}
Haocong Luo, Ataberk Olgun, Abdullah~Giray Ya\u{g}l{\i}kc{\i}, Yahya~Can Tu\u{g}rul, Steve Rhyner, Meryem~Banu Cavlak, Joël Lindegger, Mohammad Sadrosadati, and Onur Mutlu.
\newblock {RowPress: Amplifying Read Disturbance in Modern DRAM Chips}.
\newblock In {\em ISCA}, 2023.

\bibitem{jedec2024jesd795c}
{JEDEC}.
\newblock {\em {JESD79-5c: DDR5 SDRAM Standard}}, 2024.

\bibitem{seyedzadeh2017counterbased}
Seyed~Mohammad Seyedzadeh, Alex~K. Jones, and Rami Melhem.
\newblock {Counter-Based Tree Structure for Row Hammering Mitigation in DRAM}.
\newblock {\em CAL}, 2017.

\bibitem{seyedzadeh2018mitigating}
S.~M. {Seyedzadeh}, A.~K. {Jones}, and R.~{Melhem}.
\newblock {Mitigating Wordline Crosstalk Using Adaptive Trees of Counters}.
\newblock In {\em ISCA}, 2018.

\bibitem{yaglikci2021security}
A.~Giray Ya{\u{g}}l{\i}k{\c{c}}{\i}, Jeremie~S. Kim, Fabrice Devaux, and Onur Mutlu.
\newblock {Security Analysis of the Silver Bullet Technique for RowHammer Prevention}.
\newblock arXiv:2106.07084 [cs.CR], 2021.

\bibitem{jedec2020jesd795}
{JEDEC}.
\newblock {\em {JESD79-5: DDR5 SDRAM Standard}}, 2020.

\bibitem{saroiu2024ddr5}
Stefan Saroiu.
\newblock {DDR5 Spec Update Has All It Needs to End Rowhammer: Will It?}
\newblock \url{https://stefan.t8k2.com/rh/PRAC/index.html}.

\bibitem{hassan2024self}
Hasan Hassan, Ataberk Olgun, A.~Giray Ya{\u{g}}l{\i}k{\c{c}}{\i}, Haocong Luo, and Onur Mutlu.
\newblock {Self-Managing DRAM: A Low-Cost Framework for Enabling Autonomous and Efficient DRAM Maintenance Operations}.
\newblock In {\em MICRO}, 2024.

\bibitem{luo2023ramulator2}
Haocong Luo, Yahya~Can Tu\u{g}rul, F.~Nisa Bostancı, Ataberk Olgun, A.~Giray Ya\u{g}l{\i}k\c{c}{\i}, and Onur Mutlu.
\newblock {Ramulator 2.0: A Modern, Modular, and Extensible DRAM Simulator}.
\newblock CAL, 2023.

\bibitem{ramulator2github}
{SAFARI Research Group}.
\newblock {Ramulator V2.0}.
\newblock \url{https://github.com/CMU-SAFARI/ramulator2}.

\bibitem{graf1993greek}
Fritz Graf.
\newblock {\em {Greek Mythology: An Introduction}}.
\newblock JHU Press, 1993.

\bibitem{chronusgithub}
{SAFARI Research Group}.
\newblock {Chronus}.
\newblock \url{https://github.com/CMU-SAFARI/Chronus}.

\bibitem{jedec2020ddr5}
{JEDEC}.
\newblock {\em {JESD79-5: DDR5 SDRAM Standard}}, 2020.

\bibitem{jedec2017ddr4}
{JEDEC}.
\newblock {\em {JESD79-4C: DDR4 SDRAM Standard}}, 2020.

\bibitem{kim2012acase}
Yoongu Kim, Vivek Seshadri, Donghyuk Lee, Jamie Liu, Onur Mutlu, Yoongu Kim, Vivek Seshadri, Donghyuk Lee, Jamie Liu, and Onur Mutlu.
\newblock {A Case for Exploiting Subarray-Level Parallelism (SALP) in DRAM}.
\newblock In {\em ISCA}, 2012.

\bibitem{lee2013tiered}
Donghyuk Lee, Yoongu Kim, Vivek Seshadri, Jamie Liu, Lavanya Subramanian, and Onur Mutlu.
\newblock {Tiered-Latency DRAM: A Low Latency and Low Cost DRAM Architecture}.
\newblock In {\em HPCA}, 2013.

\bibitem{lee2015adaptive}
Donghyuk Lee, Yoongu Kim, Gennady Pekhimenko, Samira Khan, Vivek Seshadri, Kevin Chang, and Onur Mutlu.
\newblock {Adaptive-Latency DRAM: Optimizing DRAM Timing for the Common-Case}.
\newblock In {\em HPCA}, 2015.

\bibitem{redeker2002investigation}
Michael Redeker, Bruce~F Cockburn, and Duncan~G Elliott.
\newblock {An Investigation into Crosstalk Noise in DRAM Structures}.
\newblock In {\em MTDT}, 2002.

\bibitem{park2014active}
Kyungbae Park, Sanghyeon Baeg, ShiJie Wen, and Richard Wong.
\newblock {Active-Precharge Hammering on a Row-Induced Failure in DDR3 SDRAMs Under 3x nm Technology}.
\newblock In {\em IIRW}, 2014.

\bibitem{lim2018study}
Chulseung Lim, Kyungbae Park, Geunyong Bak, Donghyuk Yun, Myungsang Park, Sanghyeon Baeg, Shi-Jie Wen, and Richard Wong.
\newblock {Study of Proton Radiation Effect to Row Hammer Fault in DDR4 SDRAMs}.
\newblock {\em Microelectronics Reliability}, 2018.

\bibitem{jiang2021quantifying}
Yichen Jiang, Huifeng Zhu, Dean Sullivan, Xiaolong Guo, Xuan Zhang, and Yier Jin.
\newblock {Quantifying RowHammer Vulnerability for DRAM Security}.
\newblock In {\em DAC}, 2021.

\bibitem{he2023whistleblower}
Wei He, Zhi Zhang, Yueqiang Cheng, Wenhao Wang, Wei Song, Yansong Gao, Qifei Zhang, Kang Li, Dongxi Liu, and Surya Nepal.
\newblock {WhistleBlower: A System-level Empirical Study on RowHammer}.
\newblock {\em TC}, 2023.

\bibitem{baeg2022estimation}
Sanghyeon Baeg, Donghyuk Yun, Myungsun Chun, and Shi-Jie Wen.
\newblock {Estimation of the Trap Energy Characteristics of Row Hammer-Affected Cells in Gamma-Irradiated DDR4 DRAM}.
\newblock {\em IEEE TNS}, 2022.

\bibitem{mutlu2018rowhammer}
Onur Mutlu.
\newblock {RowHammer}.
\newblock Top Picks in Hardware and Embedded Security, 2018.

\bibitem{olgun2023hbm}
Ataberk Olgun, Majd Osseiran, Abdullah~Giray Yaglikci, Yahya~Can Tugrul, Haocong Luo, Steve Rhyner, Behzad Salami, Juan Gomez~Luna, and Onur Mutlu.
\newblock {An Experimental Analysis of RowHammer in HBM2 DRAM Chips}.
\newblock In {\em DSN Disrupt}, 2023.

\bibitem{olgun2023drambender}
Ataberk Olgun, Hasan Hassan, A.~Giray Ya{\u{g}}l{\i}k{c}{\i}, Yahya~Can Tu{\u{g}}rul, Lois Orosa, Haocong Luo, Minesh Patel, Ergin O{\u{g}}uz, and Onur Mutlu.
\newblock {DRAM Bender: An Extensible and Versatile FPGA-Based Infrastructure to Easily Test State-of-the-Art DRAM Chips}.
\newblock {\em {TCAD}}, 2023.

\bibitem{zhou2023double}
Longda Zhou, Jie Li, Zheng Qiao, Pengpeng Ren, Zixuan Sun, Jianping Wang, Blacksmith Wu, Zhigang Ji, Runsheng Wang, Kanyu Cao, and Ru~Huang.
\newblock {Double-sided Row Hammer Effect in Sub-20 nm DRAM: Physical Mechanism, Key Features and Mitigation}.
\newblock In {\em IRPS}, 2023.

\bibitem{lang2023blaster}
Zhenrong Lang, Patrick Jattke, Michele Marazzi, and Kaveh Razavi.
\newblock {BLASTER: Characterizing the Blast Radius of Rowhammer}.
\newblock DRAMSec, 2023.

\bibitem{baek2025marionette}
Seungmin Baek, Minbok Wi, Seonyong Park, Hwayong Nam, Michael~Jaemin Kim, Nam~Sung Kim, and Jung~Ho Ahn.
\newblock Marionette: A rowhammer attack via row coupling.
\newblock In {\em ASPLOS}, 2025.

\bibitem{hp2015rowhammer}
{Hewlett-Packard Enterprise}.
\newblock {HP Moonshot Component Pack V2015.05.0}, 2015.

\bibitem{lenovo2015rowhammer}
{Lenovo Group Ltd.}
\newblock {Row Hammer Privilege Escalation}.
\newblock \url{https://support.lenovo.com/us/en/product_security/row_hammer}, 2015.

\bibitem{hong2023dsac}
Seungki Hong, Dongha Kim, Jaehyung Lee, Reum Oh, Changsik Yoo, Sangjoon Hwang, and Jooyoung Lee.
\newblock {DSAC: Low-Cost Rowhammer Mitigation Using In-DRAM Stochastic and Approximate Counting Algorithm}.
\newblock arXiv:2302.03591 [cs.CR], 2023.

\bibitem{marazzi2023rega}
M.~Marazzi, F.~Solt, P.~Jattke, K.~Takashi, and K.~Razavi.
\newblock {REGA: Scalable Rowhammer Mitigation with Refresh-Generating Activations}.
\newblock In {\em SP}, 2023.

\bibitem{canpolat2024understanding}
O{\u{g}}uzhan Canpolat, A~Giray Ya{\u{g}}l{\i}k{\c{c}}{\i}, Geraldo~F Oliveira, Ataberk Olgun, O{\u{g}}uz Ergin, and Onur Mutlu.
\newblock {Understanding the Security Benefits and Overheads of Emerging Industry Solutions to DRAM Read Disturbance}.
\newblock {\em DRAMSec}, 2024.

\bibitem{chang2016understanding}
Kevin~K Chang, Abhijith Kashyap, Hasan Hassan, Saugata Ghose, Kevin Hsieh, Donghyuk Lee, Tianshi Li, Gennady Pekhimenko, Samira Khan, and Onur Mutlu.
\newblock {Understanding Latency Variation in Modern DRAM Chips: Experimental Characterization, Analysis, and Optimization}.
\newblock In {\em SIGMETRICS}, 2016.

\bibitem{chang2017understanding}
Kevin~K Chang, A~Giray Ya{\u{g}}l{\i}k{\c{c}}{\i}, Saugata Ghose, Aditya Agrawal, Niladrish Chatterjee, Abhijith Kashyap, Donghyuk Lee, Mike O'Connor, Hasan Hassan, and Onur Mutlu.
\newblock {Understanding Reduced-Voltage Operation in Modern DRAM Devices: Experimental Characterization, Analysis, and Mechanisms}.
\newblock In {\em SIGMETRICS}, 2017.

\bibitem{chang2017understandingphd}
Kevin~K Chang.
\newblock {\em {Understanding and Improving the Latency of DRAM-Based Memory Systems}}.
\newblock PhD thesis, Carnegie Mellon University, 2017.

\bibitem{kim2018solar}
Jeremie~S Kim, Minesh Patel, Hasan Hassan, and Onur Mutlu.
\newblock {Solar-DRAM: Reducing DRAM Access Latency by Exploiting the Variation in Local Bitlines}.
\newblock In {\em ICCD}, 2018.

\bibitem{mathew2017using}
Deepak~M Mathew, {\'E}der~F Zulian, Matthias Jung, Kira Kraft, Christian Weis, Bruce Jacob, and Norbert Wehn.
\newblock {Using Run-Time Reverse-Engineering to Optimize DRAM Refresh}.
\newblock In {\em MEMSYS}, 2017.

\bibitem{tugrul2025understanding}
Yahya~Can Tu{\u{g}}rul, A.~Giray Ya{\u{g}}l{\i}k{\c{c}}{\i}, Ismail~Emir Yuksel, Oguzhan Canpolat, Nisa Bostanci, Mohammad Sadrosadati, Oguz Ergin, and Onur Mutlu.
\newblock {Understanding RowHammer Under Reduced Refresh Latency: Experimental Analysis of Real DRAM Chips and Implications on Future Solutions}.
\newblock In {\em HPCA}, 2025.

\bibitem{olgun2023understanding}
Ataberk Olgun, Majd Osseiran, Abdullah~Giray Yaglikci, Yahya~Can Tugrul, Haocong Luo, Steve Rhyner, Behzad Salami, Juan~Gomez Luna, and Onur Mutlu.
\newblock {Understanding Read Disturbance in High Bandwidth Memory: An Experimental Analysis of Real HBM2 DRAM Chips}.
\newblock arXiv:2310.14665 [cs.AR], 2023.

\bibitem{olgun2024read}
Ataberk Olgun, Majd Osseiran, A~Giray Ya{\u{g}}l{\i}k{\c{c}}{\i}, Yahya~Can Tu{\u{g}}rul, Haocong Luo, Steve Rhyner, Behzad Salami, Juan~Gomez Luna, and Onur Mutlu.
\newblock {Read Disturbance in High Bandwidth Memory: A Detailed Experimental Study on HBM2 DRAM Chips}.
\newblock In {\em DSN}, 2024.

\bibitem{zhou2023threshold}
Ranyang Zhou, Jacqueline Liu, Sabbir Ahmed, Nakul Kochar, Adnan~Siraj Rakin, and Shaahin Angizi.
\newblock {Threshold Breaker: Can Counter-Based RowHammer Prevention Mechanisms Truly Safeguard DRAM?}
\newblock arXiv:2311.16460 [cs.AR], 2023.

\bibitem{olgun2025variable}
Ataberk Olgun, F.~Nisa Bostanci, Ismail~Emir Yuksel, Oguzhan Canpolat, Haocong Luo, Geraldo~F. Oliveira, A~Giray Ya{\u{g}}l{\i}k{\c{c}}{\i}, Minesh Patel, and Onur Mutlu.
\newblock {Variable Read Disturbance: An Experimental Analysis of Temporal Variation in DRAM Read Disturbance}.
\newblock In {\em HPCA}, 2025.

\bibitem{kim2016ramulator}
Yoongu Kim, Weikun Yang, and Onur Mutlu.
\newblock {Ramulator: A Fast and Extensible DRAM Simulator}.
\newblock {\em CAL}, 2016.

\bibitem{ramulatorgithub}
{SAFARI Research Group}.
\newblock {Ramulator --- GitHub Repository}.
\newblock \url{https://github.com/CMU-SAFARI/ramulator}, 2021.

\bibitem{eyerman2008systemlevel}
Stijn Eyerman and Lieven Eeckhout.
\newblock {System-Level Performance Metrics for Multiprogram Workloads}.
\newblock {\em IEEE Micro}, 2008.

\bibitem{snavely2000symbiotic}
Allan Snavely and Dean~M Tullsen.
\newblock {Symbiotic Job Scheduling for A Simultaneous Multithreaded Processor}.
\newblock In {\em ASPLOS}, 2000.

\bibitem{frfcfs}
Scott Rixner, William~J. Dally, Ujval~J. Kapasi, Peter Mattson, and John~D. Owens.
\newblock {Memory Access Scheduling}.
\newblock In {\em ISCA}, 2000.

\bibitem{zuravleff1997controller}
William~K Zuravleff and Timothy Robinson.
\newblock {Controller for a Synchronous DRAM That Maximizes Throughput by Allowing Memory Requests and Commands to Be Issued Out of Order}, 1997.
\newblock {U.S.}\ Patent 5,630,096.

\bibitem{mutlu2007stall}
Onur Mutlu and Thomas Moscibroda.
\newblock {Stall-Time Fair Memory Access Scheduling for Chip Multiprocessors}.
\newblock In {\em MICRO}, 2007.

\bibitem{kaseridis2011minimalistic}
Dimitris Kaseridis, Jeffrey Stuecheli, and Lizy~Kurian John.
\newblock {Minimalist Open-Page: A DRAM Page-Mode Scheduling Policy for the Many-Core Era}.
\newblock In {\em MICRO}, 2011.

\bibitem{spec2006}
{Standard Perf. Eval. Corp.}
\newblock {SPEC 2006}.
\newblock \url{http://www.spec.org/cpu2006/}, 2006.

\bibitem{spec2017}
{Standard Perf. Eval. Corp.}
\newblock {SPEC 2017}.
\newblock \url{http://www.spec.org/cpu2017}, 2017.

\bibitem{tpc}
{Transaction Processing Performance Council}.
\newblock {TPC Benchmarks}.
\newblock \url{http://tpc.org/}.

\bibitem{fritts2009media}
Jason~E. Fritts, Frederick~W. Steiling, Joseph~A. Tucek, and Wayne Wolf.
\newblock {MediaBench II Video: Expediting the Next Generation of Video Systems Research}.
\newblock {\em MICPRO}, 2009.

\bibitem{ycsb}
Brian Cooper, Adam Silberstein, Erwin Tam, Raghu Ramakrishnan, and Russell Sears.
\newblock {Benchmarking Cloud Serving Systems with {YCSB}}.
\newblock In {\em SoCC}, 2010.

\bibitem{chang2014improving}
Kevin~K Chang, Donghyuk Lee, Zeshan Chishti, Alaa~R Alameldeen, Chris Wilkerson, Yoongu Kim, and Onur Mutlu.
\newblock {Improving DRAM Performance by Parallelizing Refreshes with Accesses}.
\newblock In {\em HPCA}, 2014.

\bibitem{yuksel2024functionally}
Ismail~Emir Yuksel, Yahya~Can Tugrul, Ataberk Olgun, F.~Nisa Bostanci, A.~Giray Yaglikci, Geraldo~F. de~Oliveira, Haocong Luo, Juan~Gomez Luna, Mohammad Sadrosadati, and Onur Mutlu.
\newblock {Functionally-Complete Boolean Logic in Real DRAM Chips: Experimental Characterization and Analysis}.
\newblock In {\em {HPCA}}, 2024.

\bibitem{yuksel2024simultaneous}
Ismail~Emir Yuksel, Yahya~Can Tugrul, F~Bostanci, Geraldo~F Oliveira, A~Giray Yaglikci, Ataberk Olgun, Melina Soysal, Haocong Luo, Juan G{\'o}mez-Luna, Mohammad Sadrosadati, et~al.
\newblock {Simultaneous Many-Row Activation in Off-the-Shelf DRAM Chips: Experimental Characterization and Analysis}.
\newblock In {\em {DSN}}, 2024.

\bibitem{synopsys}
{Synopsys, Inc.}
\newblock {Synopsys Design Compiler}.
\newblock \url{https://www.synopsys.com/implementation-and-signoff/}.

\bibitem{carter201622nm}
Rick Carter, J~Mazurier, L~Pirro, JU~Sachse, P~Baars, J~Faul, C~Grass, G~Grasshoff, P~Javorka, T~Kammler, et~al.
\newblock {22nm FDSOI Technology for Emerging Mobile, Internet-of-Things, and RF Applications}.
\newblock In {\em IEDM}, 2016.

\bibitem{chen1996Assessing}
Yong-Bin Kim and T.~Chen.
\newblock {Assessing Merged DRAM/Logic Technology}.
\newblock In {\em {ISCAS}}, 1996.

\bibitem{cacti}
Rajeev Balasubramonian, Andrew~B. Kahng, Naveen Muralimanohar, Ali Shafiee, and Vaishnav Srinivas.
\newblock {CACTI 7: New Tools for Interconnect Exploration in Innovative Off-Chip Memories}.
\newblock {\em ACM TACO}, 2017.

\bibitem{canpolat2024breakhammer}
O{\u{g}}uzhan Canpolat, A~Giray Ya{\u{g}}l{\i}k{\c{c}}{\i}, Ataberk Olgun, {\.I}smail~Emir Y{\"u}ksel, Yahya~Can Tu{\u{g}}rul, Konstantinos Kanellopoulos, O{\u{g}}uz Ergin, and Onur Mutlu.
\newblock {BreakHammer: Enhancing RowHammer Mitigations by Carefully Throttling Suspect Threads}.
\newblock {\em MICRO}, 2024.

\bibitem{nagel1973spice}
Laurence Nagel and Donald~O Pederson.
\newblock {SPICE (Simulation Program with Integrated Circuit Emphasis)}.
\newblock 1973.

\bibitem{misra1982finding}
Jayadev Misra and David Gries.
\newblock {Finding Repeated Elements}.
\newblock {\em {Science of Computer Programming}}, 1982.

\bibitem{drampower}
Karthik Chandrasekar, Benny Akesson, and Kees Goossens.
\newblock {Improved Power Modeling of DDR SDRAMs}.
\newblock In {\em DSD}, 2011.

\bibitem{altman2005standard}
Douglas~G Altman and J~Martin Bland.
\newblock {Standard Deviations and Standard Errors}.
\newblock {\em BMJ}, 2005.

\bibitem{mutlu2007memory}
Thomas Moscibroda and Onur Mutlu.
\newblock {Memory Performance Attacks: Denial of Memory Service in Multi-Core Systems}.
\newblock In {\em USENIX Security}, 2007.

\bibitem{kim2010thread}
Yoongu Kim, Michael Papamichael, Onur Mutlu, and Mor Harchol-Balter.
\newblock {Thread Cluster Memory Scheduling: Exploiting Differences in Memory Access Behavior}.
\newblock In {\em MICRO}, 2010.

\bibitem{luo2024experimental}
Haocong Luo, Ismail~Emir Y{\"u}ksel, Ataberk Olgun, A~Giray Ya{\u{g}}l{\i}k{\c{c}}{\i}, Mohammad Sadrosadati, and Onur Mutlu.
\newblock {An Experimental Characterization of Combined RowHammer and RowPress Read Disturbance in Modern DRAM Chips}.
\newblock DSN Disrupt, 2024.

\bibitem{kim2021hammerfilter}
Kwangrae Kim, Jeonghyun Woo, Junsu Kim, and Ki-Seok Chung.
\newblock {HammerFilter: Robust Protection and Low Hardware Overhead Method for RowHammer}.
\newblock In {\em ICCD}, 2021.

\bibitem{bock2019riprh}
Carsten Bock, Ferdinand Brasser, David Gens, Christopher Liebchen, and Ahamd-Reza Sadeghi.
\newblock {RIP-RH: Preventing Rowhammer-Based Inter-Process Attacks}.
\newblock In {\em ASIACCS}, 2019.

\bibitem{zhang2019telehammer}
Zhi Zhang, Yueqiang Cheng, Dongxi Liu, Surya Nepal, and Zhi Wang.
\newblock {TeleHammer: A Stealthy Cross-Boundary Rowhammer Technique}.
\newblock arXiv:1912.03076 [cs.CR], 2019.

\bibitem{dell1997white}
Timothy~J Dell.
\newblock {A White Paper on the Benefits of Chipkill-Correct ECC for PC Server Main Memory}.
\newblock {\em IBM Microelectronics Division}, 1997.

\bibitem{huang2010ivec}
Ruirui Huang and G.~Edward Suh.
\newblock {IVEC: Off-Chip Memory Integrity Protection for Both Security and Reliability}.
\newblock In {\em ISCA}, 2010.

\bibitem{saileshwar2018synergy}
Gururaj Saileshwar, Prashant~J. Nair, Prakash Ramrakhyani, Wendy Elsasser, and Moinuddin~K. Qureshi.
\newblock {SYNERGY: Rethinking Secure-Memory Design for Error-Correcting Memories}.
\newblock In {\em HPCA}, 2018.

\bibitem{chen2014memguard}
Long Chen and Zhao Zhang.
\newblock {MemGuard: A Low Cost and Energy Efficient Design to Support and Enhance Memory System Reliability}.
\newblock In {\em ISCA}, 2014.

\bibitem{qureshi2021rethinking}
Moinuddin Qureshi.
\newblock {Rethinking ECC in the Era of Row-Hammer}.
\newblock {\em {DRAMSec}}, 2021.

\bibitem{canpolat2025chronus}
O{\u{g}}uzhan Canpolat, A~Giray Ya{\u{g}}l{\i}k{\c{c}}{\i}, Geraldo~F Oliveira, Ataberk Olgun, Nisa Bostanc{\i}, Ismail~Emir Yuksel, Haocong Luo, O{\u{g}}uz Ergin, and Onur Mutlu.
\newblock Chronus: Understanding and securing the cutting-edge industry solutions to dram read disturbance.
\newblock In {\em HPCA}, 2025.

\bibitem{kim2025per}
Jumin Kim, Seungmin Baek, Minbok Wi, Hwayong Nam, Michael~Jaemin Kim, Sukhan Lee, Kyomin Sohn, and Jung~Ho Ahn.
\newblock Per-row activation counting on real hardware: Demystifying performance overheads.
\newblock {\em CAL}, 2025.

\bibitem{mutlu2023retrospective}
Onur Mutlu.
\newblock {Retrospective: Flipping Bits in Memory Without Accessing Them: An Experimental Study of DRAM Disturbance Errors}.
\newblock Retrospective Issue for ISCA-50, 2023.

\bibitem{rosen2019discrete}
Kenneth~H. Rosen.
\newblock {\em Discrete Mathematics and Its Applications}.
\newblock McGraw-Hill Education, 8th edition, 2019.

\bibitem{enderton2001mathematical}
Herbert~B Enderton.
\newblock {\em A mathematical introduction to logic}.
\newblock Elsevier, 2001.

\bibitem{singh2019memory}
Sarabjeet Singh and Manu Awasthi.
\newblock Memory centric characterization and analysis of spec cpu2017 suite.
\newblock In {\em ICPE}, 2019.

\bibitem{qureshi2025autorfm}
Moinuddin Qureshi.
\newblock Autorfm: Scaling low-cost in-dram trackers to ultra-low rowhammer thresholds.
\newblock In {\em HPCA}, 2025.

\bibitem{vittal2025mopac}
Suhas Vittal, Salman Qazi, Poulami Das, and Moinuddin Qureshi.
\newblock Mopac: Efficiently mitigating rowhammer with probabilistic activation counting.
\newblock In {\em ISCA}, 2025.

\end{thebibliography}

\newpage{}
\nobalance{}
\appendix
\section{Decrementer Circuit of \X{}}
\label{apx:decrementer}

\ous{4}{
\X{} uses custom circuitry to update row activation counters.
To do so, we implement a circuit that decrements an 8-bit number (i.e., size of a \X{} activation counter, as explained in~\secref{sec:mechanism}) by 1.
\tabref{tab:decrementer} shows the pseudo hardware description of the decrementer and the resources to implement the circuit.
In the table, $x$ and $y$ respectively show the 8-bit row activation counter as input and 8-bit updated value as output of the circuit, where $y=x-1$.
The subscripted numbers denote the bit index of the input and the output, e.g., $x_1$ and $y_1$ respectively show the first bit of the 8-bit input and 8-bit output of the circuit.
Rows of the table show
1) the logical expression to obtain each bit of $y$,
2) the logic gates needed to implement the expression, and
3) the number of transistors needed to implement the logic gates.}

\begin{table}[ht]
    \centering
    \footnotesize
    \vspace{1em}
    \caption{Gate-level implementation of the circuitry that decrements an 8-bit number by 1}
    \begin{tabular}{l@{\hspace{2pt}}|@{\hspace{2pt}}ccccr}
    Logical expression & NOT & MUX & NAND & NOR & \#Ts\\
    \hline
    $y_{0}$ = $\overline{x_{0}}$                                               & 1 & 0 & 0 & 0 &  2 \\
    $y_{1}$ = \ous{0}{${x_{0}}$} ? $x_{1}$ : $\overline{x_{1}}$                & 1 & 1 & 0 & 0 & 10 \\
    $y_{2}$ = nor($x_{0}$, $x_{1}$) ? $\overline{x_{2}}$ : $x_{2}$             & 1 & 1 & 0 & 1 & 14 \\
    for $i = 3\rightarrow7$: & & & & & \\
    \hspace*{0.2cm} $y_{i}$ = nand($y_{i-1}$, $\overline{x_{i-1}}$) ? $x_{i}$ : $\overline{x_{i}}$ & 1 & 1 & 1 & 0 & 14 \\
    \hline
    \multicolumn{1}{r}{Total:} & 8 & 7 & 5 & 1 & 96 \\

    \end{tabular}
    \label{tab:decrementer}
\end{table}
%
%
%
%
%


\section{Artifact Appendix}

\subsection{Abstract}

Our artifact contains the data, source code, and scripts needed to reproduce our results.
We provide: 1) the source code of our simulation infrastructure based on Ramulator2 and 2) all evaluated memory access traces and all major evaluation results.
We provide Bash and Python scripts to analyze and plot the results automatically.

\subsection{Artifact Check-list (meta-information)}

\begin{table}[H]
  \centering
  \scriptsize
  \setlength{\tabcolsep}{0.7\tabcolsep}
  \captionsetup{justification=centering, singlelinecheck=false, labelsep=colon}
    \begin{tabular}{ll}
        {{\bf Parameter}} & \textbf{Value} \\
        \hline
                        &  C++ program \\
        Program         &  Python3 scripts \\
                        &  Shell scripts \\
        \hline
        Compilation     &  C++ compiler with c++20 features \\
        \hline
                             &  Ubuntu 20.04 (or similar) Linux \\
                             &  C++20 build toolchain (tested with GCC 10) \\
        Run-time environment &  Python 3.10+ \\
                             &  Podman 4.5+ \\
                             &  Git \\
        \hline
        Metrics  &  \makecell[l]{Weighted speedup\\DRAM energy} \\
        \hline
        Experiment workflow & \makecell[l]{Perform simulations, aggregate results, and\\run analysis scripts on the result} \\
        \hline
        Experiment customization & Possible. See \secref{sec:expcustom} \\
        \hline
        Disk space requirement & $\approx$ 30GiB \\
        \hline
        Workflow preparation time & $\approx$ 30 minutes \\
        \hline
        Experiment completion time & $\approx$ 1 day (on a compute cluster with 250 cores) \\
        \hline
                                          &  Benchmarks (\url{https://zenodo.org/record/14281771})  \\
        \makecell[l]{Publicly available?} &  Zenodo (\url{https://zenodo.org/record/14741186}) \\
                                          &  GitHub (\url{https://github.com/CMU-SAFARI/Chronus})  \\
        \hline
    \end{tabular}
    \label{tab:artifact_table}
\end{table}

\subsection{Description}
 
\noindent\emph{We highly recommend using Slurm with a cluster that can run experiments in bulk.}

\head{How To Access}
Source code and scripts are available at \url{https://github.com/CMU-SAFARI/Chronus}.

\head{Hardware Dependencies}
We recommend using a PC with 32 GiB of main memory.
Approximately 30 GiB of disk space is needed to store intermediate and final evaluation results.

\head{Software Dependencies}
\begin{itemize}
    \item GNU Make, CMake 3.20+
    \item C++20 build toolchain (tested with GCC 10)
    \item Python 3.9+
    \item pip: matplotlib, pandas, seaborn, pyyaml, wget, scipy
    \item Ubuntu 22.04
    \item (Optional) Slurm 20+
    \item (Optional) Podman 4.5+
\end{itemize}

\head{Benchmarks}
We use workload memory traces collected from SPEC2006, SPEC2017, TPC, MediaBench, and YCSB benchmark suites.
These traces are available at \url{https://zenodo.org/records/14281771}.
Install scripts will download and extract the traces.

\subsection{Installation}

\lstset{
    backgroundcolor=\color{gray!20}, 
    basicstyle=\ttfamily\bfseries\footnotesize,
    columns=fullflexible,
    frame=single,
    breaklines=true,
    postbreak=\mbox{\textcolor{red}{$\hookrightarrow$}\space},
    showstringspaces=false,
    numbersep=5pt,
    xleftmargin=6pt,
    xrightmargin=4pt,
    numbers=none,
    keywordstyle=\color{black},  
    identifierstyle=\color{black},  
    commentstyle=\color{black},  
    stringstyle=\color{black}  
}

The following steps will download and prepare the repository for the main experiments:
\begin{enumerate}
    \item Clone the git repository.
    \begin{lstlisting}[language=bash]
$ git clone \
git@github.com:CMU-SAFARI/Chronus.git
    \end{lstlisting}
    \item (Optional) Build the Podman container to run the scripts.
    \begin{lstlisting}[language=bash]
$ podman build . -t chronus_artifact
    \end{lstlisting}
    The following command runs a script using the container:
    \begin{lstlisting}[language=bash]
$ podman run --rm -v $PWD:/app \
    chronus_artifact <script>
    \end{lstlisting}
    \item Install Python dependencies, compile Ramulator2, download workload traces, and run a small sanity check.
    \begin{lstlisting}[language=bash]
$ ./run_simple_test.sh
    \end{lstlisting}
\end{enumerate}

\subsection{Evaluation and Expected Results}

\head{Claim 1 (C1)}
Latest industry solutions to read disturbance induce prohibitively large system performance overhead for both modern (i.e., \gls{nrh}$ = $1K) and future (i.e., \gls{nrh}$\leq$1K) DRAM chips.
This property is proven by evaluating the effect of state-of-the-art read disturbance mitigation mechanisms on system performance (E1) as described in \secref{sec:sensitivity} whose results are illustrated in \figref{fig:sensitivity_performance}.

\head{Claim 2 (C2)}
\X{} outperforms both 1) state-of-the-art industry solutions to read disturbance (i.e., \gls{prac}~\cite{jedec2024jesd795c} and \gls{prfm}~\cite{jedec2020jesd795}) and 2) state-of-the-art academic solutions to RowHammer (i.e., Graphene~\cite{park2020graphene}, Hydra~\cite{qureshi2022hydra}, and PARA~\cite{kim2014flipping}) in terms of system performance (\figsref{fig:benign_singlecore},~\ref{fig:benign_scaling}, and~\ref{fig:benign_workloads}), DRAM energy (\figref{fig:benign_energy}), and storage (\figref{fig:benign_storage}).
This property is proven by evaluating and comparing the impact of \X{} and prior state-of-the-art read disturbance mitigation mechanisms on system performance and DRAM energy on single-core and multi-core workloads (E2) as described in \secref{sec:methodology}.

\head{Experiments (E1 and E2)}
[Ramulator2 simulations]
[10 human-minutes + 20 compute-hours (assuming $\sim$500 Ramulator2 simulations run in parallel) + 30GiB disk]

We prove our claims in two steps:
1) Execute Ramulator2 simulations to generate data supporting C1 and C2;
2) Plot all figures that prove C1 and C2.

\begin{enumerate}
    \item Launch all Ramulator2 simulation jobs.\footnote{Slurm job partition and the maximum number of jobs are configurable via the \textit{AE\_SLURM\_PART\_NAME} variable in \texttt{./run\_with\_slurm.sh} and \textit{MAX\_SLURM\_JOBS} variable in \texttt{scripts/run\_config.py}, respectively.}
    \begin{lstlisting}[language=bash]
$ ./run_with_personalcomputer.sh
(or ./run_with_slurm.sh if Slurm is available)
    \end{lstlisting}
    \item Wait for the simulations to end. The following displays the status and generates scripts to restart failed runs:
    \begin{lstlisting}[language=bash]
$ ./check_run_status.sh
    \end{lstlisting}
    \item Parse simulation results and collect statistics.
    \begin{lstlisting}[language=bash]
$ ./parse_results.sh
    \end{lstlisting}
    \item Generate all figures that support C1 and C2.
    \begin{lstlisting}[language=bash]
$ ./plot_all_figures.sh
    \end{lstlisting}
\end{enumerate}

\subsection{Experiment Customization}
\label{sec:expcustom}
Our scripts provide easy configuration of the 1) evaluated read disturbance mitigation mechanisms, 2) tested read disturbance thresholds, 3) simulation duration, and 4) simulated workload combinations.
The run parameters are configurable in \texttt{scripts/run\_config.py} with 1) \textit{mitigation\_list}, 2) \textit{tRH\_list}, and 3) \textit{NUM\_EXPECTED\_INSTS} or \textit{NUM\_MAX\_CYCLES}, respectively.
Simulated single-core and multi-core workload combinations can be updated in \texttt{mixes/hpcasingle.mix} and \texttt{mixes/hpcabenign.mix}, respectively.

\subsection{Methodology}

Submission, reviewing and badging methodology:

\begin{itemize}
  \item \url{https://www.acm.org/publications/policies/artifact-review-and-badging-current}
  \item \url{https://cTuning.org/ae}
\end{itemize}

\section{Comparison to ABACuS}
\label{apx:abacuscomparison}

We compare \X{} to ABACuS~\cite{olgun2024abacus}, a storage-optimized deterministic mechanism that uses the Misra-Gries frequent item counting algorithm~\cite{misra1982finding} and maintains frequently accessed row counters completely within the memory controller.
ABACuS makes a key observation that \ous{9}{many workloads (both benign workloads and RowHammer attacks) tend to access DRAM rows with the same row address in multiple DRAM banks at \emph{around the same time} because i) modern memory address mapping schemes interleave consecutive cache blocks across different banks and ii) workloads tend to access cache blocks in close proximity around the same time due to the spatial locality in their memory accesses~\cite{olgun2024abacus}}.\omcomment{9}{Odd writing. Is this correct?}\ouscomment{9}{It is correct. But changed writing for better clarity and less ambiguity.}
ABACuS exploits this observation by maintaining a single counter across \om{9}{\emph{all}} banks instead of maintaining a counter \om{9}{\emph{per}} bank (e.g., ~\cite{park2020graphene,bostanci2024comet}).
By doing so, ABACuS accurately tracks activation counts at significantly reduced area overhead~\cite{olgun2024abacus}.
We implement ABACuS in Ramulator 2.0~\cite{ramulator2github, chronusgithub} and follow the same methodology in \secref{sec:methodology}.
We use the address mapping described in the ABACuS paper \om{9}{(see §9 of~\cite{olgun2024abacus})} instead of RoBaRaCoCh used in \secref{sec:evaluation}.

\begin{figure}[h]
\centering
\includegraphics[width=\linewidth]{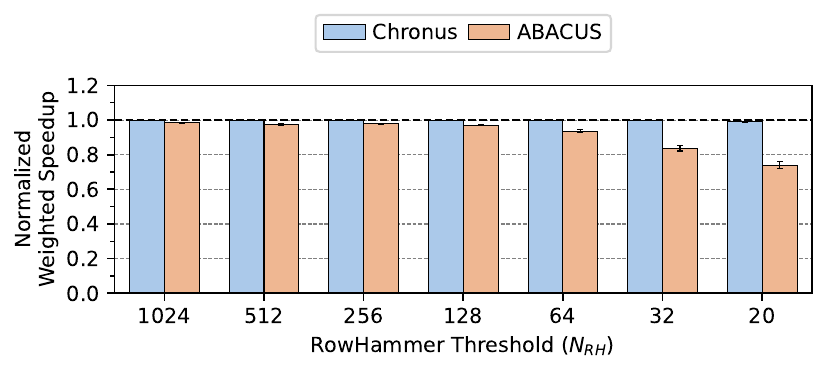}
\caption{Performance impact of \X{} and ABACuS on 60 benign four-core workloads}
\label{fig:abacusperf}
\end{figure}

\head{System Performance}
\figref{fig:abacusperf} presents the performance overheads of \X{} and ABACuS across \param{60} benign \param{four}-core workloads for \gls{nrh} values from \param{1K} to \param{20}.
x and y axes respectively show the \gls{nrh} values and system performance in terms of weighted speedup normalized to a baseline with \emph{no} read disturbance mitigation (higher y value is better).
Each colored bar depicts the mean system performance of a mechanism across 60 four-core workloads and error bars show the standard error of the mean across 60 four-core workloads.

We make \param{three} observations from \figref{fig:abacusperf}.
First, \X{} outperforms ABACuS across all evaluated \gls{nrh} values.
Second, as \gls{nrh} decreases from 1K to 20, ABACuS's performance overhead increases from \param{1.7}\% to \param{26.4}\%.
In contrast, \X{}'s performance overhead increases from \param{<0.1}\% to \om{9}{only} \param{3.2}\%.
Third, \X{}'s system performance overhead decreases with ABACuS's address mapping.
For example, at $\nrh{} =$ 20, \X{}'s performance overhead decreases from \param{17.9}\% (\ous{9}{shown in \figref{fig:benign_scaling}}) to \param{3.2}\% when the address mapping changes from RoBaRaCoCh to ABACuS's mapping.
This is likely due to ABACuS's mapping \om{9}{leading to a lower row conflict rate} by better interleaving rows across banks.\footnote{\ous{9}{Reduced row conflict rate in ABACuS's mapping increases baseline system performance. When a read disturbance mitigation mechanism is present, the benefits can be further improved because DRAM rows are activated fewer times on average, and thus the mitigation mechanism can perform fewer preventive refreshes. We do \emph{not} use ABACuS's mapping in our main evaluation (\secref{sec:evaluation}) because each mapping differently affects different read disturbance mechanisms. For example, Hydra's~\cite{qureshi2022hydra} system performance overhead significantly (\param{>10}\%) increases with ABACuS's mapping compared to RoBaRaCoCh (not shown). We leave a rigorous evaluation of the memory address mapping's effect on different read disturbance mitigation mechanisms to future work.}}

\begin{figure}[h]
\centering
\includegraphics[width=\linewidth]{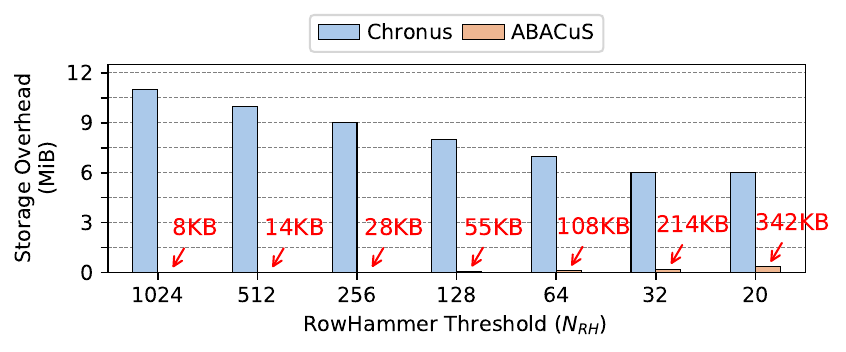}
\caption{Storage used by \X{} (DRAM) and \ous{9}{ABACuS (CAM+SRAM in CPU)} as a function of RowHammer threshold for a DRAM module with \param{64} banks and \param{128K} rows per bank}
\label{fig:abacusstorage}
\end{figure}

\head{Storage}
\figref{fig:abacusstorage} shows the storage overhead in MiB (y axis) of the evaluated read disturbance mitigation mechanisms as \gls{nrh} decreases (x axis).
We evaluate the storage usage of \ous{9}{\X{} (DRAM) and ABACuS (CAM+SRAM in CPU)} as a function of \gls{nrh} for a DRAM module with \param{64} banks and \param{128K} rows per bank.

\ous{9}{We make \param{two} observations from \figref{fig:abacusstorage}.
First, ABACuS's storage overhead in CPU is significantly lower compared to \X{}'s storage overhead in DRAM.
However, \X{} overhead is completely within the DRAM chip where the per-bit storage hardware complexity is relatively low.
Second, as \gls{nrh} decreases from \param{1K} to \param{20}, ABACuS’s storage overhead in CPU increases from 8KB to 324KB.
This is because as \gls{nrh} decreases, ABACuS needs to track many more rows and thereby requires more counters (implemented as CAM and SRAM)}.
\section{Worst-Case Access Pattern Analysis}
\label{apx:adversarialproof}

We prove that the access pattern used in \secref{sec:evaluation_dos} yields the maximum theoretical DRAM bandwidth consumption \ous{9}{of preventive refreshes} in a \X{}- or \gls{prac}-protected system.

\head{Properties}
We build our proof on \param{three} properties of \gls{prac} and \X{}:
$P1$)~a row's activation count increases \om{9}{by one} when the row is opened and closed;
$P2$)~a back-off is triggered when a row's activation count reaches \gls{aboth};
$P3$)~a back-off refreshes $i$)~$\bonrefs{}$ rows with the highest activation count in \gls{prac} and $ii$)~all rows that exceed \gls{aboth} in \X{} for each bank.

\head{Maximum Bandwidth Consumption \ous{9}{of Preventive Refreshes}}
\agy{9}{For a given access pattern, we calculate the fraction of DRAM bandwidth \ous{9}{consumed} for performing preventive refreshes.
Our \ous{9}{\emph{DRAM Bandwidth Consumption} (\dbc{})} function is formally defined as $\dbc{}: P \rightarrow [0, 1]$, where $P$ denotes the set of all possible access patterns and $[0,1]$ is the fraction of DRAM bandwidth consumed by preventive refreshes.
For example, the worst-case access pattern evaluated in \secref{sec:evaluation_dos} triggers a preventive refresh operation $\bonrefs{}$ times and each refresh operation takes $\trfm{}$ (i.e., $\bonrefs{}\times\trfm{}$).
To trigger a refresh, this access pattern includes \gls{aboth} row activations, each of which takes at least $\trc{}$ (\ous{9}{i.e.}, $\aboth{}\times{}\trc{}$)}.
As such, the adversarial pattern evaluated in \secref{sec:evaluation_dos} (i.e., $\advanal{}$) has the DRAM bandwidth consumption presented in Expression~\ref{eqn:dosanal}.
\begin{equation}
\label{eqn:dosanal}
\dbc{}(\advanal{})
=
\frac{
(\bonrefs{}\times{}\trfm{})
}{
(\bonrefs{}\times{}\trfm{})+(\aboth{}\times{}\trc{})
}
\end{equation}

\head{Proof Overview}
\ous{9}{We prove that $\advanal{}$ yields the maximum DRAM bandwidth consumption in \param{four} steps \agy{9}{using proof-by-contradiction}~\cite{rosen2019discrete}.
First, we assume that there exists an adversarial pattern $\advhyp{}$ (i.e., hypothetical adversarial pattern) that yields a greater DRAM bandwidth consumption than $\advanal{}$ within an arbitrary time window $T$.\footnote{We choose an arbitrary $T$ with no assumptions. Therefore, the following steps of the proof apply for all $T$ by the \emph{Universal Introduction} rule~\cite{enderton2001mathematical}.}
Second, we calculate the number of back-offs triggered by $\advanal{}$ within $T$ as a lower bound for $\advhyp{}$.
Third, we calculate the time needed to perform the preventive refreshes of back-offs caused by $\advhyp{}$.
Fourth, we show that time remaining after $\advhyp{}$'s preventive refreshes \emph{cannot} be used to trigger $\advhyp{}$'s back-offs within $T$.
Therefore, a pattern that yields greater DRAM bandwidth availability \emph{cannot} exist within the constraints of \gls{prac} and \X{}}.

\head{Step 1: Adversarial memory access pattern}
Assume there exists an adversarial access pattern $\advhyp{}$ that yields a greater DRAM bandwidth consumption than $\advanal{}$ within an arbitrary time window $T$ (i.e., $\dbc{}(\advhyp{}) > \dbc{}(\advanal{})$).
Since $\advhyp{}$ consumes more DRAM bandwidth than $\advanal{}$, $\advhyp{}$ must trigger at least one more back-off than $\advanal{}$.
\ous{9}{For a given access pattern $P$, we calculate the \emph{Back-Offs Triggered} (\anbot{}) within a time window of length $T$.
To do so, we find the DRAM bandwidth consumed by access pattern $P$ (i.e., $T \times{} \dbc{}(P)$) and divide it by the duration of a back-off (i.e., $\bonrefs{}\times{}\trfm{}$) as shown in Expression~\ref{eqn:botdef}}.
\begin{equation}
\label{eqn:botdef}
\begin{aligned}
\anbot{}(P, T) = T \times{} \dbc{}(P) / (\bonrefs{}\times{}\trfm{})
\end{aligned}
\end{equation}

\head{Step 2: Lower bound of the number of back-offs}
\agy{9}{Based on the assumption in Step~1, $\anbot{}(\advhyp{}, T)$ should be larger than $\anbot{}(\advanal{}, T)$.}
\agy{9}{By placing Expressions~\ref{eqn:dosanal} and~\ref{eqn:botdef} in the restriction $\anbot{}(\advhyp{}, T) > \anbot{}(\advanal{}, T)$, we \om{9}{derive} Expression~\ref{eqn:hypbotlowerbound}.}
\begin{equation}
\label{eqn:hypbotlowerbound}
\begin{aligned}
\anbot{}(\advhyp{}, T) > \frac{ T }{ \bonrefs{}\times{}\trfm{}+\aboth{}\times{}\trc{} }
\end{aligned}
\end{equation} 


\head{Step 3: Time taken to perform preventive refreshes}
\ous{9}{By knowing the lower bound for the number of back-offs triggered by $\advhyp{}$, we calculate} the time taken to perform the preventive refreshes of $\advhyp{}$'s back-offs (i.e., $\textit{PR}_{HYP}$) by multiplying the number of back-offs (i.e., $\anbot{}(\advhyp{}, T)$) with the duration of a back-off (i.e., $\bonrefs{}\times{}\trfm{}$) \ous{9}{in Expression~\ref{eqn:timetorefresh}}.
\begin{equation}
\label{eqn:timetorefresh}
\begin{aligned}
\textit{PR}_{HYP} & > \frac{T\times{}\bonrefs{}\times{}\trfm{}}{\bonrefs{}\times{}\trfm{} + \aboth{}\times{}\trc{}}
\end{aligned}
\end{equation}

\head{Step 4: Remaining time after the preventive refreshes}
\ous{9}{The time taken to both trigger and perform the preventive refreshes of $\advhyp{}$ should fit within $T$}.
We \agy{9}{calculate} the remaining time after performing the preventive refreshes of $\advhyp{}$'s back-offs (i.e., $\textit{RT}_{HYP}$).
\ous{9}{Expression~\ref{eqn:timetorefresh} provides a lower bound for $\textit{PR}_{HYP}$.
Therefore, subtracting $\textit{PR}_{HYP}$ from $T$ yields an upper bound for $\textit{RT}_{HYP}$ \agy{9}{(i.e., $\textit{RT}_{HYP} < T - \textit{PR}_{HYP}$).
Solving this expression for $\textit{RT}_{HYP}$, we} \ous{9}{derive} Expression~\ref{eqn:remainingtime}}.\agycomment{9}{Exp. 6 gives a lower bound for PRHYP, but we need the value of PRHYP in Exp 7.}
\begin{equation}
\label{eqn:remainingtime}
\begin{aligned}
\textit{RT}_{HYP}<\frac{T\times{}\aboth{}\times{}\trc{}}{\bonrefs{}\times{}\trfm{} + \aboth{}\times{}\trc{}}
\end{aligned}
\end{equation}


Triggering a back-off with a single bank takes $\aboth{}\times{}\trc{}$ (following from $P1$ and $P2$) and concurrent aggressors \emph{cannot} be used to trigger a back-off more quickly (following from $P3$).
Given $P1$, $P2$, and $P3$, \ous{9}{we calculate} time needed to trigger $\advhyp{}$'s back-offs in a \gls{prac} or \X{} protected system ($\textit{TBO}_{HYP}$).
\ous{9}{Expression~\ref{eqn:hypbotlowerbound} presents a lower bound for $\anbot{}(\advhyp{}, T)$.
Multiplying $\anbot{}(\advhyp{}, T)$ with the time taken to trigger a single back-off yields a lower bound for $\textit{TBO}_{HYP}$ (i.e.,~$\textit{TBO}_{HYP}>\anbot{}(\advhyp{}, T)\times{}\aboth{}\times{}\trc{}$)}.
Solving this expression for $\textit{TBO}_{HYP}$, we evaluate Expression~\ref{eqn:timetotrigger}.
\begin{equation}
\label{eqn:timetotrigger}
\begin{aligned}
\textit{TBO}_{HYP} & > \frac{ T\times{}\aboth{}\times{}\trc{} }{ \bonrefs{}\times{}\trfm{}+\aboth{}\times{}\trc{} }
\end{aligned}
\end{equation}

\ous{9}{By comparing Expressions~\ref{eqn:remainingtime} and~\ref{eqn:timetotrigger}}, we see that the time necessary to trigger $\advhyp{}$'s preventive refreshes ($\textit{TBO}_{HYP}$) exceeds the time remaining after $\advhyp{}$'s preventive refreshes are performed ($\textit{RT}_{HYP}$), i.e, $\textit{RT}_{HYP} < \textit{TBO}_{HYP}$.
This means that \emph{no} $\advhyp{}$ exists that obey the three properties of \gls{prac} and \X{} (i.e., $P1$, $P2$, and $P3$).
Therefore, $\advanal{}$ \agy{9}{(as used in \secref{sec:evaluation_dos})} yields the maximum DRAM bandwidth consumption, \agy{9}{and thus represents the worst possible access pattern}.
\section{Errata \om{12}{and New Results}}
\label{apx:errata}

A new study~\cite{kim2025per} pointed out a bug where the $\tras{}$, $\trtp{}$, and $\twr{}$ DRAM timing parameters for \om{14}{\gls{prac}-enabled} systems were \emph{not} being reduced correctly in the evaluations of our paper's previous version~\om{14}{\cite{canpolat2025chronus}}.
We thank the authors of~\cite{kim2025per} for providing us with the feedback.
We fixed the bug, and \ous{12}{updated all affected results and figures in this version of our paper}.
\om{13}{All results presented in this paper are collected with the bug fix included}.
We \om{13}{also} updated \X{}' open-source repository~\cite{chronusgithub} with the bug fix.

\om{13}{We now evaluate the effect the bug fix had on the performance and energy overheads of \gls{prac}}.
Table~\ref{tab:praccorrection} presents \om{14}{how the bug has changed the system performance and DRAM energy overheads of \gls{prac} in} our evaluations (see~\secref{sec:sensitivity} and~\secref{sec:evaluation} \ous{12}{for \om{14}{all} results with the bug fix}).
\ous{12}{Columns labeled as \om{13}{\emph{Old}} and \om{13}{\emph{New}} respectively present 1) the results with the bug in the previous version \om{14}{of the paper}~\ous{14}{\cite{canpolat2025chronus}} and 2) the results after fixing the bug in this \om{14}{current} version of the paper.}
We report that the difference in \gls{prac}'s results \emph{after} the bug fix is \emph{not} \emph{significant enough} to change \emph{any} of our claims or conclusions in this work (or the previous version~\cite{canpolat2025chronus}).

\begin{table}[ht]
    \centering
    \footnotesize
    \vspace{1em}
    \caption{Changes \ous{12}{in} Four-Core Results \om{12}{for \gls{prac} Performance and Energy Overheads} After Fixing \ous{12}{the} Bug. \ous{12}{Old: Results with the bug (previous version of the paper~\om{14}{\cite{canpolat2025chronus}}). New: Results after fixing the bug (this \om{14}{current} version of the paper).}}
    \setlength{\tabcolsep}{3pt}
    \begin{tabular}{l||cc||cc||cc}
    & \multicolumn{2}{c||}{\makecell[b]{Single-Core Perf. \\ Overhead}}
    & \multicolumn{2}{c||}{\makecell[b]{Four-Core Perf. \\ Overhead}}
    & \multicolumn{2}{c}{\makecell[b]{Four-Core Energy \\ Overhead}} \\
    \hline
    \makecell[b]{$\nrh{}$} & \om{13}{Old} & \om{13}{New} & \om{13}{Old} & \om{13}{New} & \om{13}{Old} & \om{13}{New} \\
    \hline\hline
    1K  & 22.0\% & 14.0\% & 9.7\% & 5.8\% & 18.4\% & 10.7\% \\
    \hline
    512  & 22.0\% & 14.0\% & 9.7\% & 5.8\%  & 18.4\% & 10.7\% \\ 
    \hline
    256  & 22.1\% & 14.1\% & 9.7\% & 5.9\%  & 18.5\% & 10.8\% \\ 
    \hline
    128  & 22.3\% & 14.4\% & 9.9\% & 6.1\%  & 18.6\% & 10.9\% \\ 
    \hline
    64  & 26.2\% & 19.1\% & 11.7\% & 8.1\%  & 20.4\% & 12.8\% \\ 
    \hline
    32  & 46.1\% & 42.5\% & 30.4\% & 29.1\%  & 48.0\% & 40.8\% \\ 
    \hline
    20  & 90.9\% & 90.3\% & 81.2\% & 78.5\%  & 6.9x & 6.6x \\ 
    \hline
    \end{tabular}
    \label{tab:praccorrection}
\end{table}

\ous{11}{From Table~\ref{tab:praccorrection}, we observe that the bug caused a smaller overhead difference at low $\nrh{}$ values (e.g., $<$128) compared to high $\nrh{}$ values (e.g., $=$1K).
\agy{11}{This is because \gls{prac} causes more back-offs as \gls{nrh} decreases and those back-offs dominate \gls{prac}'s performance overhead. As a result, the effect of increased precharge and row cycle latencies on performance degradation become less significant.}}

The \om{13}{mentioned recent}\ouscomment{13}{I may have misread this part} study~\cite{kim2025per} further argues that \gls{prac}'s system performance overhead on real hardware is lower than what our simulations show.
We identify two key methodological differences between our study and~\cite{kim2025per} that cause the lower \gls{prac} performance impact.
First, \cite{kim2025per} evaluates a relatively small set of homogeneous workloads (i.e., only 23 SPEC 2017 workloads with mostly low memory-intensity) compared to our 60 heterogeneous workloads with varying memory intensities chosen from five major benchmark suites (SPEC CPU2006~\cite{spec2006}, SPEC CPU2017~\cite{spec2017}, TPC~\cite{tpc}, MediaBench~\cite{fritts2009media}, and YCSB~\cite{ycsb}, see~\secref{sec:sensitivity}).
Second, \cite{kim2025per} evaluates systems with much larger caches.
For example, the last-level caches of systems in \cite{kim2025per} is 4.5x larger than our simulated system (see~\secref{sec:sensitivity}).
The increased cache size makes SPEC 2017 workloads mostly cache-resident~\cite{singh2019memory}.
Therefore, \gls{prac} does \emph{not} induce performance overhead on a system that is \emph{not} bottlenecked by memory.
In contrast, our simulation methodology implements smaller caches as \om{12}{a} best-effort to account for memory-bottlenecked systems \om{12}{which is} standard practice in literature (see, \om{13}{e.g.},~\ous{13}{\cite{olgun2024abacus, bostanci2024comet, qureshi2022hydra, qureshi2025autorfm, saileshwar2022randomized, vittal2025mopac, park2020graphene, lee2019twice, kim2020revisiting, yaglikci2021blockhammer, hassan2019crow, marazzi2022protrr, yaglikci2022hira, saxena2022aqua, marazzi2023rega}}), using publicly available well-established benchmark suites.

\begin{figure}[h]
\centering
\includegraphics[width=\linewidth]{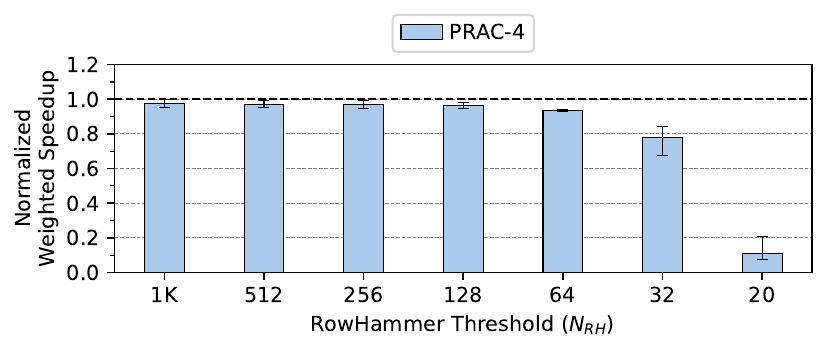}
\caption{Performance impact of \gls{prac} on 23 eight-core homogeneous SPEC 2017 workloads, \om{13}{normalized to} \ous{13}{a baseline with \emph{no} read disturbance mitigation}}
\label{fig:errata_benign}
\end{figure}

We \om{14}{now} experimentally demonstrate that these two methodological choices \agy{11}{lead to the discrepancy in \gls{prac}'s performance overheads \om{12}{between} our work and~\cite{kim2025per}.
To do so, we repeat our simulation-based evaluation using the \om{13}{\emph{same}} system configuration and \om{13}{\emph{same}} workloads used in \cite{kim2025per}.}
\ous{11}{\figref{fig:errata_benign} presents the performance overhead of \gls{prac} across \param{23} \param{eight}-core homogeneous SPEC 2017 workloads for \gls{nrh} values from \param{1K} to \param{20} (\om{12}{these are the} same 23 workloads evaluated in~\cite{kim2025per}).
x and y axes respectively show the \gls{nrh} values and performance in terms of weighted speedup normalized to a baseline with \emph{no} read disturbance mitigation (higher y value is better).
Each bar depicts the geomean system performance of \om{12}{\gls{prac}-4} across 23 eight-core workloads and error bars show the 100\% confidence interval across 23 eight-core workloads.}

\ous{11}{We make two observations from \figref{fig:errata_benign}.
First, at $\nrh{} = 1K$, \ous{12}{\gls{prac}-4} induces an average (maximum) system performance overhead of \ous{13}{2.4\% (4.6\%)}, \agy{11}{which} is \agy{11}{similar} to the average (maximum) numbers of 1.3\% (3.8\%) reported in~\cite{kim2025per} \om{12}{for a similarly configured real system}.
Second, when \gls{nrh} decreases from \param{1K} to \param{20}, \ous{12}{\gls{prac}-4}'s average (maximum) system performance overhead increases to 78.8\% (85.9\%).
Kim et al.'s real system-based evaluation methodology limits their tests to \emph{only} an \gls{nrh} value of 1K. Therefore, we \emph{cannot} compare our simulation results for sub-1K \gls{nrh} values against~\cite{kim2025per}.}

We conclude that the \agy{11}{discrepancy in \gls{prac}'s} performance impact \agy{11}{across} \ous{11}{our work} and~\cite{kim2025per} is \om{12}{very likely} an artifact of~\om{12}{\cite{kim2025per}} adopting \om{13}{a} different system configuration and \om{13}{different} workloads, as opposed to simulation-based evaluations that ``\emph{\om{14}{often} fall short of accurately modeling the performance characteristics of real-world memory systems}'' (as claimed \om{13}{very generally}\omcomment{14}{correct?}\ouscomment{14}{Yes, I think anything other than ``simulations are harder to get right when real system of what you are simulating exist'' is a far fetched claim. Also, their real system evaluation requires them putting in PRAC timing parameters by hand anyways (because their system does not really have PRAC), so they could very well mess up the same DRAM timing parameters. I don't see how simulators fall short from their methodology GIVEN the fixed bug in my code} by~\cite{kim2025per} without providing \om{14}{commensurate} evidence).

\begin{figure}[h]
\centering
\includegraphics[width=\linewidth]{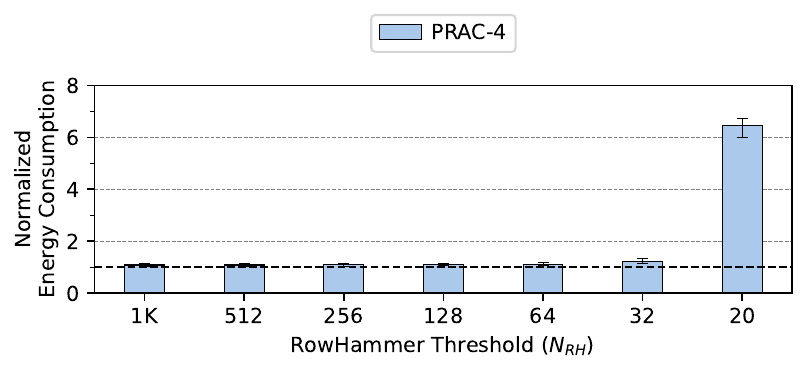}
\caption{DRAM energy impact of \gls{prac} on 23 eight-core homogeneous SPEC 2017 workloads, normalized to a baseline with \emph{no} read disturbance mitigation}
\label{fig:errata_energy}
\end{figure}

\ous{14}{We also evaluate the DRAM energy overhead of \gls{prac} using the DRAMPower~\cite{drampower} integration of Ramulator 2.0 under the \emph{same} system configuration and \emph{same} workloads used in \cite{kim2025per}.\omcomment{14}{I assume~\cite{kim2025per} does not have energy evaluation?}\ouscomment{14}{Correct. They dont have energy evaluation.}
\figref{fig:errata_energy} presents the energy consumption of \gls{prac} across \param{23} \param{eight}-core homogeneous SPEC 2017 workloads for \gls{nrh} values from \param{1K} to \param{20}.
x and y axes respectively show the \gls{nrh} values and energy consumption normalized to a baseline with \emph{no} read disturbance mitigation (lower y value is better).
Each bar depicts the geomean energy consumption of \gls{prac}-4 across 23 eight-core workloads and error bars show the 100\% confidence interval across 23 eight-core workloads.}

\ous{14}{We make two observations from \figref{fig:errata_energy}.
First, at $\nrh{} = 1K$, \gls{prac}-4 induces an average (maximum) energy consumption overhead of 8.9\% (14.1\%).
Second, when \gls{nrh} decreases from \param{1K} to \param{20}, \gls{prac}-4's average (maximum) energy consumption overhead increases to 5.46x (5.72x).
We \emph{cannot} compare our simulation results values against~\cite{kim2025per} because Kim et al.'s real system-based study does \emph{not} include energy evaluations.}

\ous{14}{We conclude that \gls{prac}'s energy overhead is 1) non-negligible ($\geq$8\%) at high \gls{nrh} values (i.e., $\nrh{} =$ 1K) and 2) prohibitively large ($\geq$5x) at low \gls{nrh} values (i.e., $\nrh{} <$ 32) even under system configuration and workload combinations that observe relatively lower \gls{prac} system performance overhead (e.g.,~\cite{kim2025per}).}


\end{document}